\documentclass[12pt,english]{article}
\usepackage{times}
\usepackage[T1]{fontenc}
\usepackage{geometry}
\usepackage{rotating}
\usepackage{graphicx}
\usepackage{setspace}
\usepackage{amssymb}

\makeatletter

\providecommand{\tabularnewline}{\\}

\usepackage{bm}

\usepackage{babel}
\makeatother
\begin{document}

\section*{\noindent Features and Potentialities of Static Passive \emph{EM}
Skins for \emph{NLOS} Specular Wireless Links}

\noindent ~

\noindent \vfill

\noindent G. Oliveri,$^{(1)(2)}$, M. Salucci,$^{(1)(2)}$, and A.
Massa,$^{(1)(2)(3)(4)}$ 

\noindent \vfill

\noindent ~

\noindent {\footnotesize $^{(1)}$} \emph{\footnotesize ELEDIA Research
Center} {\footnotesize (}\emph{\footnotesize ELEDIA}{\footnotesize @}\emph{\footnotesize UniTN}
{\footnotesize - University of Trento)}{\footnotesize \par}

\noindent {\footnotesize DICAM - Department of Civil, Environmental,
and Mechanical Engineering}{\footnotesize \par}

\noindent {\footnotesize Via Mesiano 77, 38123 Trento - Italy}{\footnotesize \par}

\noindent \textit{\emph{\footnotesize E-mail:}} {\footnotesize \{}\emph{\footnotesize giacomo.oliveri,
marco.salucci, andrea.massa}{\footnotesize \}@}\emph{\footnotesize unitn.it}{\footnotesize \par}

\noindent {\footnotesize Website:} \emph{\footnotesize www.eledia.Arg/eledia-unitn}{\footnotesize \par}

\noindent {\footnotesize ~}{\footnotesize \par}

\noindent {\footnotesize $^{(2)}$} \emph{\footnotesize CNIT - \char`\"{}University
of Trento\char`\"{} ELEDIA Research Unit }{\footnotesize \par}

\noindent {\footnotesize Via Sommarive 9, 38123 Trento - Italy}{\footnotesize \par}

\noindent {\footnotesize Website:} \emph{\footnotesize www.eledia.org/eledia-unitn}{\footnotesize \par}

\noindent {\footnotesize ~}{\footnotesize \par}

\noindent {\footnotesize $^{(3)}$} \emph{\footnotesize ELEDIA Research
Center} {\footnotesize (}\emph{\footnotesize ELEDIA}{\footnotesize @}\emph{\footnotesize UESTC}
{\footnotesize - UESTC)}{\footnotesize \par}

\noindent {\footnotesize School of Electronic Engineering, Chengdu
611731 - China}{\footnotesize \par}

\noindent \textit{\emph{\footnotesize E-mail:}} \emph{\footnotesize andrea.massa@uestc.edu.cn}{\footnotesize \par}

\noindent {\footnotesize Website:} \emph{\footnotesize www.eledia.org/eledia}{\footnotesize -}\emph{\footnotesize uestc}{\footnotesize \par}

\noindent {\footnotesize ~}{\footnotesize \par}

\noindent {\footnotesize $^{(4)}$} \emph{\footnotesize ELEDIA Research
Center} {\footnotesize (}\emph{\footnotesize ELEDIA@TSINGHUA} {\footnotesize -
Tsinghua University)}{\footnotesize \par}

\noindent {\footnotesize 30 Shuangqing Rd, 100084 Haidian, Beijing
- China}{\footnotesize \par}

\noindent {\footnotesize E-mail: \{}\emph{\footnotesize andrea.massa}{\footnotesize \}}\emph{\footnotesize @tsinghua.edu.cn}{\footnotesize \par}

\noindent {\footnotesize Website:} \emph{\footnotesize www.eledia.org/eledia-tsinghua}{\footnotesize \par}

\noindent \vfill

\noindent \textbf{\emph{This work has been submitted to the IEEE for
possible publication. Copyright may be transferred without notice,
after which this version may no longer be accessible.}}

\noindent \vfill

\newpage
\section*{Features and Potentialities of Static Passive \emph{EM} Skins for
\emph{NLOS} Specular Wireless Links}

~

~

~

\begin{flushleft}G. Oliveri, M. Salucci, and A. Massa\end{flushleft}

\vfill

\begin{abstract}
\noindent The ability of passive flat patterned electromagnetic skins
(\emph{EMS}s) to overcome the asymptotic limit of the total path attenuation
(\emph{TPA}) of flat metallic reflectors of arbitrary size in non-line-of-sight
(\emph{NLOS}) specular wireless links is assessed. Closed-form expressions
for the achievable \emph{TPA} in \emph{EMS}-powered \emph{NLOS} links
as well as the condition on the panel size of \emph{EMS}-screens to
improve the performance of flat passive conductive screens (\emph{PCS}s)
with the same aperture are derived and numerically validated by considering
different incidence angles, screen apertures, transmitter/receiver
distances, antenna gains, meta-atom geometries, and carrier frequencies.

\vfill
\end{abstract}
\noindent \textbf{Key words}: Static Passive \emph{EM} Skins; Metamaterials;
Inverse Scattering; Inverse Problems; Smart Electromagnetic Environment;
\emph{EM} Holography; Fixed Wireless Access; Point-to-Point Wireless
Link; Next-Generation Communications.

\newpage
\section{Introduction and Motivation\label{sec:Introduction}}

\noindent Wireless systems exploiting carriers at mmWave frequencies
and beyond are becoming more and more popular for both fixed wireless
access (e.g., \emph{802.11ay}) \cite{Chen 2019}\cite{Alubaikhy 2020}
and mobile applications (e.g., \emph{5G}, \emph{B5G}, \emph{6G}) in
outdoor \cite{Guo 2021} as well as indoor \cite{Kazim 2021} scenarios.
Indeed, increasing the carrier frequency allows one to exploit wider
portions of the spectrum to deliver high-speed broadband connections,
while avoiding complex and costly cabled infrastructures \cite{Chen 2019}-\cite{Kazim 2021}.
Unfortunately, the path loss and the attenuation experienced at higher
and higher frequencies become increasingly severe especially in non-line-of-sight
(\emph{NLOS}) conditions \cite{Du 2020}-\cite{Khawaja 2020} and
suitable countermeasures \cite{Khawaja 2020}-\cite{Zabihi 2021}
are mandatory to guarantee a suitable quality-of-service (\emph{QoS})
at the users by recovering adequate signal levels even on large distances.
Towards this end, one option is the exploitation of reflections from
suitably located flat passive conducting screens (\emph{PCS}s) \cite{Khawaja 2020}\cite{Norton 1962}.
The possibility of establishing reliable wireless links in \emph{NLOS}
conditions thanks such a technology is well known since decades \cite{Norton 1962}\cite{Ryerson 1960}.
As a matter of fact, \emph{PCS}s have been successfully used for telemetry
and voice communications \cite{Microflect 1989} and, more recently,
to enhance the indoor and the outdoor \emph{5G} coverage \cite{Khawaja 2020}\cite{Anjinappa 2021}.
The working principle of \emph{PCS}s is that of re-directing the incident
power towards the receiver through a specular reflection according
to the first Snell's law {[}Fig. 1(\emph{a}){]} \cite{Khawaja 2020}\cite{Norton 1962}\cite{Microflect 1989}\cite{Anjinappa 2021}.
Therefore, the overall attenuation {[}namely, the \emph{Total} \emph{Path
Attenuation} (\emph{TPA}){]} experienced by the \emph{EM} wave that
travels from the transmitter to the receiver, thanks to the \emph{PCS}
reflection, depends on the size of the same screen \cite{Khawaja 2020}\cite{Norton 1962}\cite{Microflect 1989}\cite{Anjinappa 2021}.
Unfortunately, the \emph{TPA} cannot be arbitrarily reduced by increasing
the \emph{PCS} aperture \cite{Khawaja 2020} and its value is known
to converge to the free space loss of a virtual transmitter-receiver
\emph{LOS} link with the same overall length {[}Fig. 1(\emph{a}){]}
\cite{Khawaja 2020}\cite{Balanis 2012} in the asymptotic case of
a \emph{PEC} screen with infinite extension. This limit, which is
as a consequence of the classical theory of images \cite{Balanis 2012},
implies that \emph{PCS}s cannot be used to overcome the free-space
attenuation in point-to-point links, but they only allow to bypass
obstacles {[}Fig. 1(\emph{a}){]}.

\noindent Flat electromagnetic skins (\emph{EMS}s) \cite{Massa 2021}-\cite{Oliveri 2022c}
may be profitably considered to address such a limitation by deriving
new practical guidelines for \emph{NLOS} point-to-point wireless communications.
As a matter of fact, \emph{EMS}s are a class of inexpensive static
passive devices that recently have enabled an advanced control of
the wireless propagation \cite{Oliveri 2021c}-\cite{Oliveri 2022c}
in several instances of the so-called \emph{Smart ElectroMagnetic
Environment} (\emph{SEME}) \cite{Massa 2021}\cite{Di Renzo 2019}-\cite{Barbuto 2022}.
Towards this end, the \emph{EMS} micro-scale physical descriptors
have been set to guarantee that the macro-scale performance of the
device at hand comply with the user-defined requirements \cite{Oliveri 2021c}\cite{Oliveri 2022}.
Up to now (i.e., current \emph{EMS} applications), anomalous reflections
towards non-Snell angular directions have been yielded thanks to \emph{EMS}s
to generate collimated \cite{Rocca 2022}\cite{Benoni 2022} or contoured
footprints \cite{Oliveri 2021c}\cite{Oliveri 2022}. On the other
hand, there are not theoretical motivations that prevent the design
of \emph{EMS}s that focus the reflected power along the standard Snell's
specular direction.

\noindent According to this line of reasoning, this paper proposes
a technological solution based on \emph{EMS} specular screens to overcome
the asymptotic limit of the \emph{TPA} of \emph{PCS}s. By using the
synthesis method presented in \cite{Oliveri 2022c}, \emph{EMS}s are
designed to perform better than \emph{PCS}s in terms of \emph{TPA}
and an upper bound for the achievable \emph{TPA} is derived. Moreover,
user-oriented guidelines are also drawn from a numerical assessment
concerned with different (i.e., incidence directions, screen apertures,
and transmitter/receiver distances) \emph{NLOS} point-to-point propagation
scenarios.

\noindent To the best of the authors' knowledge, the main methodological
innovations of the proposed research work with respect to the state-of-the-art
include

\begin{itemize}
\item a proof of the possibility to overcome with \emph{EMS}s the asymptotic
\emph{TPA} limit of standard metallic screens;
\item the derivation of the analytical expression for the upper bound of
the \emph{TPA} achievable when using \emph{EMS}-based screens as a
function of the screen aperture, the transmitter/receiver features,
and the \emph{NLOS} link setup; 
\item the derivation of operative guidelines on the design of \emph{EMS}s
that improve, in terms of \emph{NLOS} power transfer efficiency, the
performance of standard metallic panels of the same size.
\end{itemize}
\noindent The outline of the paper is as follows. Starting from the
formulation of the problem of a wireless \emph{NLOS} specular link,
the values of the \emph{TPA} theoretically achievable by \emph{PCS}s
and \emph{EMS}s are derived (Sect. \ref{sec:Problem-Formulation}).
A set of numerical results from an exhaustive numerical study is then
presented to assess the performance of the synthesized \emph{EMS}s
in comparison with those yielded with standard \emph{PCS}s under the
same operative conditions (Sect. \ref{sec:Results}). Finally, some
conclusions are drawn (Sect. \ref{sec:Conclusions-and-Remarks}).

\section{\noindent Problem Formulation\label{sec:Problem-Formulation}}

\noindent Let us consider the propagation scenario in Fig. 1(\emph{a})
where a pair of transmitting and receiving antennas, centered at $\mathbf{r}_{TX}\triangleq\left\{ r_{TX},\theta_{TX},\varphi_{TX}\right\} $
and $\mathbf{r}_{RX}\triangleq\left\{ r_{RX},\theta_{RX},\varphi_{RX}\right\} $,
respectively, operates in \emph{NLOS} conditions, the direct path
between them being obstructed by obstacles. Both antennas are oriented
so that the direction of their maximum gain (i.e., $G_{TX}$ and $G_{RX}$)
is towards a square ($L$ being the side) flat passive reflecting
screen centered in the origin of the global system of coordinates
$\left(x,y,z\right)$ {[}Fig. 1(\emph{a}){]}.

\noindent Without any other surrounding obstacle in between the transmitter/receiver
and the screen, a wireless link between the two antennas can be established
thanks to the reflection by such a screen (i.e., $\theta_{TX}=\theta_{RX}=\theta_{0}$).
This ideal assumption, which is commonly adopted in standard fixed
wireless link engineering if the first Fresnel zone clearance condition
holds true \cite{Xia 1993}\cite{Ruthroff 1971}, allows one to assess
the effect of the reflecting screen on the \emph{NLOS} propagation
link.

\noindent Under these hypotheses and considering a time-harmonic dependence
on the working frequency $f$, the problem of designing a passive
flat screen to establish a \emph{NLOS} wireless link between the transmitter
and the receiver can be formulated as that of finding the set $\mathbf{d}$
of $S$ screen descriptors, $\mathbf{d}\triangleq\left\{ d_{s};\, s=0,...,S-1\right\} $,
so that the power at the receiver (i.e., $\mathbf{r}=\mathbf{r}_{RX}$)
\cite{Balanis 2012}\begin{equation}
\mathcal{P}_{RX}\left(\mathbf{r}_{RX};\,\mathbf{d}\right)=\frac{\lambda^{2}G_{RX}\left|\mathbf{E}^{sca}\left(\mathbf{r}_{RX};\,\mathbf{d}\right)\right|^{2}}{8\pi\eta}\label{eq:received power}\end{equation}
is maximized. In (\ref{eq:received power}), $\lambda$ and $\eta$
are the wavelength ($\lambda\triangleq\frac{c}{f}$) and the impedance
($\eta\triangleq\sqrt{\frac{\mu_{0}}{\epsilon_{0}}}$) of the free-space
with permittivity and permeability equal to $\epsilon_{0}$ and $\mu_{0}$,
respectively, while $c$ is the free-space speed of light ($c\triangleq\sqrt{\frac{1}{\mu_{0}\epsilon_{0}}}$).
Moreover, $\mathbf{E}^{sca}\left(\mathbf{r}_{RX};\mathbf{d}\right)$
is the electric field reflected in $\mathbf{r}=\mathbf{r}_{RX}$ by
the flat screen, described by the vector $\mathbf{d}$, when illuminated
by the incident electric field $\mathbf{E}^{inc}\left(\mathbf{r}\right)$
generated by the transmitter, the time-dependence $\exp\left(j2\pi ft\right)$
being omitted hereinafter for the sake of notation compactness.

\noindent To determine the received power (\ref{eq:received power}),
the computation of the field reflected by the screen, $\mathbf{E}^{sca}\left(\mathbf{r};\mathbf{d}\right)$,
is carried out with the method developed in \cite{Oliveri 2022c}
for the analysis of \emph{EMS}s, but suitable for \emph{PCS}s, as
well. By discretizing the square flat screen area, $\Theta$, in $P\times Q$
($P=Q$) square cells, \{$\Theta_{pq}$; $p=0,...,P-1$; $q=0,...,Q-1$\},
of side $\Delta$ ($\Delta\triangleq\frac{L}{Q}$), it turns out that
\cite{Oliveri 2022c}\begin{equation}
\begin{array}{l}
\mathbf{E}^{sca}\left(\mathbf{r};\mathbf{d}\right)\approx-\frac{j\exp\left(-jkr\right)}{2\lambda r}\Delta^{2}\mathrm{sinc}\left(\frac{\pi\Delta\sin\theta\cos\varphi}{\lambda}\right)\mathrm{sinc}\left(\frac{\pi\Delta\sin\theta\sin\varphi}{\lambda}\right)\sum_{p=0}^{P-1}\sum_{q=0}^{Q-1}\exp\left[j\frac{2\pi}{\lambda}\beta_{pq}\left(\mathbf{r}\right)\right]\times\\
\left\{ \left[\eta\cos\theta\cos\varphi\left(J_{e}^{x}\right)_{pq}+\eta\cos\theta\sin\varphi\left(J_{e}^{y}\right)_{pq}-\sin\varphi\left(J_{m}^{x}\right)_{pq}+\cos\varphi\left(J_{m}^{y}\right)_{pq}\right]\widehat{\bm{\theta}}+\right.\\
\left.+\left[-\eta\sin\varphi\left(J_{e}^{x}\right)_{pq}+\eta\cos\varphi\left(J_{e}^{y}\right)_{pq}+\cos\theta\cos\varphi\left(J_{m}^{x}\right)_{pq}+\cos\theta\sin\varphi\left(J_{m}^{y}\right)_{pq}\right]\widehat{\bm{\varphi}}\right\} ,\end{array}\label{eq:reflected field}\end{equation}
where $\mathrm{sinc}\left(\cdot\right)\triangleq\frac{\sin\left(\cdot\right)}{\left(\cdot\right)}$
and the phase term $\beta_{pq}\left(\mathbf{r}\right)$ is given by\begin{equation}
\beta_{pq}\left(\mathbf{r}\right)=x_{p}\sin\theta\cos\varphi+y_{q}\sin\theta\sin\varphi-\frac{\cos^{2}\theta\left(x_{p}^{2}+y_{q}^{2}\right)}{2r}-\frac{\left(x_{p}\sin\theta\sin\varphi-y_{q}\sin\theta\cos\varphi\right)^{2}}{2r},\label{eq:phase term}\end{equation}
$\mathbf{J}_{\alpha}$ ($\alpha\in\left\{ e,m\right\} $) being the
discretized electric ($\alpha=e$) or magnetic ($\alpha=m$) surface
current induced on the screen surface, $\mathbf{r}\in\Theta$, by
the incident field $\mathbf{E}^{inc}$

\noindent \begin{equation}
\mathbf{J}_{\alpha}\left(\mathbf{r}\right)=\sum_{l=x,y}\sum_{p=0}^{P-1}\sum_{q=0}^{Q-1}\left(J_{\alpha}^{l}\right)_{pq}\Pi_{pq}\left(\mathbf{r}\right)\widehat{\mathbf{l}}\label{eq:W Rscugni}\end{equation}
where $\Pi_{pq}\left(\mathbf{r}\right)$ is the ($p$, $q$)-th ($p=0,...,P-1$;
$q=0,...,Q-1$) pixel basis function centered in the barycenter $\left(x_{p},y_{q}\right)$
{[}$\left(x_{p},y_{q}\right)=\left(-\frac{L}{2}+p\Delta,-\frac{L}{2}+q\Delta\right)${]}
of the ($p$, $q$)-th discretization domain with extension $\Theta_{pq}$
($\Theta=\bigcup_{p=0}^{P-1}\bigcup_{q=0}^{Q-1}\Theta_{pq}$).

\noindent It is worth remarking that (\ref{eq:reflected field}) is
accurate and reliable for predicting the received power in $\mathbf{r}=\mathbf{r}_{RX}$
(\ref{eq:received power}) as long as the Fresnel condition holds
true \cite{Oliveri 2022c}\begin{equation}
r\ge r_{FR}\label{eq:fresnel condition}\end{equation}
where $r_{FR}\triangleq\max\left\{ 10\times L\sqrt{2};\,0.62\times\sqrt{\frac{2\times L^{3}\sqrt{2}}{\lambda}};\,10\lambda\right\} $
being $r\triangleq\left|\mathbf{r}\right|$.

\noindent The reflected field $\mathbf{E}^{sca}$ is known {[}i.e.,
the expression (\ref{eq:reflected field}) is fully explicit{]} once
the $\alpha$-th ($\alpha\in\left\{ e,m\right\} $) surface current,
$\mathbf{J}_{\alpha}$, induced on the screen surface (i.e., $\mathbf{r}\in\Theta$)
is available. Towards this end, the \emph{Generalized Sheet Transition
Condition} (\emph{GSTC}) method is applied \cite{Oliveri 2022c} to
yield the following relation

\noindent \begin{equation}
\left\{ \begin{array}{l}
\mathbf{J}_{e}\left(\mathbf{r}\right)=j\omega\mathbf{F}_{e}^{\top}\left(\mathbf{r}\right)-\widehat{\mathbf{n}}\times\nabla^{\top}F_{m}^{n}\left(\mathbf{r}\right)\\
\mathbf{J}_{m}\left(\mathbf{r}\right)=j\omega\mu_{0}\mathbf{F}_{m}^{\top}\left(\mathbf{r}\right)+\frac{1}{\varepsilon_{0}}\widehat{\mathbf{n}}\times\nabla^{\top}F_{e}^{n}\left(\mathbf{r}\right)\end{array}\right.\label{eq:GSTC currents}\end{equation}
where $\widehat{\mathbf{n}}$ is the normal to the flat screen,$\nabla^{\top}$
is the transverse gradient operator, and $\mathbf{F}_{\alpha}$ is
the $\alpha$-th ($\alpha\in\left\{ e,m\right\} $) polarization surface
density, $F_{\alpha}^{n}$ and $\mathbf{F}_{\alpha}^{\top}$ being
its normal {[}$F_{\alpha}^{n}\left(\mathbf{r}\right)\triangleq\mathbf{F}_{\alpha}\left(\mathbf{r}\right)\cdot\widehat{\mathbf{n}}$,
$\mathbf{r}\in\Theta${]} and transversal {[}$\mathbf{F}_{\alpha}^{\top}\left(\mathbf{r}\right)\triangleq\mathbf{F}_{\alpha}\left(\mathbf{r}\right)-F_{\alpha}^{n}\left(\mathbf{r}\right)\widehat{\mathbf{n}}$,
$\mathbf{r}\in\Theta${]} components, respectively.

\noindent Under the hypothesis of local symmetry in each ($p$, $q$)-th
($p=0,...,P-1$; $q=0,...,Q-1$) discretization domain $\Theta_{pq}$
of the screen surface $\Theta$, the polarization surface densities
can be expressed as\begin{equation}
\left\{ \begin{array}{l}
\mathbf{F}_{e}\left(\mathbf{r}\right)\approx\sum_{p=0}^{P-1}\sum_{q=0}^{Q-1}\left[\varepsilon_{0}\left.\overline{\overline{\sigma}}_{e}\right\rfloor _{pq}\cdot\mathbf{E}_{pq}^{av}\right]\Pi_{pq}\left(\mathbf{r}\right)\\
\mathbf{F}_{m}\left(\mathbf{r}\right)\approx\sum_{p=0}^{P-1}\sum_{q=0}^{Q-1}\left[\left.\overline{\overline{\sigma}}_{m}\right\rfloor _{pq}\cdot\mathbf{H}_{pq}^{av}\right]\Pi_{pq}\left(\mathbf{r}\right)\end{array}\right.\label{eq:}\end{equation}
 where $\left.\overline{\overline{\sigma}}_{\alpha}\right\rfloor _{pq}$
is the diagonal tensor of the $\alpha$-th ($\alpha\in\left\{ e,m\right\} $)
local surface susceptibility in the $\Theta_{pq}$ sub-domain \cite{Yang 2019}\cite{Oliveri 2021c}\cite{Oliveri 2022}\cite{Oliveri 2022c},
while the surface average fields in the ($p$, $q$)-th ($p=0,...,P-1$;
$q=0,...,Q-1$) discretization domain are given by\begin{equation}
\left\{ \begin{array}{l}
\mathbf{E}_{pq}^{av}=\frac{1}{2\Delta^{2}}\int_{\Theta_{pq}}\left[1+\overline{\overline{\Gamma}}_{pq}\right]\cdot\mathbf{E}^{inc}\left(\mathbf{r}\right)d\mathbf{r}\\
\mathbf{H}_{pq}^{av}=\frac{1}{2\Delta^{2}}\int_{\Theta_{pq}}\left[1-\overline{\overline{\Gamma}}_{pq}\right]\cdot\mathbf{H}^{inc}\left(\mathbf{r}\right)d\mathbf{r},\end{array}\right.\label{eq:}\end{equation}
$\mathbf{H}^{inc}\left(\mathbf{r}\right)$ being the incident magnetic
field radiated by the transmitter in $\mathbf{r}\in\Theta$. Moreover,
$\overline{\overline{\Gamma}}_{mn}$ is the local reflection tensor/matrix
whose elements/entries can be determined from the local surface susceptibilities
as detailed in \cite{Yang 2019}.

\noindent Finally, the explicit form of the power at the receiver,
$\mathcal{P}_{RX}\left(\mathbf{r}_{RX};\,\mathbf{d}\right)$, is obtained
by substituting (\ref{eq:GSTC currents}) in (\ref{eq:reflected field})
and then this latter in (\ref{eq:received power}).

\subsection{\noindent The \emph{PCS} Case}

\noindent By considering specular reflections {[}i.e., $\varphi_{TX}=180$
{[}deg{]}, $\varphi_{RX}=0$ {[}deg{]}, $\theta_{TX}=\theta_{RX}=\theta_{0}$
- Fig. 1(\emph{a}){]} from the screen, the design of an ideal {[}i.e.,
composed by a perfect electric conductor (\emph{PEC}){]} \emph{PCS}
consists in the definition of the panel side $L$ (i.e., $S=1$, $d_{0}=L$).

\subsubsection*{Finite-Size \emph{PCS} Model}

\noindent Since the contribution to the reflection by each ($p$,
$q$)-th ($p=0,...,P-1$; $q=0,...,Q-1$) cell of an ideal \emph{PCS}
can be seen as a canonical scattering problem from a \emph{PEC} plate
with area $\Theta_{pq}$, standard approaches such as those based
on the \emph{Induction Equivalent} or the \emph{Physical Equivalent}
approximations \cite{Balanis 2012} can be reliably applied. By using
this latter, the expression (\ref{eq:GSTC currents}) of the surface
currents on the screen aperture (i.e., $\mathbf{r}\in\Theta$), can
be simplified for \emph{PCS}s as follows \cite{Balanis 2012}\begin{equation}
\left\{ \begin{array}{l}
\mathbf{J}_{e}\left(\mathbf{r}\right)=2\sum_{p=0}^{P-1}\sum_{q=0}^{Q-1}\widehat{\mathbf{n}}\times\mathbf{H}_{pq}^{av}\Pi_{pq}\left(\mathbf{r}\right)\\
\mathbf{J}_{m}\left(\mathbf{r}\right)\approx0\end{array}\right.\label{eq:correnti PEC}\end{equation}
and sequentially substituted first in (\ref{eq:reflected field})
and then in (\ref{eq:received power}) to determine the amount of
power reflected by the \emph{PCS} at the receiver, $\mathcal{P}_{RX}^{PCS}\left(\mathbf{r}_{RX};\,\mathbf{d}\right)$
{[}$\mathcal{P}_{RX}^{PCS}\left(\mathbf{r}_{RX};\,\mathbf{d}\right)\triangleq\mathcal{P}_{RX}^{PCS}\left(\mathbf{r}_{RX};\, L\right)${]},
as well as the corresponding total path attenuation, $\mathcal{A}^{PCS}\left(\mathbf{r}_{RX};\, L\right)$
{[}$\mathcal{A}^{PCS}\left(\mathbf{r}_{RX};\, L\right)\triangleq\frac{\mathcal{P}_{RX}^{PCS}\left(\mathbf{r}_{RX};\, L\right)}{\mathcal{P}_{TX}}${]},
this latter being defined as\begin{equation}
\mathcal{A}\left(\mathbf{r}_{RX};\,\mathbf{d}\right)\triangleq\frac{\mathcal{P}_{RX}\left(\mathbf{r}_{RX};\,\mathbf{d}\right)}{\mathcal{P}_{TX}}\label{eq:TPA}\end{equation}
where $\mathcal{P}_{TX}$ is the transmitted power.

\subsubsection*{\noindent Asymptotic \emph{PCS} Model}

\noindent When $L\to\infty$, the problem at hand can be studied with
the theory of images \cite{Balanis 2012} as the interaction between
a source and a flat infinite \emph{PEC}.

\noindent By introducing the \emph{virtual source} symmetrically to
the original transmitting antenna as shown in Fig. 1(\emph{a}) and
recalling that there is no direct link between the transmitter and
the receiver, the power reflected at the receiver location, $\mathbf{r}=\mathbf{r}_{RX}$,
is given by the canonical Friis' transmission equation provided that
the two antennas are mutually in far field \cite{Balanis 2012}. Accordingly,
the asymptotic value of the \emph{TPA} for a \emph{PCS} of infinite
extension ($L\to\infty$), $\mathcal{A}_{\infty}^{PCS}\left(\mathbf{r}_{RX}\right)$
{[}$\mathcal{A}_{\infty}^{PCS}\left(\mathbf{r}_{RX}\right)\triangleq\lim_{L\to\infty}\frac{\mathcal{P}_{RX}^{PCS}\left(\mathbf{r}_{RX};\, L\right)}{\mathcal{P}_{TX}}${]},
can be derived\begin{equation}
\mathcal{A}_{\infty}^{PCS}\left(\mathbf{r}_{RX}\right)=\left[\frac{\lambda}{4\pi\left(r_{RX}+r_{TX}\right)}\right]^{2}G_{RX}G_{TX}.\label{eq:limit PCS}\end{equation}
It is worth remarking that (\ref{eq:limit PCS}) models the asymptotic
behavior of the \emph{PCS} {[}i.e., $\mathcal{A}^{PCS}\left(\mathbf{r}_{RX};\, L\right)\to\mathcal{A}_{\infty}^{PCS}\left(\mathbf{r}_{RX}\right)$,
$L\to\infty${]}, but it is not an upper bound for the \emph{TPA}
of a \emph{PCS} valid whatever the \emph{PCS} side $L$. Indeed, \emph{TPA}
values greater than $\mathcal{A}_{\infty}^{PCS}\left(\mathbf{r}_{RX}\right)$
can be reached by finite aperture \emph{PCS}s as numerically proved
in Sect. \ref{sec:Results}.

\subsection{\noindent The \emph{EMS}-Screen \emph{}Case\label{sub:The-EMS-Screen-Case}}

\subsubsection*{\noindent Finite-Size \emph{EMS}-Screen \emph{}Model}

\noindent The synthesis of an \emph{EMS} screen to maximize the power
reflected towards the receiver (\ref{eq:received power}) requires
the setup of the $B$ geometrical descriptors, \{$g_{pq}^{b}$; $b=0,...,B-1$\},
of each ($p$, $q$)-th ($p=0,...,P-1$; $q=0,...,Q-1$) meta-atom
{[}e.g., Fig. 1(\emph{d}){]} within the screen aperture $\Theta$,
hence in the overall the synthesis of the $S$-dimensional ($S\triangleq1+B\times P\times Q$)
descriptor vector $\mathbf{d}$ whose entries are defined as follows\begin{equation}
\left\{ \begin{array}{l}
d_{s}=L\quad s=0\\
d_{1+b+\left(p+q\times Q\right)\times B}=g_{pq}^{b}\quad b=0,...,B-1;\, p=0,...,P-1;\, q=0,...,Q-1\end{array}\right..\label{eq:descriptors EMS}\end{equation}
The values of these descriptors (\ref{eq:descriptors EMS}) are chosen
with the method detailed in \cite{Oliveri 2022c} and briefly resumed
next, as well, since it has proved to yield \emph{EMS}s featuring
optimal and reliable focusing properties when the receiver lies in
the radiative near-field or far-field \emph{}of the \emph{EMS} itself
(\ref{eq:fresnel condition}). More specifically, the \emph{EMS} descriptors,
$\mathbf{d}^{EMS}$, are determined by performing the following procedural
steps:

\begin{itemize}
\item \noindent \textbf{Initialization} - Select a suitable meta-atom/unit-cell
{[}e.g., Fig. 1(\emph{d}){]} and identify its $B$ descriptors. Fill
the local susceptibility look-up table by computing the $\alpha$-th
($\alpha\in\left\{ e,m\right\} $) diagonal tensor of the local surface
susceptibility in the $\Theta_{pq}$ sub-domain, $\left.\overline{\overline{\sigma}}_{\alpha}\right\rfloor _{pq}$,
in correspondence with a wide set of $U$ unit-cell configurations,
$\left\{ g_{pq}^{b};\, b=0,...,B-1\right\} _{u}$ ($u=1,...,U$),
by means of a full-wave \emph{EM} simulator (e.g., Ansys HFSS \cite{HFSS 2021}\cite{Oliveri 2022c}),
$\left\{ \left.\overline{\overline{\sigma}}_{\alpha}\right\rfloor _{pq}\right\} _{u}\Leftrightarrow\left\{ g_{pq}^{b},b=0,...,B-1\right\} _{u}$;
$u=1,...,U$;
\item \textbf{Ideal Surface Current Phase Synthesis} - Compute the phase
of each $l$-th ($l\in\left\{ x,\, y\right\} $) component of the
($p$, $q$)-th ($p=0,...,P-1$; $q=0,...,Q-1$) expansion coefficient
of the $\alpha$-th ($\alpha\in\left\{ e,m\right\} $) ideal current,
$\widetilde{J}_{\alpha}^{l}$ {[}$\widetilde{\mathbf{J}}_{\alpha}\left(\mathbf{r}\right)=\sum_{l=x,y}\sum_{p=0}^{P-1}\sum_{q=0}^{Q-1}\left(\widetilde{J}_{\alpha}^{l}\right)_{pq}\Pi_{pq}\left(\mathbf{r}\right)\widehat{\mathbf{l}}${]},
by imposing the phase conjugation condition \cite{Oliveri 2022c}
in the expression (\ref{eq:reflected field}) to maximize the power
reflected at the receiver position $\mathbf{r}_{RX}$ (\ref{eq:received power})\begin{equation}
\angle\left(\widetilde{J}_{\alpha}^{l}\right)_{pq}=-\beta_{pq}\left(\mathbf{r}_{RX}\right)\label{eq:design EMS}\end{equation}
where $\angle\cdot$ stands for the argument operator;
\item \textbf{Surface Current Phase Matching} - Perform the multi-scale
iterative process described in \cite{Oliveri 2022c} and based on
the System-by-Design (\emph{SbD}) paradigm \cite{Massa 2022} to solve
the following optimization problem\begin{equation}
\mathbf{d}^{EMS}\triangleq\arg\left\{ \min_{\mathbf{d}}\left[\Phi\left(\mathbf{d}\right)\right]\right\} \label{eq:}\end{equation}
where $\Phi\left(\mathbf{d}\right)$ is the surface current mismatch
{[}$\Phi\left(\mathbf{d}\right)$ $\triangleq$ $\sum_{\alpha=e,m}$
$\sum_{l=x,y}$ $\int_{\Theta}$ $\left|\angle\widetilde{\mathbf{J}}_{\alpha}\left(\mathbf{r}\right)\right.$
$-$ $\left.\angle\mathbf{J}_{\alpha}\left(\mathbf{r};\mathbf{d}\right)\right|^{2}$
$\mathrm{d}\mathbf{r}${]}, $\mathbf{J}_{\alpha}\left(\mathbf{r};\mathbf{d}\right)$
being evaluated through (\ref{eq:GSTC currents}) by using the entries
of the pre-computed local susceptibility look-up table that correspond
to the current configuration of the $P\times Q$ meta-atoms coded
in $\mathbf{d}$.
\end{itemize}
\noindent Once $\mathbf{d}^{EMS}$ has been determined and under the
hypothesis that there is a perfect phase matching {[}i.e., $\Phi\left(\mathbf{d}^{EMS}\right)=0${]}
between the surface current supported by the \emph{EMS} screen, \{$\mathbf{J}_{\alpha}\left(\mathbf{r};\mathbf{d}\right)$;
$\mathbf{r}\in\Theta$\}, and the ideal one, \{$\widetilde{\mathbf{J}}_{\alpha}\left(\mathbf{r}\right)$;
$\mathbf{r}\in\Theta$\}, the amount of power reflected by the \emph{EMS}
screen at the receiver, $\mathcal{P}_{RX}^{EMS}\left(\mathbf{r}_{RX};\,\mathbf{d}\right)$
{[}$\mathcal{P}_{RX}^{EMS}\left(\mathbf{r}_{RX};\,\mathbf{d}\right)\triangleq\mathcal{P}_{RX}^{EMS}\left(\mathbf{r}_{RX};\,\mathbf{d}^{EMS}\right)${]}
is equal to\begin{equation}
\begin{array}{l}
\mathcal{P}_{RX}^{EMS}\left(\mathbf{r}_{RX};\,\mathbf{d}\right)\approx\frac{G_{RX}}{32\pi\eta r_{RX}^{2}}\sum_{p=1}^{P}\sum_{q=1}^{Q}\\
\left\{ \left[\eta\cos\theta\cos\varphi\left|\left(J_{e}^{x}\right)_{pq}\right|+\eta\cos\theta\sin\varphi\left|\left(J_{e}^{y}\right)_{pq}\right|-\sin\varphi\left|\left(J_{m}^{x}\right)_{pq}\right|+\cos\varphi\left|\left(J_{m}^{y}\right)_{pq}\right|\right]^{2}+\right.\\
\left.+\left[-\eta\sin\varphi\left|\left(J_{e}^{x}\right)_{pq}\right|+\eta\cos\varphi\left|\left(J_{e}^{y}\right)_{pq}\right|+\cos\theta\cos\varphi\left|\left(J_{m}^{x}\right)_{pq}\right|+\cos\theta\sin\varphi\left|\left(J_{m}^{y}\right)_{pq}\right|\right]^{2}\right\} \end{array}\label{eq:potenza RX EMS}\end{equation}
and the corresponding path attenuation, $\mathcal{A}^{EMS}\left(\mathbf{r}_{RX};\,\,\mathbf{d}^{EMS}\right)$,
is given by\begin{equation}
\mathcal{A}^{EMS}\left(\mathbf{r}_{RX};\,\,\mathbf{d}^{EMS}\right)=\frac{\mathcal{P}_{RX}^{EMS}\left(\mathbf{r}_{RX};\,\,\mathbf{d}^{EMS}\right)}{\mathcal{P}_{TX}}\label{eq:TPA EMS}\end{equation}
according to (\ref{eq:TPA}).

\noindent It is worthwhile to point out that the value of $\mathcal{P}_{RX}^{EMS}\left(\mathbf{r}_{RX};\,\mathbf{d}\right)$
and, consequently, the corresponding path attenuation, $\mathcal{A}^{EMS}\left(\mathbf{r}_{RX};\,\,\mathbf{d}^{EMS}\right)$
(\ref{eq:TPA EMS}), both depend on the meta-atom structure since
this latter affects the surface currents on the \emph{EMS} screen,
which are involved in (\ref{eq:potenza RX EMS}).

\subsubsection*{\emph{EMS}-Screen Performance Upper Bound}

\noindent While (\ref{eq:TPA EMS}) gives the actual achievable link
performance by taking into account the full-wave modeling of the finite-size
\emph{EMS} screen at hand, it is also of interest to know the performance
\emph{}bound achievable \emph{}by an optimal/ideal \emph{EMS} able
to collect and reflect to the receiver all the incident power.

\noindent By considering the working principle of \emph{EMS}s, such
an upper bound can be deduced by noting that (\emph{i}) the incident
power on the \emph{EMS}, $\mathcal{P}_{INC}$, is proportional to
both the gain and the aperture of the transmitter by also fulfilling
the condition $\mathcal{P}_{INC}\leq\frac{G_{TX}\mathcal{P}_{TX}}{4\pi r_{TX}^{2}}\cos\left(\theta_{TX}\right)L^{2}$,
(\emph{ii}) the \emph{EMS} screen behaves as a transmitting antenna
with input power $\mathcal{P}_{INC}$ and gain $\frac{4\pi\cos\left(\theta_{RX}\right)L^{2}}{\lambda^{2}}$
\cite{Balanis 2012} behaves as a transmitting antenna with input
power $\mathcal{P}_{INC}$ and gain $G_{TX}=\frac{4\pi\cos\left(\theta_{RX}\right)L^{2}}{\lambda^{2}}$
\cite{Balanis 2012} in ideal conditions since it redirects all the
incident power along the receiver direction, and (\emph{iii}) the
propagation towards the receiver can be faithfully modeled with the
Friis' transmission equation \cite{Balanis 2012}. Therefore, it is
possible to infer, as detailed in the \emph{Appendix A.1}, that the
upper bound for the \emph{TPA} of an ideal \emph{EMS} square screen
of side $L$ (i.e., $\mathcal{A}_{opt}^{EMS}\left(\mathbf{r}_{RX};\, L\right)\triangleq\max_{\left\{ d_{s}^{EMS};\, s=1,...,S\right\} }\left[\mathcal{A}^{EMS}\left(\mathbf{r}_{RX};\,\left\{ L,\left(d_{s}^{EMS};\, s=1,...,S\right)\right\} \right)\right]$)
is\begin{equation}
\mathcal{A}_{opt}^{EMS}\left(\mathbf{r}_{RX};\, L\right)=\frac{G_{TX}G_{RX}\cos^{2}\left(\theta_{0}\right)L^{4}}{\left(4\pi r_{TX}r_{RX}\right)^{2}}.\label{eq:TPA limit EMS}\end{equation}
The value of $\mathcal{A}_{opt}^{EMS}\left(\mathbf{r}_{RX};\, L\right)$
only depends on the screen size $L$, but not on the \emph{EMS} unit-cell
shape and/or material. However, one should consider that (\ref{eq:TPA limit EMS})
provides an ideal upper bound without taking into account edge effects,
local periodicity approximations, material losses, imperfect local
phase control, and illumination tapering/spillover effects. Nevertheless,
it can be a very useful tool for roughly sizing an \emph{EMS} screen
starting from the user needs in terms of \emph{QoS} at the user located
at $\mathbf{r}=\mathbf{r}_{RX}$.

\subsubsection*{\noindent \emph{EMS}-Screen Optimality Condition}

\noindent Under the assumption that $r_{RX}>10\lambda$, by combining
(\ref{eq:limit PCS}) and (\ref{eq:TPA limit EMS}) as detailed in
the \emph{Appendix A.2}, it is possible to yield the following condition
(denoted as {}``\emph{optimality condition}'') on the size $\Theta$
($\Theta\triangleq L\times L$) of an \emph{EMS} so that $\mathcal{A}_{opt}^{EMS}\left(\mathbf{r}_{RX};\, L\right)\ge\mathcal{A}_{\infty}^{PCS}\left(\mathbf{r}_{RX}\right)$
\begin{equation}
L_{TH}\leq L\leq L_{FR}.\label{eq: L optimality condition}\end{equation}
In (\ref{eq: L optimality condition}), $L_{TH}$ and $L_{FR}$ are
the threshold and the Fresnel side lengths, respectively, equal to\begin{equation}
L_{TH}\triangleq\sqrt{\frac{\lambda}{\cos\left(\theta_{0}\right)}\left(\frac{r_{TX}\times r_{RX}}{r_{TX}+r_{RX}}\right)}\label{eq: L_TH}\end{equation}
and\begin{equation}
L_{FR}\triangleq\min\left\{ \frac{r_{RX}}{10\sqrt{2}};\sqrt[3]{\frac{\lambda}{2\sqrt{2}}\left(\frac{r_{RX}}{0.62}\right)^{2}}\right\} .\label{eq: L_FR}\end{equation}

\section{\noindent Numerical Results\label{sec:Results}}

\noindent The objective of this Section is manifold. On the one had,
it is aimed at comparing the specular reflection efficiency of \emph{PCS}s
and \emph{EMS}s in terms of \emph{TPA} under the same operative conditions
as well as to check the reliability and the accuracy of the \emph{EMS}-screen
optimality condition (\ref{eq: L optimality condition}) to overcome
the end-to-end free-space \emph{TPA}. On the other hand, it is devoted
to assess the actual \emph{NLOS} link performance of different class
of screen panels, with respect to the theoretical derivations of Sect.
\ref{sec:Problem-Formulation}, namely (\ref{eq:limit PCS}) and (\ref{eq:TPA limit EMS}),
as well. Towards this end, the benchmark scenario models the specular
reflection {[}i.e., $\varphi_{TX}=180$ {[}deg{]}, $\varphi_{RX}=0$
{[}deg{]}, $\theta_{TX}=\theta_{RX}=\theta_{0}$ - Fig. 1(\emph{a}){]}
from a flat screen of an \emph{EM} wave radiated by a transmitter
equipped with the same linearly-polarized pyramidal horn antenna {[}Fig.
2(\emph{a}){]} of the receiver%
\footnote{\noindent Although in this numerical analysis both the transmitter
and the receiver have been modeled with the same elementary radiator,
the same outcomes can be drawn by changing such an assumption.%
}.

\noindent The first numerical experiment compares the performance
of a $L=0.8$ {[}m{]}-sided square \emph{PCS} with those from an equal-size
\emph{EMS} screen when illuminated by the $G_{TX}=15.4$ {[}dBi{]}
horn antenna in Fig. 2(\emph{b}) and features in Tab. I, which is
located $r_{TX}=15$ {[}m{]} far from $\Theta$ along the direction
$\theta_{TX}=\theta_{0}=30$ {[}deg{]}. The transmitter has been supposed
to operate at $f=27$ {[}GHz{]} and to radiate a power of $P_{TX}=20$
{[}dBm{]} for generating the incident $y$-polarized electric field
with magnitude and phase distributions shown in Fig. 3(\emph{b}) and
Fig. 3(\emph{c}), respectively. The aperture $\Theta$ of the patterned
panel has been discretized in a half-wavelength uniform lattice (i.e.,
$\Delta=5.556\times10^{-3}$ {[}m{]}) and the design of the \emph{EMS}
screen has been carried out by considering the $B=1$ meta-atom in
Fig. 1(\emph{d}) composed by a canonical square-shaped patch of side
$\left.g^{b}\right\rfloor _{b=0}=\left.g^{b}\right\rfloor _{b=1}=g^{\left(1\right)}$
{[}$\to$ $d_{1+\left(p+q\times Q\right)\times B}=d_{2+\left(p+q\times Q\right)\times B}=g_{pq}^{\left(1\right)}$
($p=0,...,P-1$; $q=0,...,Q-1$), while $d_{0}=L${]} printed on a
single-layer Rogers \emph{RO4350} substrate with thickness $\tau=5.08\times10^{-4}$
{[}m{]}. Moreover, the receiver has been located, along the specular
direction $\theta_{RX}=\theta_{0}$ {[}deg{]}, $r_{RX}=15$ {[}m{]}
far from the panel to comply with (\ref{eq: L optimality condition}).

\noindent The layout of the \emph{EMS} screen, designed with the method
summarized in Sect. \ref{sub:The-EMS-Screen-Case} and fulfilling
(\ref{eq:design EMS}), features a concentric pattern (Fig. 4) as
actually expected since a {}``lensing'' effect {[}Fig. 1(\emph{b}){]},
which cannot be realized with a standard \emph{PCS} {[}Fig. 1(\emph{c}){]},
has to be afforded to focus the power at the receiver. Such a behavior
is pictorially highlighted by the plots of the magnitude of the dominant
component of the reflected field (i.e., $\left|E_{\varphi}^{sca}\left(\mathbf{r}\right)\right|$)
in the {}``transversal cut'' {[}Fig. 5(\emph{d}) vs. Fig. 5(\emph{c}){]},
which is the $z''=0$ plane in the receiver local coordinate system
$\left(x'',y'',z''\right)$ {[}Fig. 1(\emph{b}){]}. Thanks to a proper
control of the surface currents induced on the panel $\Theta$ {[}Fig.
5(\emph{b}) vs. Fig. 5(\emph{a}){]}, the \emph{EMS} screen focuses
the power towards the receiver region better than a standard \emph{PCS}
of the same size {[}Fig. 5(\emph{d}) vs. Fig. 5(\emph{c}){]}. This
is also pointed out by the comparison of the corresponding distributions
of the field magnitude in the {}``longitudinal cut'' {[}i.e., the
$y''=0$ plane in the receiver coordinate system - Fig. 1(\emph{b}){]}.

\noindent As for the \emph{TPA}, Figure 6(\emph{a}) shows that such
a $L=0.8$ {[}m{]}-sided \emph{EMS} panel is significantly more effective
than the \emph{PCS} with the same size {[}i.e., $\left.\Delta\mathcal{A}_{PCS}^{EMS}\left(\mathbf{r}_{RX};\, L\right)\right]_{L=0.8\,[m]}^{r_{RX}=15\,[m]}\approx15$
{[}dB{]} being $\Delta\mathcal{A}_{PCS}^{EMS}\left(\mathbf{r}_{RX};\, L\right)\triangleq\mathcal{A}^{EMS}\left(\mathbf{r}_{RX};\, L\right)-\mathcal{A}^{PCS}\left(\mathbf{r}_{RX};\, L\right)${]}
and it reaches a \emph{TPA} value considerably better than that from
an infinitely extended \emph{PCS} (i.e., $\left.\mathcal{A}^{EMS}\left(\mathbf{r}_{RX};\, L\right)\right]_{L=0.8\,[m]}^{r_{RX}=15\,[m]}=-48.5$
{[}dB{]} vs. $\left.\mathcal{A}_{\infty}^{PCS}\left(\mathbf{r}_{RX}\right)\right]^{r_{RX}=15\,[m]}=-59.8$
{[}dB{]} $\to$ $\left.\Delta\mathcal{A}_{PCS_{\infty}}^{EMS}\left(\mathbf{r}_{RX};\, L\right)\right]_{L=0.8\,[m]}^{r_{RX}=15\,[m]}\approx11.3$
{[}dB{]}). This latter result is not surprising since the \emph{EMS}-screen
optimality condition (\ref{eq: L optimality condition}) holds true,
the values of the panel side, $L$, the side-length threshold, $L_{TH}$,
and the Fresnel side length, $L_{FR}$ being $L=0.8$ {[}m{]}, $L_{TH}=3.10\times10^{-1}$
{[}m{]}, and $L_{FR}=1.06$ {[}m{]}, respectively. Furthermore, it
is worth noticing that, despite the simple (i.e., single-layer and
standard cheap off-the-shelf materials) meta-atom at hand {[}Fig.
1(\emph{d}){]}, the synthesized \emph{EMS} screen affords a \emph{TPA}
which is only $\approx5$ {[}dB{]} below the theoretical upper bound
for such an aperture size (i.e., $\left.\mathcal{A}_{opt}^{EMS}\left(\mathbf{r}_{RX};\, L\right)\right]_{L=0.8\,[m]}^{r_{RX}=15\,[m]}=-43.4$
{[}dB{]}).

\noindent The same scenario of the first experiment has been dealt
with next by varying the panel side $L$ to account for different
screen sizes, but always fulfilling the Fresnel condition $L\leq L_{FR}$.
Figure 6(\emph{a}) shows the plots of the \emph{TPA} in the range
$0.1$ {[}m{]} $\le$$L$ $\le$ $1.0$ {[}m{]}. As it can be observed,
the \emph{PCS} and the \emph{EMS} screen behave similarly, analogously
to the {}``ideal'' patterned panel, until $L<L_{TH}$ by yielding
close \emph{TPA} values, which are also below the free-space asymptotic
limit $\left.\mathcal{A}_{\infty}^{PCS}\left(\mathbf{r}_{RX}\right)\right]^{r_{RX}=15\,[m]}=-59.8$
{[}dB{]}. The fact that $\left.\mathcal{A}^{EMS}\left(\mathbf{r}_{RX};\, L\right)\right]^{r_{RX}=15\,[m]}$
$\approx$ $\left.\mathcal{A}^{PCS}\left(\mathbf{r}_{RX};\, L\right)\right]^{r_{RX}=15\,[m]}$
$\approx$ $\left.\mathcal{A}_{opt}^{EMS}\left(\mathbf{r}_{RX};\, L\right)\right]^{r_{RX}=15\,[m]}$
when $L<L_{TH}$ suggests that (\ref{eq:TPA limit EMS}) can be used
to predict the \emph{TPA} of small panels (i.e., $L<L_{TX}$) regardless
of the design technology (e.g., patterned or metallic) and implementation.
On the other hand, it turns out that an \emph{EMS} screen of side
$L$ slightly above the marker $L_{TH}^{EMS}$ ($L_{TH}^{EMS}$ $\triangleq$
$\arg_{L}$ \{$\mathcal{A}^{EMS}\left(\mathbf{r}_{RX};\, L\right)$
$=$ $\mathcal{A}_{\infty}^{PCS}\left(\mathbf{r}_{RX}\right)$\}),
which is here very close to $L_{TH}$ (i.e., $L_{TH}^{EMS}=3.39\times10^{-1}$
{[}m{]} vs. $L_{TH}=3.10\times10^{-1}$ {[}m{]}), performs better
than an infinite \emph{PEC} screen (i.e.,$\left.\mathcal{A}^{EMS}\left(\mathbf{r}_{RX};\, L\right)\right]^{r_{RX}=15\,[m]}$
$>$ $\left.\mathcal{A}_{\infty}^{PCS}\left(\mathbf{r}_{RX}\right)\right]^{r_{RX}=15\,[m]}$)
despite the simplicity of the chosen meta-atom {[}Fig. 1(\emph{d}){]}.
This result confirms that it is possible to overcome the limit dictated
by the theory of images by recurring to the wave manipulation properties
of \emph{EMS}s. Moreover, the \emph{TPA} improvement enabled by the
patterned screen over the asymptotic \emph{PCS} limit is greater and
greater widening more and more the panel dimension $\Theta$ {[}e.g.,
$\left.\Delta\mathcal{A}_{PCS_{\infty}}^{EMS}\left(\mathbf{r}_{RX};\, L\right)\right]_{L=0.4\,[m]}^{r_{RX}=15\,[m]}\approx1.76$
{[}dB{]} vs. $\left.\Delta\mathcal{A}_{PCS_{\infty}}^{EMS}\left(\mathbf{r}_{RX};\, L\right)\right]_{L=1.0\,[m]}^{r_{RX}=15\,[m]}\approx15.19$
{[}dB{]} as well as solid red vs. dashed green lines in Fig. 6(\emph{a}){]}
until the Fresnel threshold ($L=L_{FR}$). On the other hand, the
use of an \emph{EMS} screen, instead of a \emph{PCS} of the same finite
size, is also more and more convenient as $L$ grows {[}e.g., $\left.\Delta\mathcal{A}_{PCS}^{EMS}\left(\mathbf{r}_{RX};\, L\right)\right]_{L=0.4\,[m]}^{r_{RX}=15\,[m]}\approx0.0$
{[}dB{]} vs. $\left.\Delta\mathcal{A}_{PCS}^{EMS}\left(\mathbf{r}_{RX};\, L\right)\right]_{L=1.0\,[m]}^{r_{RX}=15\,[m]}\approx14.91$
{[}dB{]} as well as red vs. blue solid lines in Fig. 6(\emph{a}){]})
provided that $L>L_{PCS}^{EMS}$ ($L_{PCS}^{EMS}=0.5$ {[}m{]}), $L_{PCS}^{EMS}$
being the threshold side for which the equality $\mathcal{A}^{EMS}\left(\mathbf{r}_{RX};\, L\right)=\mathcal{A}^{PCS}\left(\mathbf{r}_{RX};\, L\right)$
is fulfilled (i.e., $L_{PCS}^{EMS}\triangleq\arg_{L}\left\{ \mathcal{A}^{EMS}\left(\mathbf{r}_{RX};\, L\right)=\mathcal{A}^{PCS}\left(\mathbf{r}_{RX};\, L\right)\right\} $).
Of course, it is generally possible to reduce the value of $L_{PCS}^{EMS}$
(i.e., $L_{PCS}^{EMS}\to L_{TH}$) by implementing more sophisticated
and performing meta-atoms.

\noindent For the sake of completeness and as representative examples,
the plots of the distributions in the transverse (Fig. 7) and the
horizontal (Fig. 8) cuts of the field reflected by the \emph{PCS}
and the \emph{EMS} layouts when setting the panel side to $L=0.4$
{[}m{]} {[}Fig. 6(\emph{b}){]} or $L=1.0$ {[}m{]} {[}Fig. 6(\emph{c}){]}
are provided in the left or in the right column of Figs. 7-8, respectively.
As expected from Fig. 6(\emph{a}), the \emph{PCS} and the \emph{EMS}
screen perform similarly as focusing devices for small apertures (e.g.,
Fig. 7(\emph{a}) vs. Fig. 7(\emph{b}) and Fig. 8(\emph{a}) vs. Fig.
8(\emph{b}) - $L=0.4$ {[}m{]} $\to$$L\approx L_{PCS}^{EMS}$), while
the efficiency of the \emph{EMS} in reflecting the power towards the
receiver strongly enhances when $L\gg L_{TH}$ (e.g., Fig. 7(\emph{d})
vs. Fig. 7(\emph{c}) and Fig. 8(\emph{d}) vs. Fig. 8(\emph{c}) - $L=1.0$
{[}m{]}).

\noindent The third experiment is then aimed at evaluating the \emph{EMS}
screen performance when the wireless link is established over longer
distances. More specifically, the transmitter and the receiver have
been placed $r_{TX}=r_{RX}=50$ {[}m{]} from the panel so that the
magnitude and the phase of the incident field over $\Theta$ are those
in Fig. 3(\emph{c}) and Fig. 3(\emph{d}), respectively. As expected,
the plots of the \emph{TPA} versus the screen side $L$ {[}Fig. 9(\emph{a}){]}
confirm the outcomes from the previous benchmark with shorter distances
(e.g., $\left.\mathcal{A}^{EMS}\left(\mathbf{r}_{RX};\, L\right)\right]^{r_{RX}=50\,[m]}$
$\approx$ $\left.\mathcal{A}^{PCS}\left(\mathbf{r}_{RX};\, L\right)\right]^{r_{RX}=50\,[m]}$
$\approx$ $\left.\mathcal{A}_{opt}^{EMS}\left(\mathbf{r}_{RX};\, L\right)\right]^{r_{RX}=50\,[m]}$
when $L<L_{TH}$, while $\left.\Delta\mathcal{A}_{PCS_{\infty}}^{EMS}\left(\mathbf{r}_{RX};\, L\right)\right]^{r_{RX}=50\,[m]}\ge0$
{[}dB{]} and $\left.\Delta\mathcal{A}_{PCS}^{EMS}\left(\mathbf{r}_{RX};\, L\right)\right]_{L=1.0\,[m]}^{r_{RX}=50\,[m]}\ge0$
{[}dB{]} as long as $L>L_{TH}^{EMS}$ and $L>L_{PCS}^{EMS}$, respectively),
but widening both the theoretical thresholds (Tab. II). This means
that the greater the receiver distance, $r_{RX}$, the wider must
be the patterned panel to yield improved performance with respect
to the equivalent-aperture \emph{PCS} as well as to overcome the free-space
path attenuation limit.

\noindent Unless the higher thresholds, the lensing behavior of the
\emph{EMS} screens turns out to be analogous to that of the previous
test case where the receiver/transmitter are closer to the panel as
pointed out by the maps of the reflected field in the transverse cut
(Fig. 10) when $L_{TH}<L<L_{TH}^{EMS}$ {[}$L=0.5$ {[}m{]} - Fig.
10(\emph{b}){]} and $L>L_{PCS}^{EMS}$ {[}$L=2.9$ {[}m{]} - Fig.
10(\emph{c}){]} if compared to Fig. 7(\emph{b}) and Fig. 7(\emph{d}),
respectively.

\noindent To check the independence of previous outcomes on the transmitter
type, the numerical analysis has been repeated with higher gain transmitting/receiving
devices {[}i.e., $G_{TX}=G_{RX}=25.5$ {[}dBi{]} - Fig. 2(\emph{c})
and Tab. I{]}. As expected, except for a scaling factor due to the
different gains, the behaviour of the \emph{TPA} versus $L$ turns
out to be almost identical to that of the lower gain horn in both
$r_{TX}=r_{RX}=15$ {[}m{]} {[}Fig. 11(\emph{a}) vs. Fig. 6(\emph{a}){]}
and $r_{TX}=r_{RX}=50$ {[}m{]} {[}Fig. 11(\emph{b}) vs. Fig. 9(\emph{a}){]}
cases. Moreover, the values of the \emph{TPA} markers are very close
(Tab. II), as well, the thresholds being identical by definition (\ref{eq: L_TH})(\ref{eq: L_FR}).

\noindent On the contrary, while the maps of the reflected field magnitude
in the transversal cut look similar (e.g., Fig. 12(\emph{e}) vs. Fig.
7(\emph{c}) and Fig. 12(\emph{c}) vs. Fig. 7(\emph{d}) - $r_{TX}=r_{RX}=15$
{[}m{]} and $L=1.0$ {[}m{]}; Fig. 12(\emph{f}) vs. Fig. 10(\emph{c})
and Fig. 12(\emph{d}) vs. Fig. 10(\emph{d}) - $r_{TX}=r_{RX}=50$
{[}m{]} and $L=1.0$ {[}m{]}), the patterns of the synthesized layouts
when illuminated by the two different horns are significantly different
{[}i.e., Fig. 12(\emph{a}) vs. Fig. 6(\emph{c}) and Fig. 12(\emph{b})
vs. Fig. 9(\emph{c}){]} since the design of an \emph{EMS} screen depends
on the incident field. Nevertheless, the fact that \emph{EMS} layouts
designed for different scenarios and working conditions yield satisfactory
lensing effects and fulfil challenging targets in terms of \emph{TPA}
values (i.e., \emph{QoS} at the receiver) is also an important and
further proof of the reliability and the robustness of the synthesis
process in Sect. \ref{sub:The-EMS-Screen-Case}.

\noindent The next experiment has been performed to assess whether
there is still an improvement in the \emph{TPA} using \emph{EMS} panels
instead of \emph{PCS}s and with respect to the free-space limit if
moving the receiver farther and farther away from the screen. Towards
this purpose, the receiver distance has been varied from $r_{RX}=20$
{[}m{]} up to $r_{RX}=300$ {[}m{]} by keeping the same antenna type
for the transmitter (i.e., $G_{TX}=G_{RX}=25.5$ {[}dBi{]}) located
at $r_{TX}=15$ {[}m{]}. Figure 13 summarizes the results of this
study by showing the plots of the \emph{TPA} for different aperture
sizes (i.e., $L\in\left\{ 0.8,1,1.2\right\} $ {[}m{]}) always above
the $L_{TH}$ threshold ($L_{TH}=0.428$ {[}m{]}). One can observe
that properly-designed patterned screens give superior wireless link
performance with respect to standard \emph{PCS}s (i.e., red vs. blue
plots - Fig. 13), by also overcoming the asymptotic free-space limit
$\mathcal{A}_{\infty}^{PCS}\left(\mathbf{r}_{RX}\right)$ (i.e., red
plots vs. green line - Fig. 13), whatever the $r_{RX}$ distance.
Moreover, as expected, the benefit of installing an \emph{EMS} for
a user located in a fixed position $r_{RX}=r_{0}$ increases as the
size of the panel widens (Tab. III).

\noindent It is also interesting to note that the same target \emph{TPA}
of a standard metallic panel can be reached by an \emph{EMS} screen
of the same side $L$ (i.e., equal aperture $\Theta$) over longer
end-to-end distances (i.e., $\left.\mathcal{A}^{EMS}\left(\mathbf{r}_{RX};\, L\right)\right]_{L=1.2\,[m]}^{r_{RX}=300\,[m]}$
$\approx$ $\left.\mathcal{A}^{PCS}\left(\mathbf{r}_{RX};\, L\right)\right]_{L=1.2\,[m]}^{r_{RX}=55\,[m]}$
- Fig. 13). This suggests that wireless links considerably longer
than those achievable with traditional flat metallic reflectors can
be established by using \emph{EMS} screens without increasing the
transmitter/receiver gains, the input power, or the panel size.

\noindent In order to analyze the dependence of the performance of
\emph{EMS}-based screens on the angle of the specular reflection,
two other ($\theta_{0}\ne30$) angular directions have been considered,
namely $\theta_{0}=20$ {[}deg{]} and $\theta_{0}=45$ {[}deg{]},
while the other scenario parameters have been kept unaltered from
those of Fig. 11(a). Once again, it turns out {[}Fig. 14(\emph{a}){]}
that it is advantageous (i.e., $\left.\Delta\mathcal{A}_{PCS_{\infty}}^{EMS}\left(\mathbf{r}_{RX};\, L\right)\right]^{r_{RX}=15\,[m]}>0$
and $\left.\Delta\mathcal{A}_{PCS}^{EMS}\left(\mathbf{r}_{RX};\, L\right)\right]_{L=1.0\,[m]}^{r_{RX}=50\,[m]}\ge0$)
adopting a suitably designed \emph{EMS} screen if the panel aperture
$\Theta$ fulfils the condition $L\ge L_{TH}^{EMS}$, being $L_{TH}^{EMS}\approx L_{TH}$,
whatever the $\theta_{0}$ value. However, the \emph{TPA} value afforded
by the \emph{EMS} screen gets worse and worse as $\theta_{0}$ increases
towards end-fire ($\to$ $\theta_{0}=90$ {[}deg{]}). For instance,
$\left.\mathcal{A}^{EMS}\left(\mathbf{r}_{RX};\, L\right)\right]_{\theta_{0}=20\,[deg]}^{r_{RX}=15\,[m]}$
$>$ $\left.\mathcal{A}^{EMS}\left(\mathbf{r}_{RX};\, L\right)\right]_{\theta_{0}=30\,[deg]}^{r_{RX}=15\,[m]}$
$>$ $\left.\mathcal{A}^{EMS}\left(\mathbf{r}_{RX};\, L\right)\right]_{\theta_{0}=45\,[deg]}^{r_{RX}=15\,[m]}$
regardless of the aperture side, $L$. Analogously, the performance
enhancement granted by the \emph{EMS}s reduces when tilting more and
more the incidence/reflection angle as shown in the representative
plots in Fig. 14(\emph{c}) for an \emph{EMS} screen of side $L=1$
{[}m{]} ($L\ge L_{TH}^{EMS}$). Such behaviors are actually the result
of two concurring effects. On the one hand, the amount of power impinging
on the panel aperture, $\mathcal{P}_{INC}^{opt}$, grows as $\theta_{0}\to0$
{[}deg{]} since the effective area intercepting the incident beams
is maximum when the source is at broadside (i.e., $\theta_{0}=0$
{[}deg{]}). On the other hand, there in an unavoidable degradation
of the effectiveness of the unit-cell {[}Fig. 1(\emph{d}){]} as $\theta_{0}\to90$
{[}deg{]} as it is well-known from \emph{EM} surface engineering \cite{Yang 2019}.
As a matter of fact, larger incidence angles require more abrupt phase
variations within the $\Theta$ aperture, which are more difficult
to be realized when using simple unit cells with inherently reduced
phase-control capabilities \cite{Yang 2019}. In turns, this implies
a more and more non-uniform/non-homogeneous distribution of the \emph{EMS}
meta-atoms over the flat panel {[}e.g., Fig. 15(\emph{a}) vs. 15(\emph{b}){]}
that unavoidably results in a lower focusing efficiency. Vice-versa,
there is a closer and closer fitting between the ideal \emph{TPA},
$\mathcal{A}_{opt}^{EMS}\left(\mathbf{r}_{RX};\, L\right)$, and the
actual one, $\mathcal{A}^{EMS}\left(\mathbf{r}_{RX};\, L\right)$,
{[}Fig. 14(\emph{a}){]}, as well as smaller and smaller deviations
of $L_{TH}^{EMS}$ from $L_{TH}$ {[}Fig. 14(\emph{b}){]} as $\theta_{0}$
approaches broadside. For instance, Figure 14(\emph{a}) shows that
$\left.\Delta\mathcal{A}_{opt}^{EMS}\left(\mathbf{r}_{RX};\, L\right)\right]_{\theta_{0}=20\,[deg]}^{r_{RX}=15\,[m]}\approx2.3$
{[}dB{]} vs. $\left.\Delta\mathcal{A}_{opt}^{EMS}\left(\mathbf{r}_{RX};\, L\right)\right]_{\theta_{0}=45\,[deg]}^{r_{RX}=15\,[m]}\approx9.7$
{[}dB{]} being $\Delta\mathcal{A}_{opt}^{EMS}\left(\mathbf{r}_{RX};\, L\right)\triangleq\mathcal{A}^{EMS}\left(\mathbf{r}_{RX};\, L\right)-\mathcal{A}_{opt}^{EMS}\left(\mathbf{r}_{RX};\, L\right)$.
This means that simpler meta-atoms can yield a power focusing efficiency
close to the ideal one as the incidence on the screen tends to the
normal to the surface.

\noindent The last test case is aimed at assessing the feasibility
and the benefit of \emph{NLOS} specular links with \emph{EMS} screens
when increasing the end-to-end distance $\rho$ ($\rho\triangleq r_{TX}+r_{RX}$).
Therefore, the behavior of the \emph{TPA} versus the screen side $L$
has been analyzed when $\rho=400$ {[}m{]} and $\rho=2000$ {[}m{]}
by setting $r_{TX}=r_{RX}=\frac{\rho}{2}$. As expected, Figures 16(\emph{a})-16(\emph{b})
show that wider panels are necessary to overcome the asymptotic free-space
limit as the link length $\rho$ grows {[}e.g., $\left.L_{TH}\right]_{\rho=400\,[m]}=1.132$
{[}m{]} vs. $\left.L_{TH}\right]_{\rho=2000\,[m]}=2.532$ {[}m{]}
- Fig. 16(\emph{b}){]}, the threshold value $L_{TH}$ being proportional
to the term ($\frac{r_{TX}\times r_{RX}}{\rho}$) in (\ref{eq: L_TH}).
However, it is worth pointing out that such a dependence is on the
square-root of the end-to-end distance $\rho$ {[}i.e., $L_{TH}\propto L_{TH}\left(\sqrt{\rho}\right)${]}
and not directly on $\rho$, as pointed out by the corresponding line-plot
in Fig. 16(\emph{b}) being $L_{TH}\approx0.056\times\sqrt{\rho}$.
Moreover, Figure 16(\emph{b}) as well as Figure 14(\emph{b}) show
that the value of the \emph{TAP} marker $L_{TH}^{EMS}$ is generally
very close to the threshold $L_{TH}$ with very similar dependence
on both the end-to-end distance, $\rho$, and the specular reflection
angle, $\theta_{0}$. This is not a trivial outcome since it implies
that the use of the closed-form expression of the threshold $L_{TH}$
(\ref{eq: L_TH}) is a reliable rule-of-thumb to quickly (i.e., without
any synthesis process and/or full-wave numerical simulations) predict,
for a given wireless link and incidence angle, the smallest panel
size of an \emph{EMS} screen that overcomes the free-space attenuation.

\noindent For completeness, Figure 16(\emph{c}) gives some insights
on the efficiency of an \emph{EMS} reflective panel of side $L=6$
{[}m{]} ($L\ge L_{TH}^{EMS}$) versus the link length $\rho$, while
the layouts of two $L$-sided \emph{EMS} screens are shown in Fig.
17.

\section{\noindent Conclusions\label{sec:Conclusions-and-Remarks}}

\noindent The feasibility of flat passive \emph{EMS} screens to overcome
the asymptotic \emph{TPA} limit of flat \emph{PEC} reflectors in \emph{NLOS}
specular conditions has been investigated. An extensive numerical
assessment has been performed to (\emph{a}) check the reliability
and the accuracy of the theoretical deductions, (\emph{b}) illustrate
the \emph{EM} features of \emph{EMS}s for \emph{NLOS} specular wireless
links, (\emph{c}) give some insights on the potentialities of such
a technological solution for \emph{NLOS} point-to-point propagation,
and (\emph{d}) deduce user-oriented guidelines.

\noindent To the best of the authors' knowledge, the main innovative
methodological outcomes of this work include

\begin{itemize}
\item \noindent the use of flat static \emph{EMS}s as reflective panels
in point-to-point \emph{NLOS} wireless links to yield a {}``lensing''
effect towards the receiver {[}Fig. 1(\emph{b}){]} instead of the
simple reflection of the incident beam {[}Fig. 1(c){]} as done by
\emph{PCS}s;
\item \noindent the (numerically validated) proof that, beyond well-known
anomalous reflection applications \cite{Oliveri 2021c}\cite{Oliveri 2022},
flat patterned passive panels can be profitably used to improve the
performance of traditional metallic screens in end-to-end \emph{NLOS}
wireless communications as well as the free-space limit derived by
the theory of images;
\item the analytic derivation of a closed-form upper bound (\ref{eq:TPA limit EMS})
for the achievable \emph{TPA} in an \emph{EMS}-powered point-to-point
reflective wireless links;
\item the definition of the optimality condition on the \emph{EMS} screen
aperture $\Theta$ of side $L$ (\ref{eq: L optimality condition})
so that the corresponding \emph{TPA} value is better than the asymptotic
one from a \emph{PCS} of infinite extension ($L\to\infty$).
\end{itemize}
\noindent From the numerical validation and performance assessment,
the following user-guidelines can be drawn

\begin{itemize}
\item \noindent subject to the \emph{EMS}-screen optimality condition (\ref{eq: L optimality condition}),
an $L$-side \emph{EMS}-screen outperforms a \emph{PCS} with the same
size in terms of \emph{TPA} and it also overcomes the asymptotic end-to-end
free-space \emph{TPA} limit;
\item \noindent the advantage of using an \emph{EMS} screen with respect
to a \emph{PCS} with the same aperture $\Theta$ becomes greater and
greater widening the panel side $L$;
\item the greater the receiver distance, $r_{RX}$, the wider must be the
patterned panel to yield improved performance with respect to the
equivalent-aperture \emph{PCS} as well as to overcome the free-space
path attenuation limit. On the other hand, the same target \emph{TPA}
of a standard metallic panel can be reached by an \emph{EMS} screen
of the same side $L$ (i.e., equal aperture $\Theta$) over longer
end-to-end distances;
\item wider \emph{EMS} panels are necessary to overcome the asymptotic free-space
limit as the end-to-end distance of \emph{NLOS} links, $\rho$, grows,
but the side $L$ of the squared \emph{EMS} aperture increases proportionally
to the square-root of $\rho$;
\item \noindent the performance enhancement granted by the \emph{EMS}-screens
with respect to \emph{PCS}s reduces when tilting more and more the
specular reflection angle $\theta_{0}$ towards end-fire ($\to$ $\theta_{0}=90$
{[}deg{]}). Such an efficiency reduction depends on the meta-atom
features, as well;
\item \noindent the use of simple \emph{EMS} meta-atoms as unit cells for
the patterned reflective panel does not prevent the synthesized \emph{EMS}
screen either to yield \emph{TPA} values better than the asymptotic
limit for \emph{PCS}s or/and to get better performance than the equal-aperture
\emph{PCS}. On the other hand, it is possible to reduce the \emph{TPA}
gap from the upper bound value (\ref{eq:TPA limit EMS}) by recurring
to more sophisticated and performing meta-atoms;
\item as a by-product, the synthesis method \cite{Oliveri 2022c} used for
the design of the \emph{EMS} screens proved further to be reliable,
effective, and computationally efficient.
\end{itemize}
Moreover, the following {}``tools'' for the design of an \emph{EMS}
screen have been derived:

\begin{itemize}
\item a minority(majority) relationship between the threshold length, $L_{TH}$,
and the Fresnel one, $L_{FR}$, {[}i.e., $L_{FR}>L_{TH}$($L_{FR}<L_{TH}$){]}
is a condition of existence(non-existence) of an $L$-size \emph{EMS}-screen
better than a \emph{PCS} of infinite extension ($L\to\infty$) for
a \emph{NLOS} wireless link at the frequency $f$ between a transmitter
and a receiver located at a distance $r_{TX}$ and $r_{RX}$ from
the reflective panel;
\item the closed-form expression (\ref{eq:TPA limit EMS}) can be used to
roughly size an \emph{EMS} screen given the end-to-end \emph{NLOS}
wireless scenario (i.e., $G_{RX}$, $r_{RX}$, $G_{TX}$, $r_{TX}$,
and $\theta_{0}$) or, vice-versa, to derive the end-to-end \emph{NLOS}
wireless scenario given the $L$-sided \emph{EMS} screen, starting
from the user needs in terms of \emph{QoS} (i.e., a \emph{TAP} target);
\item the expression (\ref{eq: L_TH}) is a reliable rule-of-thumb to quickly
(i.e., without any synthesis process and/or full-wave numerical simulations)
predict, for a given wireless link and incidence angle, the smallest
panel size $L$ of an \emph{EMS} screen that overcomes the free-space
end-to-end attenuation limit;
\item the difference between (\ref{eq:TPA limit EMS}) and (\ref{eq:limit PCS})
gives an expression for a rough estimate of the maximum margin of
improvement of an \emph{EMS} screen with respect to a \emph{PCS};
\end{itemize}
\noindent Future works, beyond the scope of the current manuscript,
will be aimed at extending the formulation and the theoretical study
to reconfigurable \emph{EMS}s for managing dynamic scenarios as well
as to shaped-beam reflecting \emph{EMS}s for advanced holographic
purposes. Moreover, the synthesis of more complex and efficient \emph{EMS}
meta-atoms is on the agenda to reach the \emph{TPA} upper bound of
an ideal \emph{EMS} screen.

\section*{\noindent Appendix}

\subsection*{A.1 - Proof of (\ref{eq:TPA limit EMS})}

\noindent With reference to the scenario in Fig. 1(\emph{b}), the
maximum incident power on an $L$-side square \emph{EMS} screen is
equal to\begin{equation}
\mathcal{P}_{INC}^{opt}=\frac{G_{TX}\mathcal{P}_{TX}}{4\pi r_{TX}^{2}}\cos\left(\theta_{TX}\right)L^{2}\label{eq: A1.1}\end{equation}
when the transmitter is in far field and it has a uniform gain within
the solid angle between the same transmitter and the \emph{EMS} surface
$\Theta$ \cite{Balanis 2012}.

\noindent In ideal conditions, the \emph{EMS} operates as an aperture
antenna with gain\begin{equation}
G_{opt}^{EMS}=\frac{4\pi\cos\left(\theta_{RX}\right)L^{2}}{\lambda^{2}}\label{eq:A1.2}\end{equation}
 and input power $\mathcal{P}_{INC}^{opt}$ since all the incident
power is sent along to the receiver direction.

\noindent Accordingly, the Friis' transmission equation reads as \cite{Balanis 2012}\begin{equation}
\mathcal{P}_{RX}^{EMS_{opt}}=\mathcal{P}_{INC}^{opt}\times G_{opt}^{EMS}\times G_{RX}\times\frac{\lambda^{2}}{4\pi r_{RX}^{2}}\label{eq: A1.3}\end{equation}
where $G_{RX}$ is the gain of the antenna at the receiver location
$\mathbf{r}=\mathbf{r}_{RX}$.

\noindent By simple substitution of (\ref{eq: A1.1}) and (\ref{eq:A1.2})
in (\ref{eq: A1.3}), it turns out that\begin{equation}
\begin{array}{r}
\mathcal{P}_{RX}^{EMS_{opt}}=\mathcal{P}_{TX}\frac{G_{RX}G_{TX}}{\left(4\pi r_{TX}r_{RX}\right)^{2}}\cos\left(\theta_{TX}\right)\cos\left(\theta_{RX}\right)L^{4}\end{array}\label{eq:A1.4}\end{equation}
and the expression (\ref{eq:TPA limit EMS}) for the upper bound of
an ideal \emph{EMS} square screen of side $L$ is yielded by using
(\ref{eq:A1.4}) in (\ref{eq:TPA}).

\subsection*{\noindent A.2 - Proof of (\ref{eq: L optimality condition})}

\noindent By combining (\ref{eq:limit PCS}) and (\ref{eq:TPA limit EMS}),
one obtains\begin{equation}
\frac{G_{TX}G_{RX}\cos\left(\theta_{TX}\right)\cos\left(\theta_{RX}\right)L^{4}}{\left(4\pi r_{TX}r_{RX}\right)^{2}}\geq\left(\frac{\lambda}{4\pi\left(r_{RX}+r_{TX}\right)}\right)^{2}G_{RX}G_{TX}\label{eq:A2.1}\end{equation}
that, solved with respect to $L$, becomes\begin{equation}
L^{4}\geq\frac{\lambda^{2}}{\cos\left(\theta_{RX}\right)\cos\left(\theta_{TX}\right)}\left(\frac{r_{TX}\times r_{RX}}{r_{RX}+r_{TX}}\right)^{2}.\label{eq:A2.2}\end{equation}
The expression (\ref{eq:A2.2}) can be further simplified into\begin{equation}
L\geq L_{TH}\label{eq:A2.3}\end{equation}
by taking into account that $\theta_{RX}=\theta_{TX}=\theta_{0}$.

\noindent Since the above considerations hold true if $r_{RX}$ complies
with (\ref{eq:fresnel condition}), let us rewrite this latter as
a function of $L$\begin{equation}
\left\{ \begin{array}{l}
L\leq\frac{r_{RX}}{10\sqrt{2}}\\
L\leq\left[\left(\frac{r_{RX}}{0.62}\right)^{2}\frac{\lambda}{2\sqrt{2}}\right]^{\frac{1}{3}}\end{array}\right.\label{eq:A2.4}\end{equation}
still under the hypothesis that $r_{RX}\geq10\lambda$ to fulfil (\ref{eq:fresnel condition}).

\noindent The optimality condition (\ref{eq: L optimality condition})
is then derived by combining (\ref{eq:A2.3}) and (\ref{eq:A2.4}).

\section*{\noindent Acknowledgements}

\noindent This work benefited from the networking activities carried
out within the Project {}``Cloaking Metasurfaces for a New Generation
of Intelligent Antenna Systems (MANTLES)'' (Grant No. 2017BHFZKH)
funded by the Italian Ministry of Education, University, and Research
under the PRIN2017 Program (CUP: E64I19000560001). Moreover, it benefited
from the networking activities carried out within the Project {}``SPEED''
(Grant No. 61721001) funded by National Science Foundation of China
under the Chang-Jiang Visiting Professorship Program, the Project
'Inversion Design Method of Structural Factors of Conformal Load-bearing
Antenna Structure based on Desired EM Performance Interval' (Grant
no. 2017HZJXSZ) funded by the National Natural Science Foundation
of China. A. Massa wishes to thank E. Vico for her never-ending inspiration,
support, guidance, and help.

\newpage
\section*{FIGURE CAPTIONS}

\begin{itemize}
\item \textbf{Figure 1.} \emph{Problem Formulation}. Sketch of (\emph{a})
the problem scenario, (\emph{b})(\emph{c}) the principle of operation
of (\emph{b}) the \emph{EMS} screen and (\emph{c}) the \emph{PCS},
and (\emph{d}) the layout of the \emph{EMS} meta-atom.
\item \textbf{Figure 2.} \emph{Problem Formulation.} Sketch of (\emph{a})
reference pyramidal horn antenna and (\emph{b})(\emph{c}) layouts
featuring a gain of (\emph{b}) $G=15.4$ {[}dBi{]} and (\emph{c})
$G=25.5$ {[}dBi{]} at $f=27$ {[}GHz{]}.
\item \textbf{Figure 3.} \emph{Numerical Validation} ($f=27$ {[}GHz{]},
$\mathcal{P}_{TX}=20$ {[}dBm{]}, $G_{TX}=15.4$ {[}dBi{]}, $\theta_{TX}=30$
{[}deg{]}) - Plots of the distribution of (\emph{a})(\emph{b}) the
magnitude and (\emph{c})(\emph{d}) the phase of $E_{y}^{inc}$ within
the screen aperture $\Theta$ when the transmitter in located (\emph{a})(\emph{c})
$r_{TX}=15$ {[}m{]} and (\emph{b})(\emph{d}) $r_{TX}=50$ {[}m{]}
far from the reflective panel.
\item \textbf{Figure 4.} \emph{Numerical Validation} ($f=27$ {[}GHz{]},
$\mathcal{P}_{TX}=20$ {[}dBm{]}, $G_{TX}=G_{RX}=15.4$ {[}dBi{]},
$r_{TX}=r_{RX}=15$ {[}m{]}, $\theta_{0}=30$ {[}deg{]}, $L=0.8$
{[}m{]}) - \emph{EMS} layout.
\item \textbf{Figure 5.} \emph{Numerical Validation} ($f=27$ {[}GHz{]},
$\mathcal{P}_{TX}=20$ {[}dBm{]}, $G_{TX}=G_{RX}=15.4$ {[}dBi{]},
$r_{TX}=r_{RX}=15$ {[}m{]}, $\theta_{0}=30$ {[}deg{]}, $L=0.8$
{[}m{]}) - Maps of (\emph{a})(\emph{b}) the phase of the electric
surface current, $\angle J_{e}^{y}$, induced on the panel aperture,
$\Theta$, and (\emph{c})-(\emph{f}) the magnitude of the dominant
component of the field reflected (i.e., $\left|E_{\varphi}^{sca}\right|$)
in (\emph{c})(\emph{d}) the {}``transversal cut'' and (\emph{e})(\emph{f})
the {}``longitudinal cut'' {[}Fig. 1(\emph{b}){]} by (\emph{a})(\emph{c})(\emph{e})
the \emph{PCS} and (\emph{b})(\emph{d})(\emph{f}) the \emph{EMS}-screen
in Fig. 4.
\item \textbf{Figure 6.} \emph{Numerical Validation} ($f=27$ {[}GHz{]},
$G_{TX}=G_{RX}=15.4$ {[}dBi{]}, $r_{TX}=r_{RX}=15$ {[}m{]}, $\theta_{0}=30$
{[}deg{]}) - Plots of (\emph{a}) the \emph{TPA} versus the side $L$
of the reflective panel and (\emph{b})(\emph{c}) the layouts of the
synthesized $L$-sided square \emph{EMS}-screens: (\emph{b}) $L=0.4$
{[}m{]} and (\emph{c}) $L=1.0$ {[}m{]}.
\item \textbf{Figure 7.} \emph{Numerical Validation} ($f=27$ {[}GHz{]},
$G_{TX}=G_{RX}=15.4$ {[}dBi{]}, $r_{TX}=r_{RX}=15$ {[}m{]}, $\theta_{0}=30$
{[}deg{]}) - Maps of (\emph{a})-(\emph{d}) the magnitude of the dominant
component of the field reflected (i.e., $\left|E_{\varphi}^{sca}\right|$)
in the {}``transversal cut'' {[}Fig. 1(\emph{b}){]} by (\emph{a})(\emph{c})
the \emph{PCS} and (\emph{b})(\emph{d}) the \emph{EMS}-screen in Figs.
6(\emph{b})-6(\emph{c}) of side length (\emph{a})(\emph{b}) $L=0.4$
{[}m{]} and (\emph{b})(\emph{c}) $L=1.0$ {[}m{]}.
\item \textbf{Figure 8.} \emph{Numerical Validation} ($f=27$ {[}GHz{]},
$G_{TX}=G_{RX}=15.4$ {[}dBi{]}, $r_{TX}=r_{RX}=15$ {[}m{]}, $\theta_{0}=30$
{[}deg{]}) - Maps of (\emph{a})-(\emph{d}) the magnitude of the dominant
component of the field reflected (i.e., $\left|E_{\varphi}^{sca}\right|$)
in the {}``longitudinal cut'' {[}Fig. 1(\emph{b}){]} by (\emph{a})(\emph{c})
the \emph{PCS} and (\emph{b})(\emph{d}) the \emph{EMS}-screen in Figs.
6(\emph{b})-6(\emph{c}) of side length (\emph{a})(\emph{b}) $L=0.4$
{[}m{]} and (\emph{b})(\emph{c}) $L=1.0$ {[}m{]}.
\item \textbf{Figure 9.} \emph{Numerical Validation} ($f=27$ {[}GHz{]},
$G_{TX}=G_{RX}=15.4$ {[}dBi{]}, $r_{TX}=r_{RX}=50$ {[}m{]}, $\theta_{0}=30$
{[}deg{]}) - Plots of (\emph{a}) the \emph{TPA} versus the side $L$
of the reflective panel and (\emph{b})(\emph{c}) the layouts of the
synthesized $L$-sided square \emph{EMS}-screens: (\emph{b}) $L=0.5$
{[}m{]} and (\emph{c}) $L=2.9$ {[}m{]}.
\item \textbf{Figure 10.} \emph{Numerical Validation} ($f=27$ {[}GHz{]},
$G_{TX}=G_{RX}=15.4$ {[}dBi{]}, $r_{TX}=r_{RX}=50$ {[}m{]}, $\theta_{0}=30$
{[}deg{]}) - Maps of (\emph{a})-(\emph{d}) the magnitude of the dominant
component of the field reflected (i.e., $\left|E_{\varphi}^{sca}\right|$)
in the {}``transversal cut'' {[}Fig. 1(\emph{b}){]} by (\emph{a})(\emph{c})
the \emph{PCS} and (\emph{b})(\emph{d}) the \emph{EMS}-screen in Figs.
9(\emph{b})-9(\emph{c}) of side length (\emph{a})(\emph{b}) $L=0.5$
{[}m{]} and (\emph{b})(\emph{c}) $L=2.9$ {[}m{]}.
\item \textbf{Figure 11.} \emph{Numerical Validation} ($f=27$ {[}GHz{]},
$G_{TX}=G_{RX}=25.5$ {[}dBi{]}, $r_{TX}=r_{RX}=r_{0}$, $\theta_{0}=30$
{[}deg{]}) - Plots of the \emph{TPA} versus the side $L$ of the reflective
panel when (\emph{a}) $r_{0}=15$ {[}m{]} and (\emph{b}) $r_{0}=50$
{[}m{]}.
\item \textbf{Figure 12.} \emph{Numerical Validation} ($f=27$ {[}GHz{]},
$G_{TX}=G_{RX}=25.5$ {[}dBi{]}, $r_{TX}=r_{RX}=r_{0}$, $\theta_{0}=30$
{[}deg{]}) - Plots of (\emph{a})(\emph{b}) the layouts of the \emph{EMS}-screens
and (\emph{a})(\emph{b}) the magnitude of the dominant component of
the field reflected (i.e., $\left|E_{\varphi}^{sca}\right|$) in the
{}``transversal cut'' {[}Fig. 1(\emph{b}){]} at (\emph{c})(\emph{e})
$r_{0}=15$ {[}m{]} and (\emph{d})(\emph{f}) $r_{0}=50$ {[}m{]} far
from the reflective aperture by (\emph{a})(\emph{c}) the \emph{EMS}-screens
in Figs. 12(\emph{a})-12(\emph{b}) and (\emph{b})(\emph{d}) the \emph{PCS}
with side length (\emph{c})(\emph{e}) $L=1.0$ {[}m{]} and (\emph{d})(\emph{f})
$L=2.9$ {[}m{]}.
\item \textbf{Figure 13.} \emph{Numerical Validation} ($f=27$ {[}GHz{]},
$G_{TX}=G_{RX}=25.5$ {[}dBi{]}, $r_{TX}=15$ {[}m{]}, $\theta_{0}=30$
{[}deg{]}) - Plot of the \emph{TPA} versus the receiver distance from
the screen, $r_{RX}$.
\item \textbf{Figure 14.} \emph{Numerical Validation} ($f=27$ {[}GHz{]},
$G_{TX}=G_{RX}=25.5$ {[}dBi{]}, $r_{TX}=r_{RX}=15$ {[}m{]}) - Plot
of (\emph{a}) the \emph{TPA} versus the side $L$ of the reflective
panel along with (\emph{b})(\emph{c}) the behavior of the side-length
thresholds/markers and (\emph{c}) the margins of \emph{TPA} improvement
of the synthesized $L=1$ {[}m{]}-sized \emph{EMS}s versus $\theta_{0}$.
\item \textbf{Figure 15.} \emph{Numerical Validation} ($f=27$ {[}GHz{]},
$G_{TX}=G_{RX}=25.5$ {[}dBi{]}, $r_{TX}=r_{RX}=15$ {[}m{]}, $L=1.0$
{[}m{]}) - Layouts of the \emph{EMS}-screens synthesized when (\emph{a})
$\theta_{0}=20$ {[}deg{]} and (\emph{b}) $\theta_{0}=45$ {[}deg{]}.
\item \textbf{Figure 16.} \emph{Numerical Validation} ($f=27$ {[}GHz{]},
$G_{TX}=G_{RX}=25.5$ {[}dBi{]}, $r_{TX}=r_{RX}=\frac{\rho}{2}$,
$\theta_{0}=30$ {[}deg{]}) - Plot of (\emph{a}) the \emph{TPA} versus
the side $L$ of the reflective panel along with (\emph{b})(\emph{c})
the behavior of the side-length thresholds/markers and (\emph{c})
the margins of \emph{TPA} improvement of the synthesized $L=6$ {[}m{]}-sized
\emph{EMS}s versus the end-to-end distance $\rho$.
\item \textbf{Figure 17.} \emph{Numerical Validation} ($f=27$ {[}GHz{]},
$G_{TX}=G_{RX}=25.5$ {[}dBi{]}, $r_{TX}=r_{RX}=\frac{\rho}{2}$,
$\theta_{0}=30$ {[}deg{]}, $L=6.0$ {[}m{]}) - Layouts of the \emph{EMS}-screens
synthesized when (\emph{a}) $\rho=400$ {[}m{]} and (\emph{b}) $\rho=2000$
{[}m{]}.
\end{itemize}

\section*{TABLE CAPTIONS}

\begin{itemize}
\item \textbf{Table I.} \emph{Numerical Validation} - Horn antenna descriptors.
\item \textbf{Table II.} \emph{Numerical Validation} - Values of the side-length
thresholds/markers.
\item \textbf{Table III.} \emph{Numerical Validation} - Values of the margins
of \emph{TPA} improvement.
\end{itemize}
\newpage
\begin{center}~\vfill\end{center}

\begin{center}\begin{tabular}{cc}
\multicolumn{2}{c}{\includegraphics[%
  width=0.70\columnwidth]{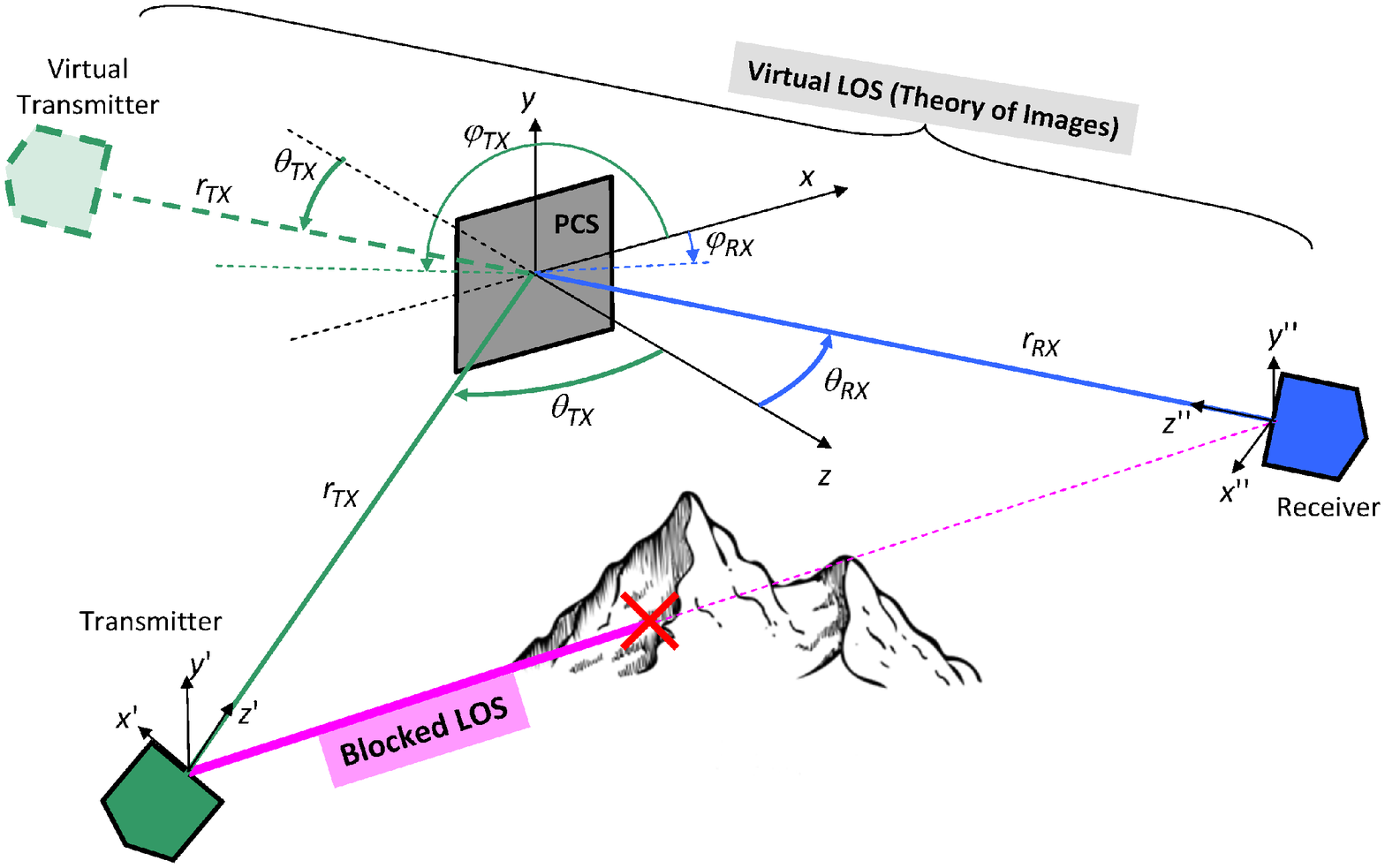}}\tabularnewline
\multicolumn{2}{c}{(\emph{a})}\tabularnewline
\includegraphics[%
  width=0.45\columnwidth]{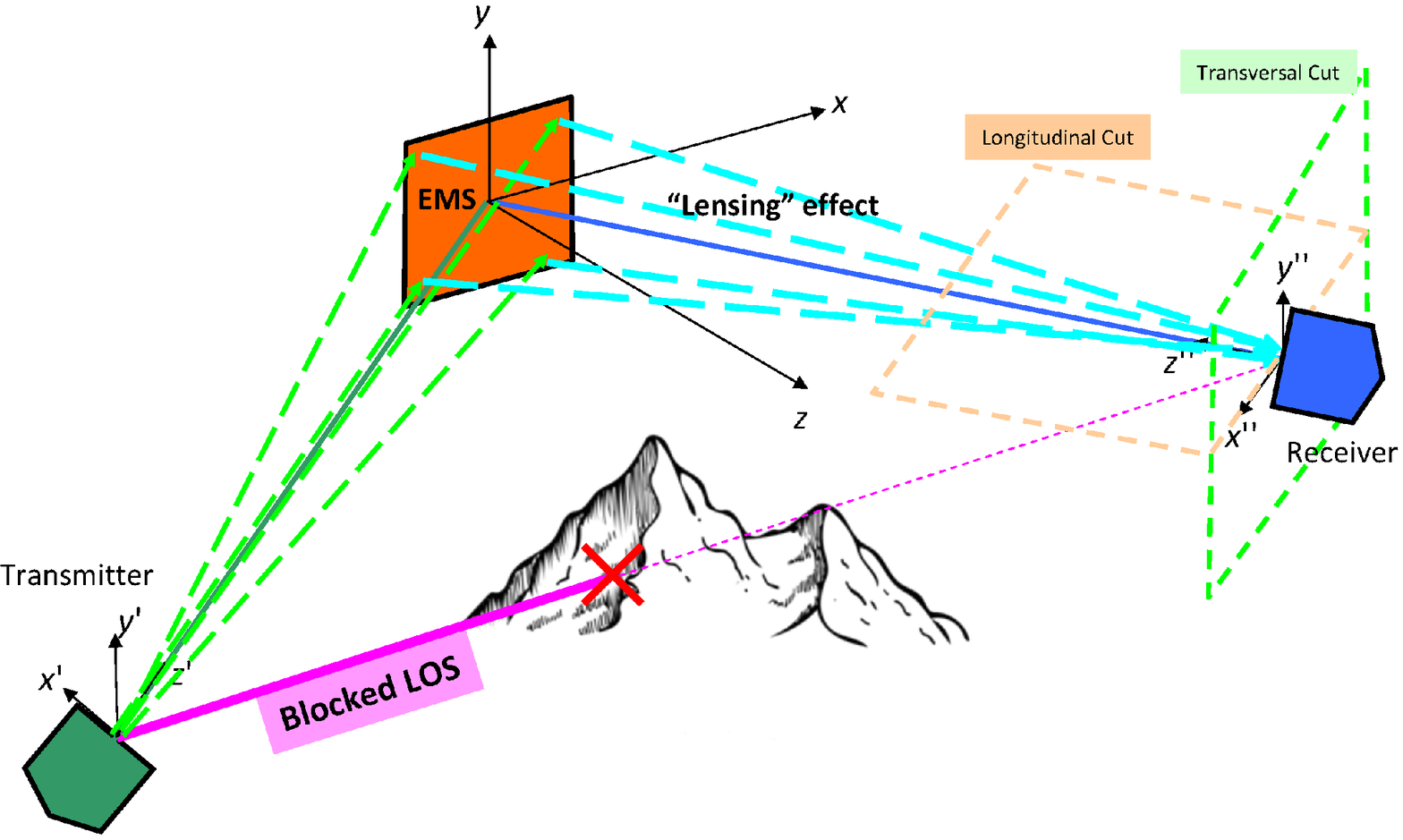}&
\includegraphics[%
  width=0.45\columnwidth]{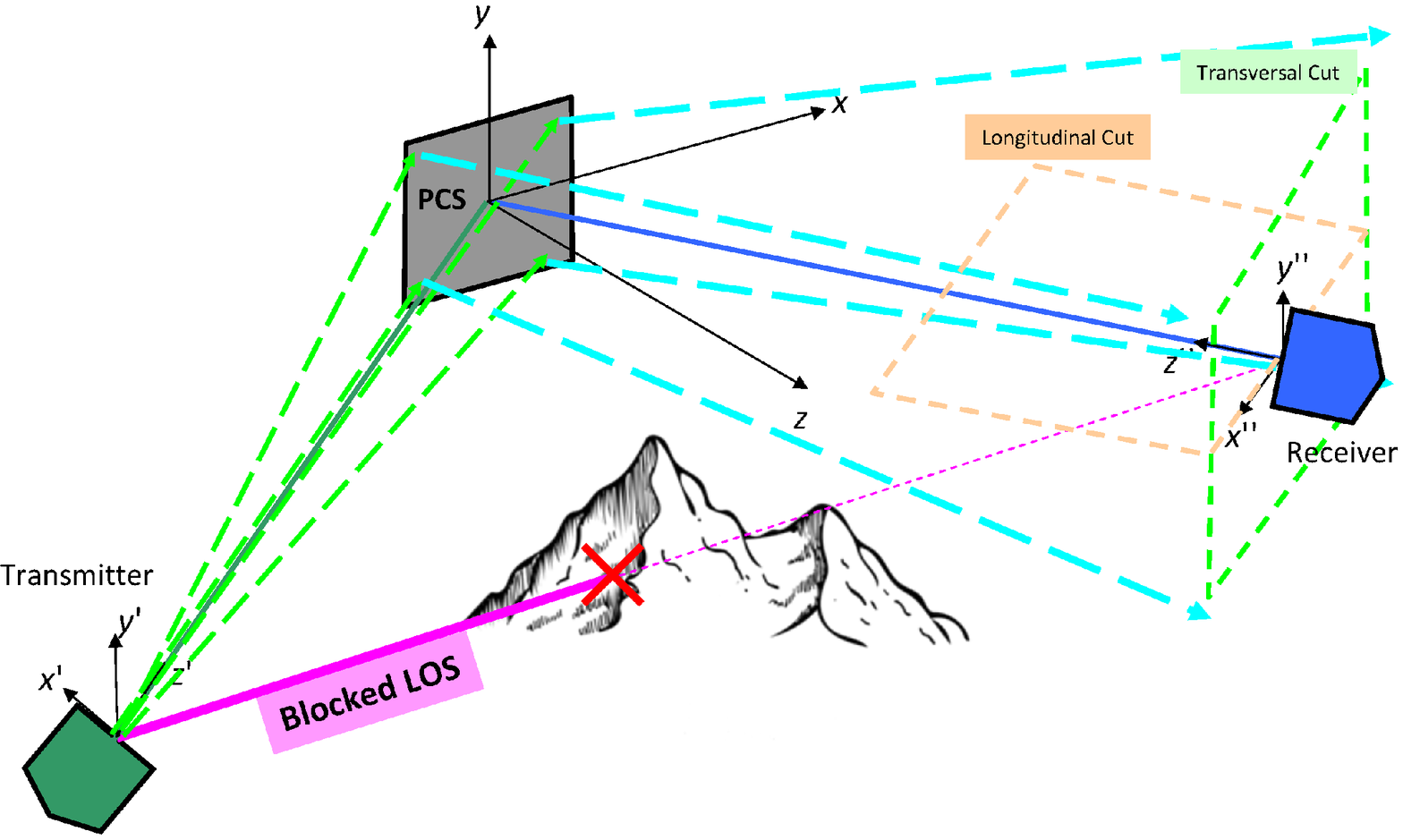}\tabularnewline
(\emph{b})&
(\emph{c})\tabularnewline
\multicolumn{2}{c}{\includegraphics[%
  width=0.45\columnwidth]{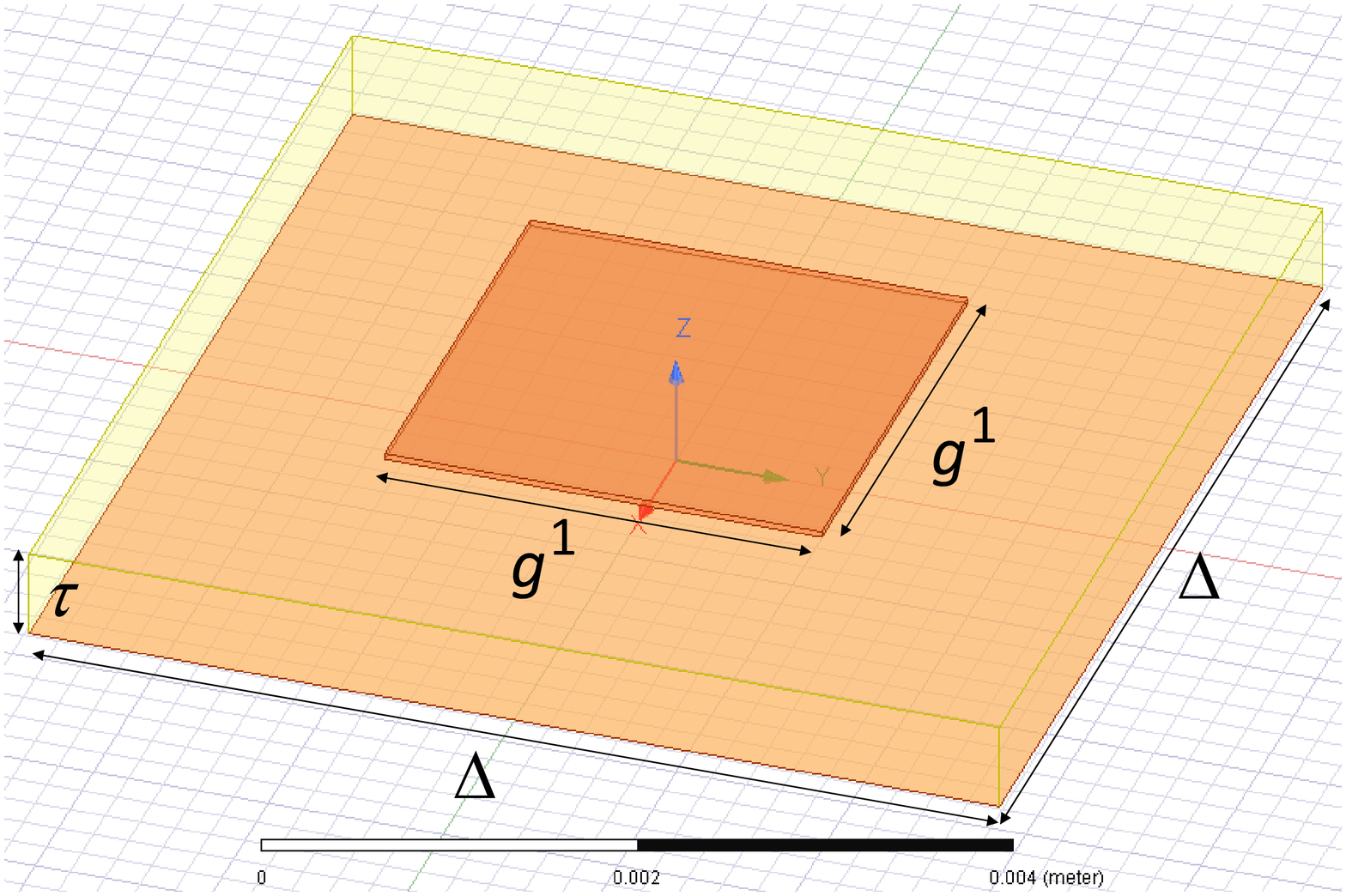}}\tabularnewline
\multicolumn{2}{c}{(\emph{d})}\tabularnewline
\end{tabular}\end{center}

\begin{center}~\vfill\end{center}

\begin{center}\textbf{Fig. 1 - G. Oliveri et} \textbf{\emph{al.}}\textbf{,}
\textbf{\emph{{}``}}Features and Potentialities of ... ''\end{center}

\newpage
\begin{center}~\vfill\end{center}

\begin{center}\begin{tabular}{cc}
\multicolumn{2}{c}{\includegraphics[%
  width=0.50\columnwidth]{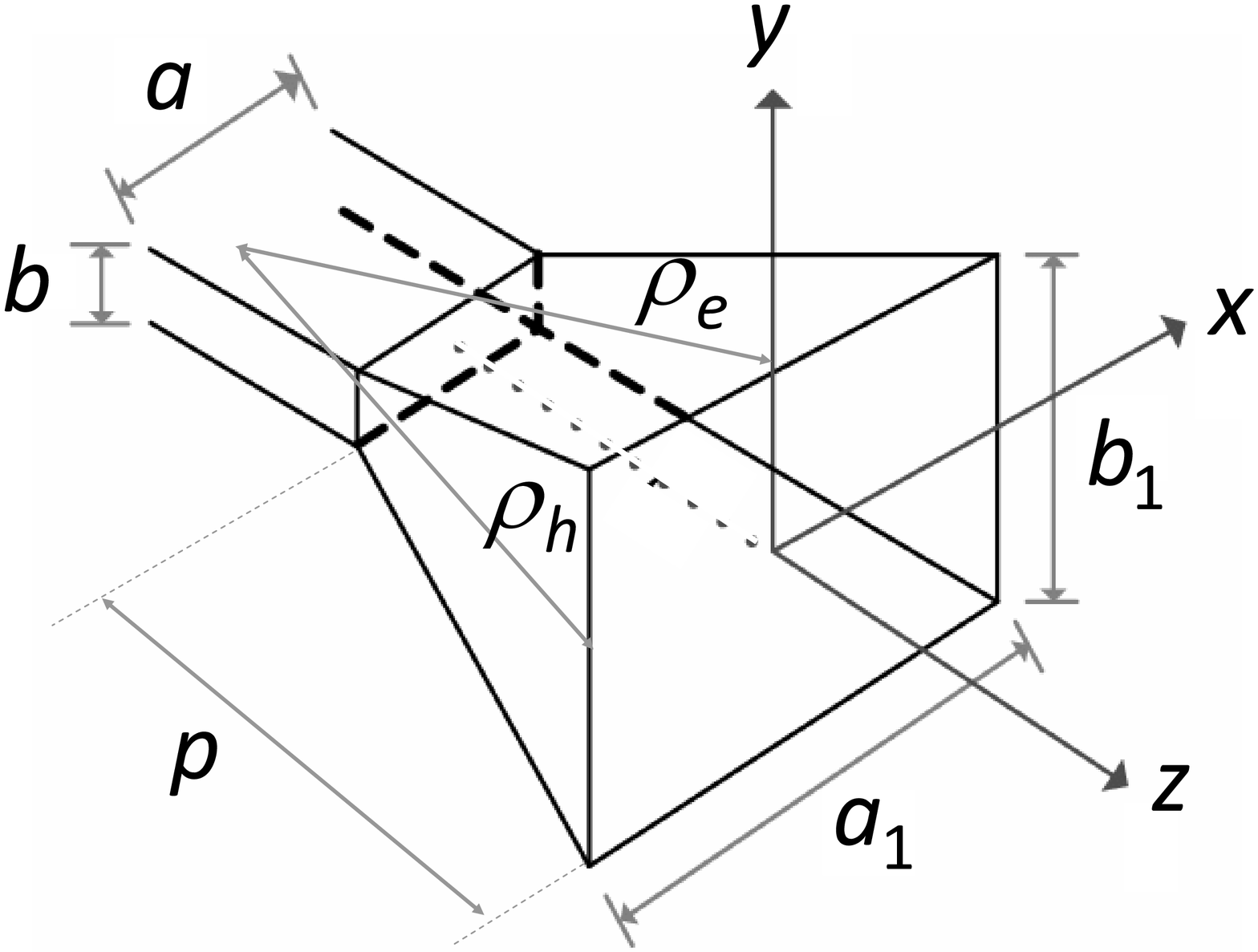}}\tabularnewline
\multicolumn{2}{c}{(\emph{a}) }\tabularnewline
\includegraphics[%
  width=0.35\columnwidth]{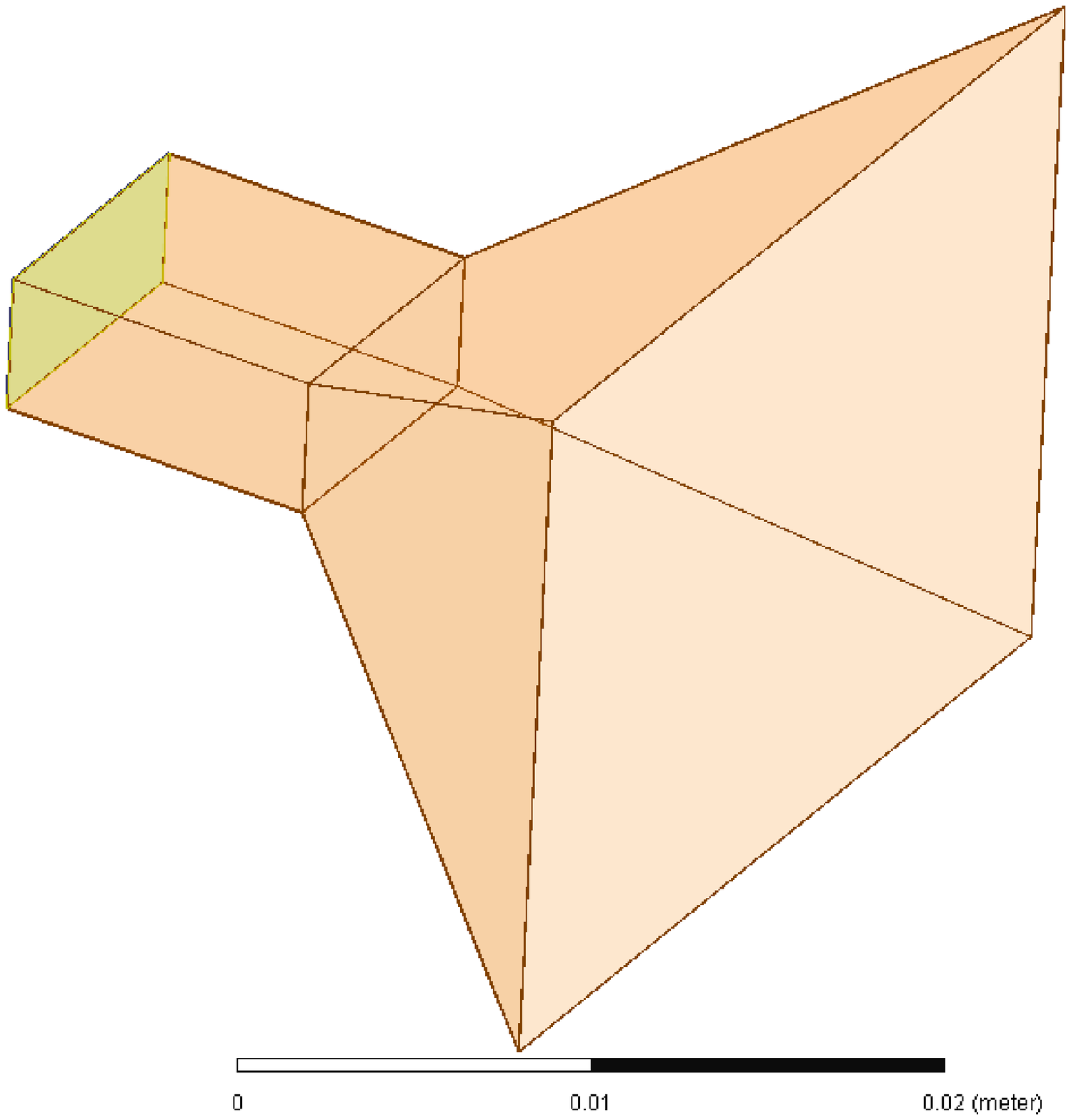}&
\includegraphics[%
  width=0.48\columnwidth]{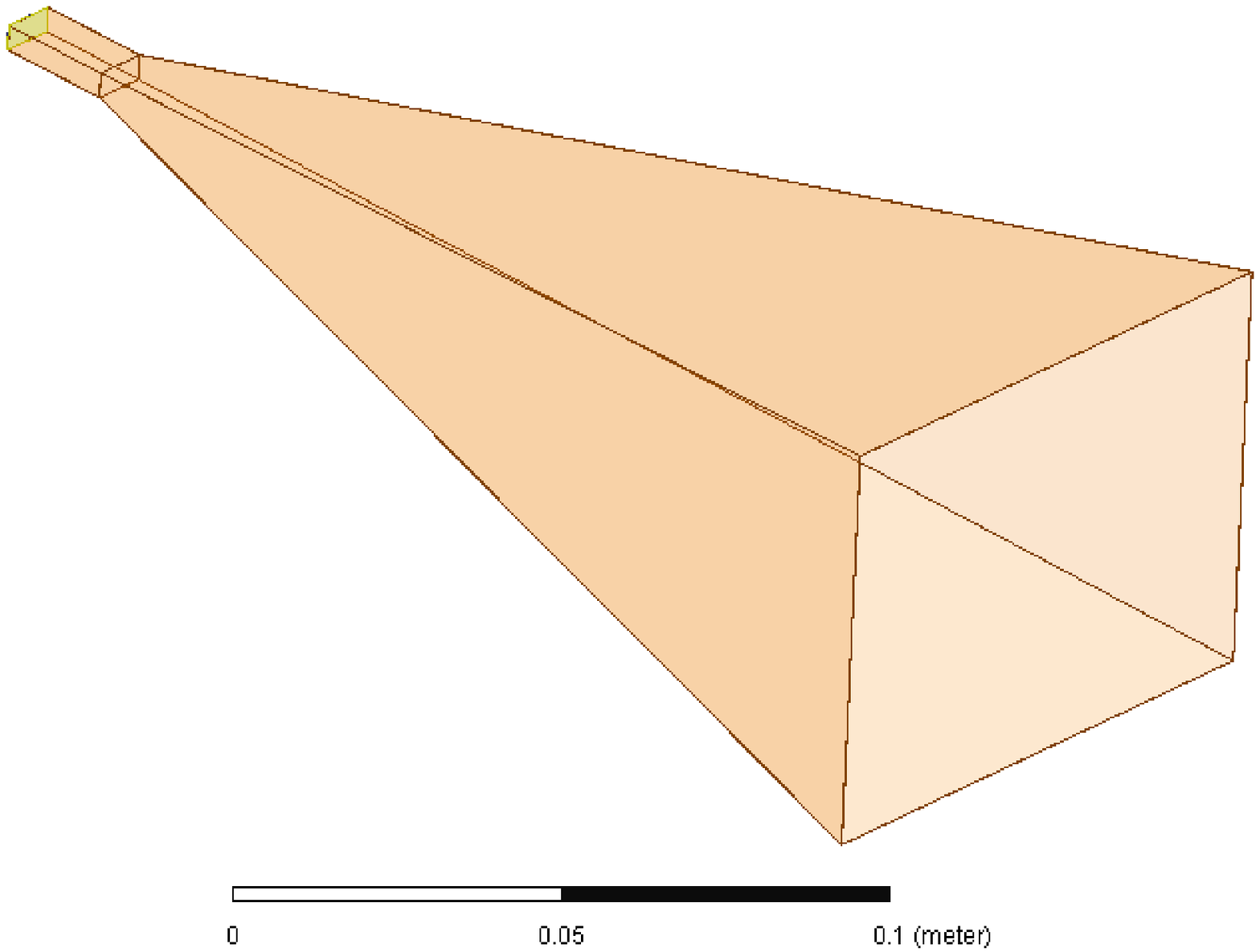}\tabularnewline
(\emph{b})&
(\emph{c})\tabularnewline
\end{tabular}\end{center}

\begin{center}~\vfill\end{center}

\begin{center}\textbf{Fig. 2 - G. Oliveri et} \textbf{\emph{al.}}\textbf{,}
\textbf{\emph{{}``}}Features and Potentialities of ... ''\end{center}

\newpage
\begin{center}~\vfill\end{center}

\begin{center}\begin{tabular}{cc}
\includegraphics[%
  width=0.45\columnwidth]{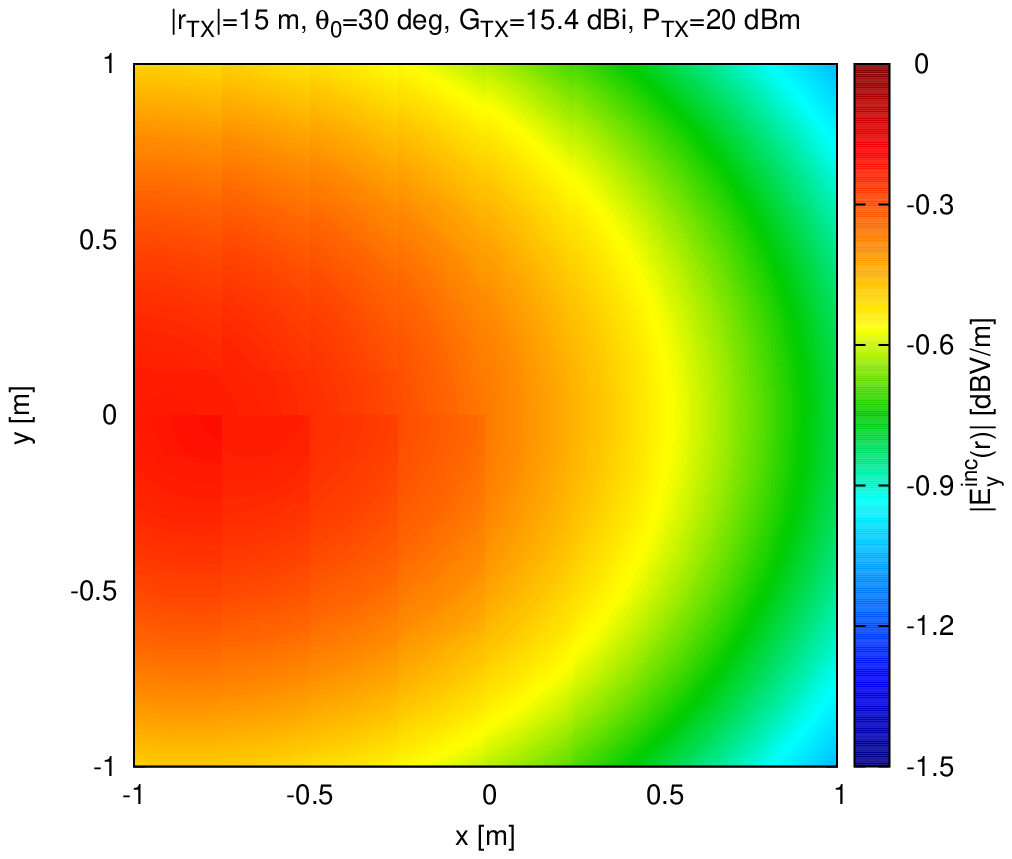}&
\includegraphics[%
  width=0.45\columnwidth]{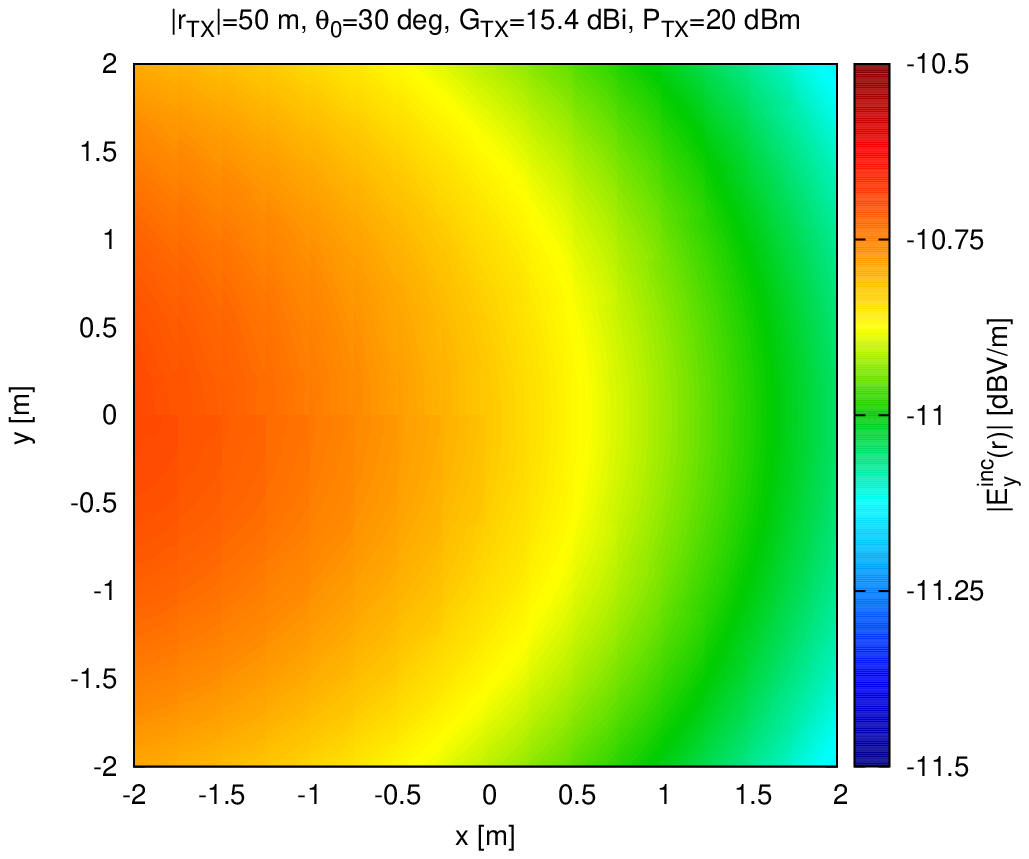}\tabularnewline
(\emph{a})&
(\emph{b})\tabularnewline
\includegraphics[%
  width=0.45\columnwidth]{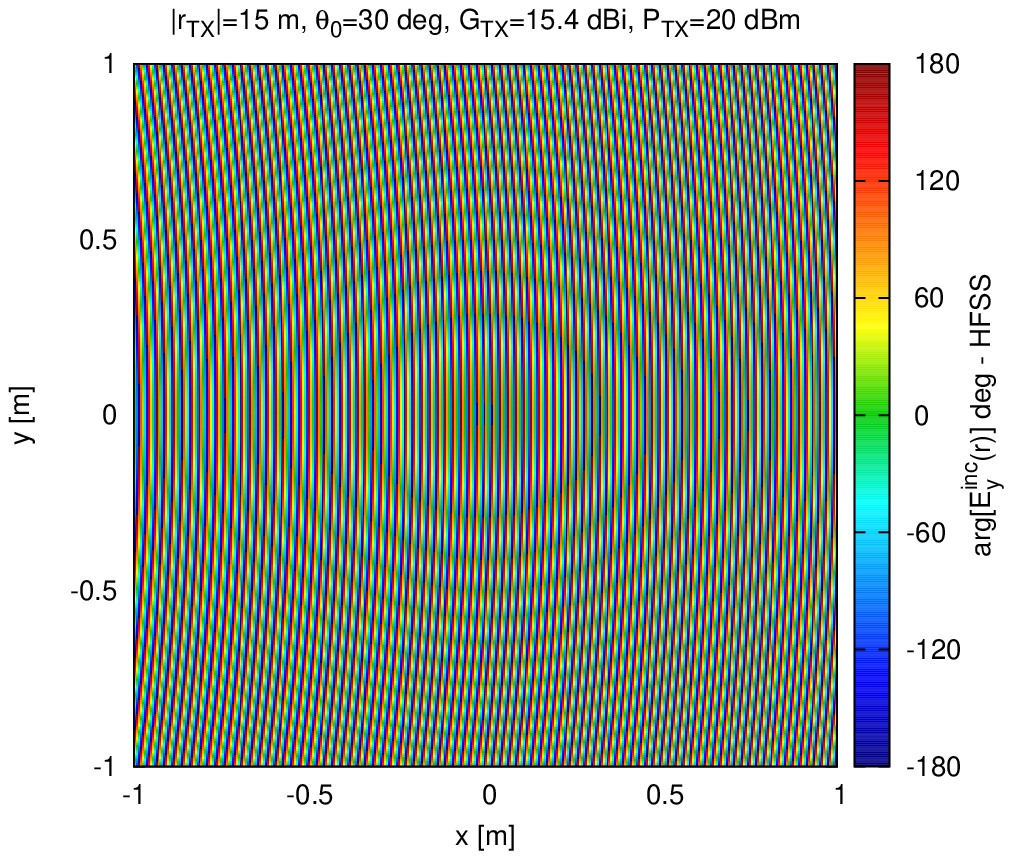}&
\includegraphics[%
  width=0.45\columnwidth]{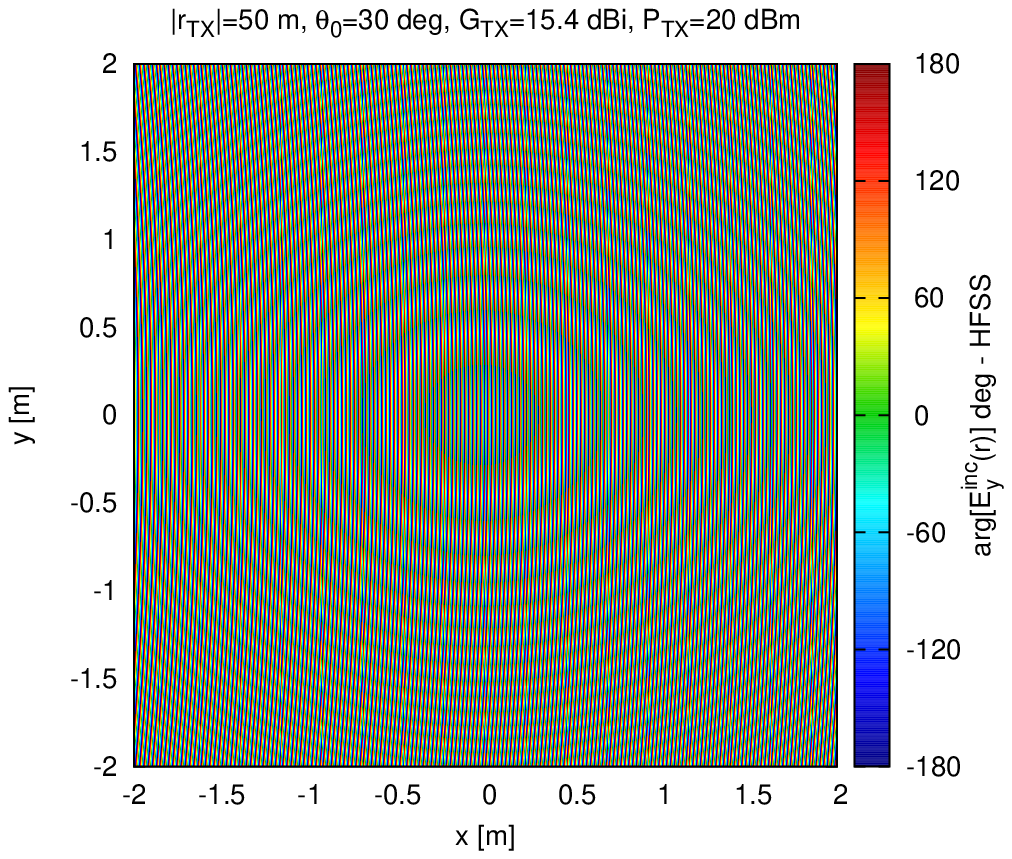}\tabularnewline
(\emph{c})&
(\emph{d})\tabularnewline
\end{tabular}\end{center}

\begin{center}~\vfill\end{center}

\begin{center}\textbf{Fig. 3 - G. Oliveri et} \textbf{\emph{al.}}\textbf{,}
\textbf{\emph{{}``}}Features and Potentialities of ... ''\end{center}

\newpage
\begin{center}~\vfill\end{center}

\begin{center}\includegraphics[%
  width=0.95\columnwidth]{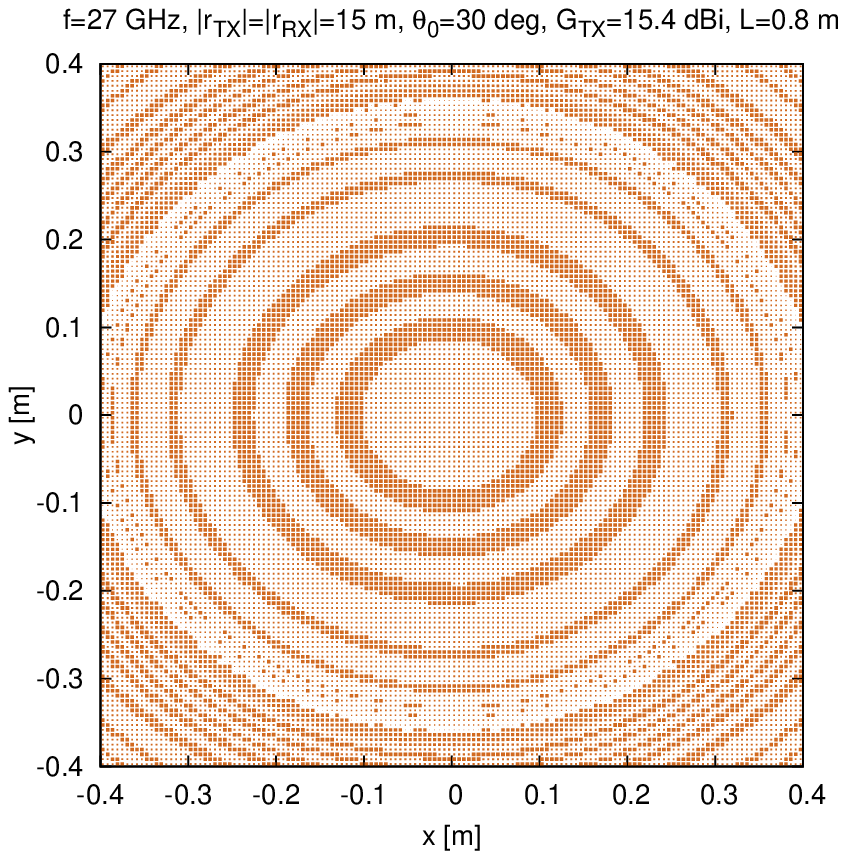}\end{center}

\begin{center}~\vfill\end{center}

\begin{center}\textbf{Fig. 4 - G. Oliveri et} \textbf{\emph{al.}}\textbf{,}
\textbf{\emph{{}``}}Features and Potentialities of ... ''\end{center}

\newpage
\begin{center}~\vfill\end{center}

\begin{center}\begin{tabular}{cc}
\emph{PCS}&
\emph{EMS}\tabularnewline
\includegraphics[%
  width=0.45\columnwidth]{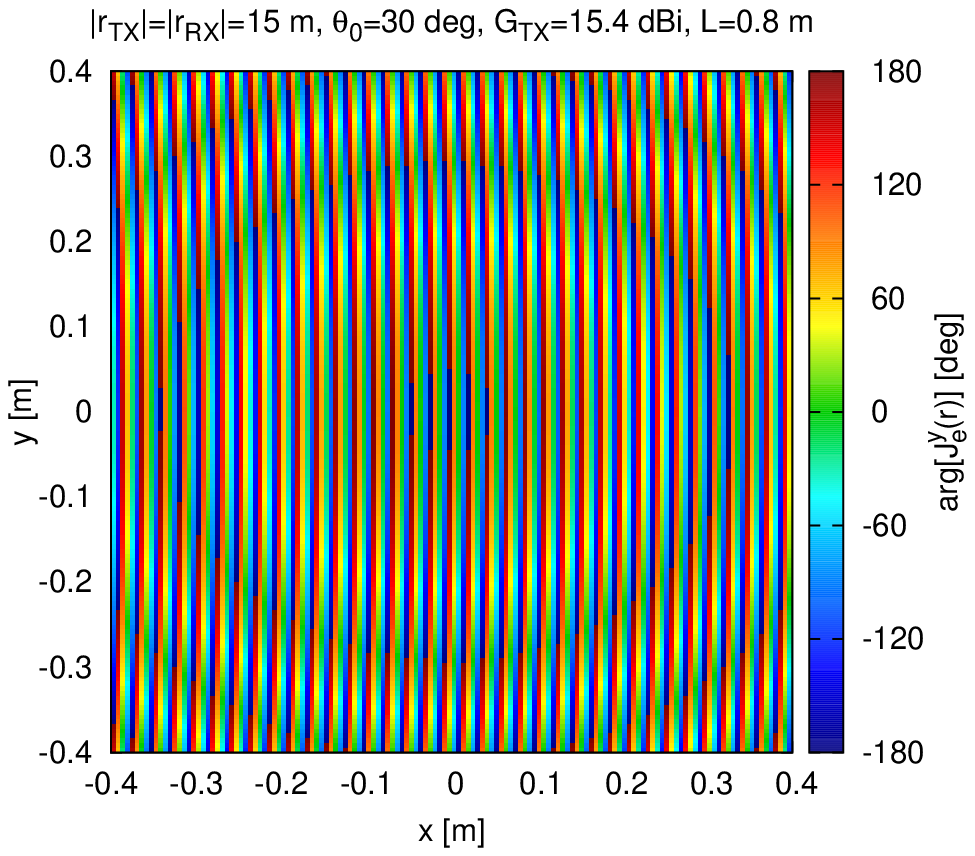}&
\includegraphics[%
  width=0.45\columnwidth]{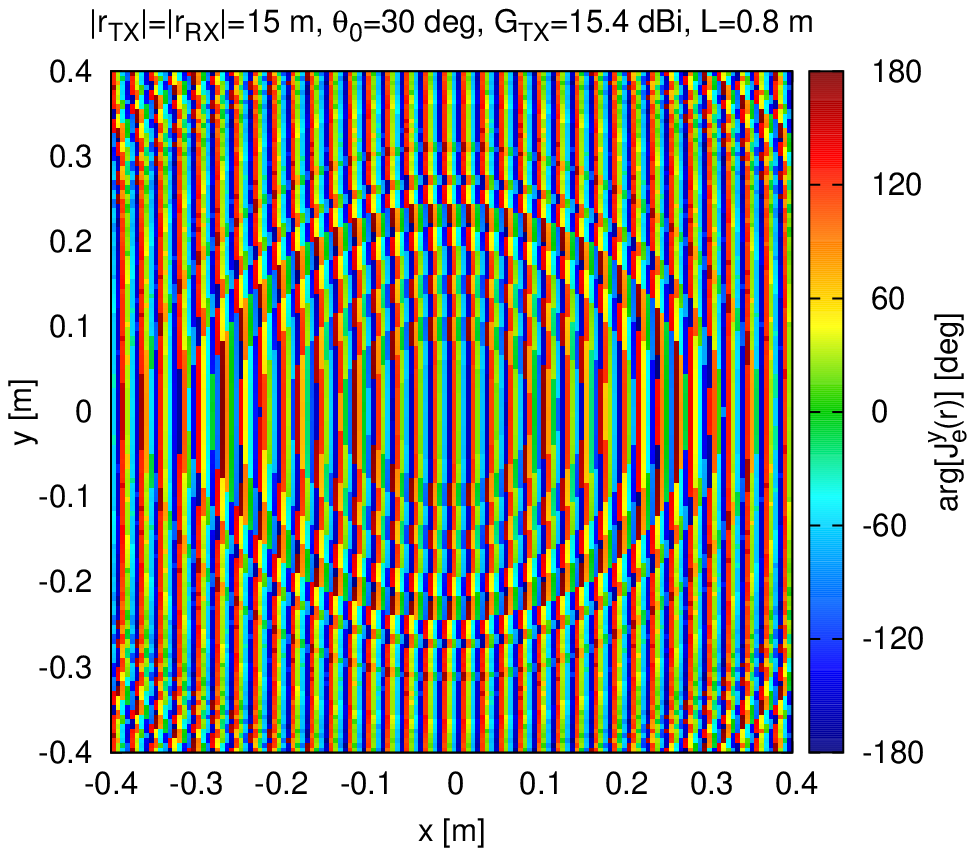}\tabularnewline
(\emph{a})&
(\emph{b})\tabularnewline
\includegraphics[%
  width=0.45\columnwidth]{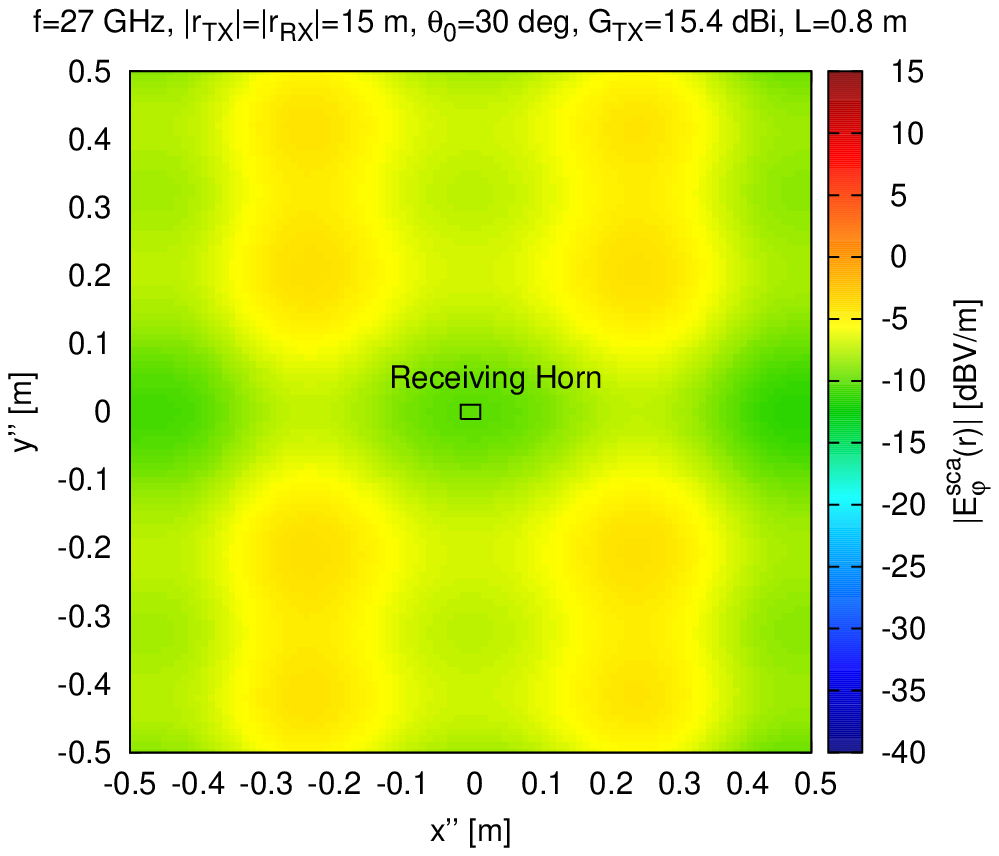}&
\includegraphics[%
  width=0.45\columnwidth]{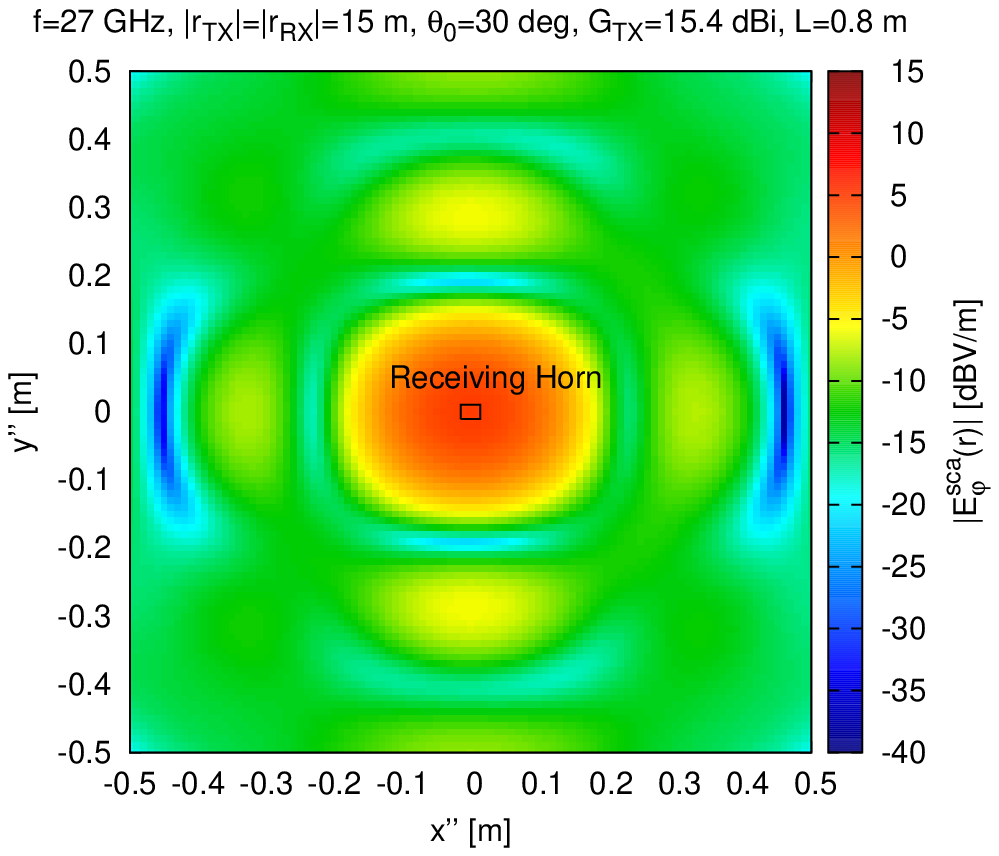}\tabularnewline
(\emph{c})&
(\emph{d})\tabularnewline
\includegraphics[%
  width=0.45\columnwidth]{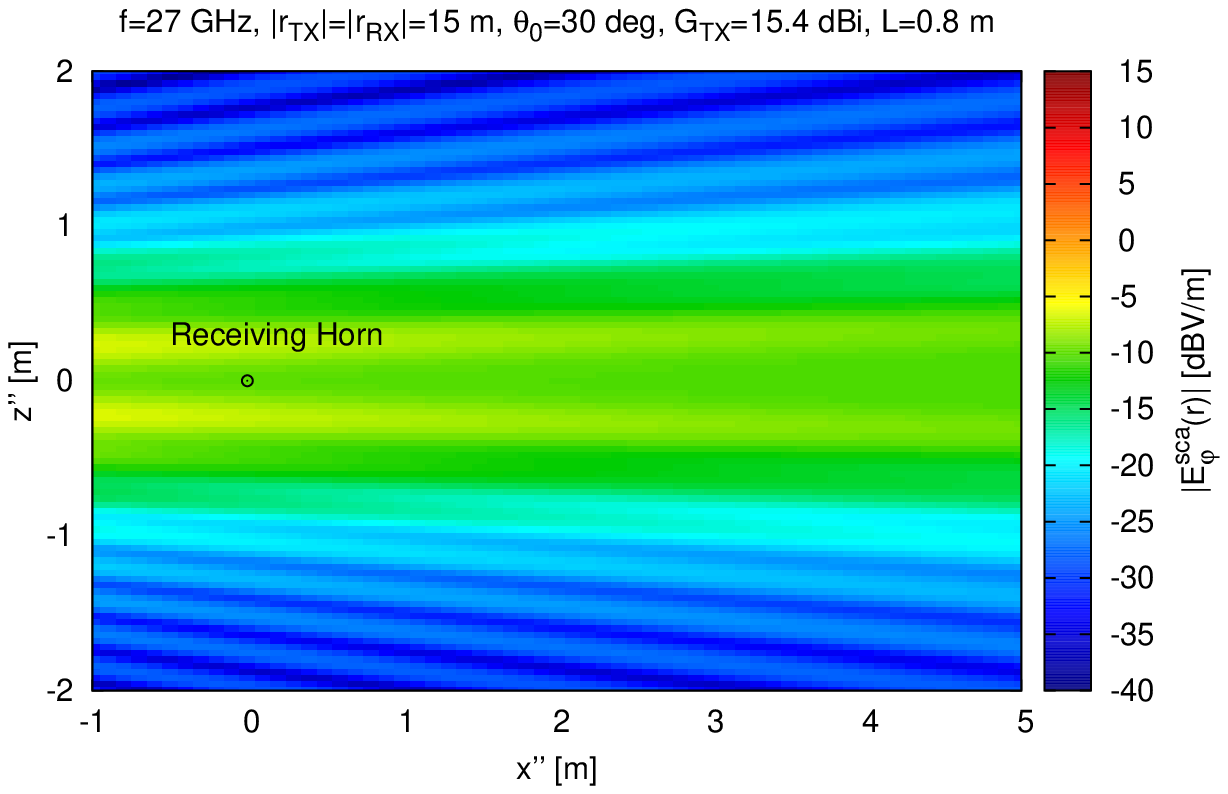}&
\includegraphics[%
  width=0.45\columnwidth]{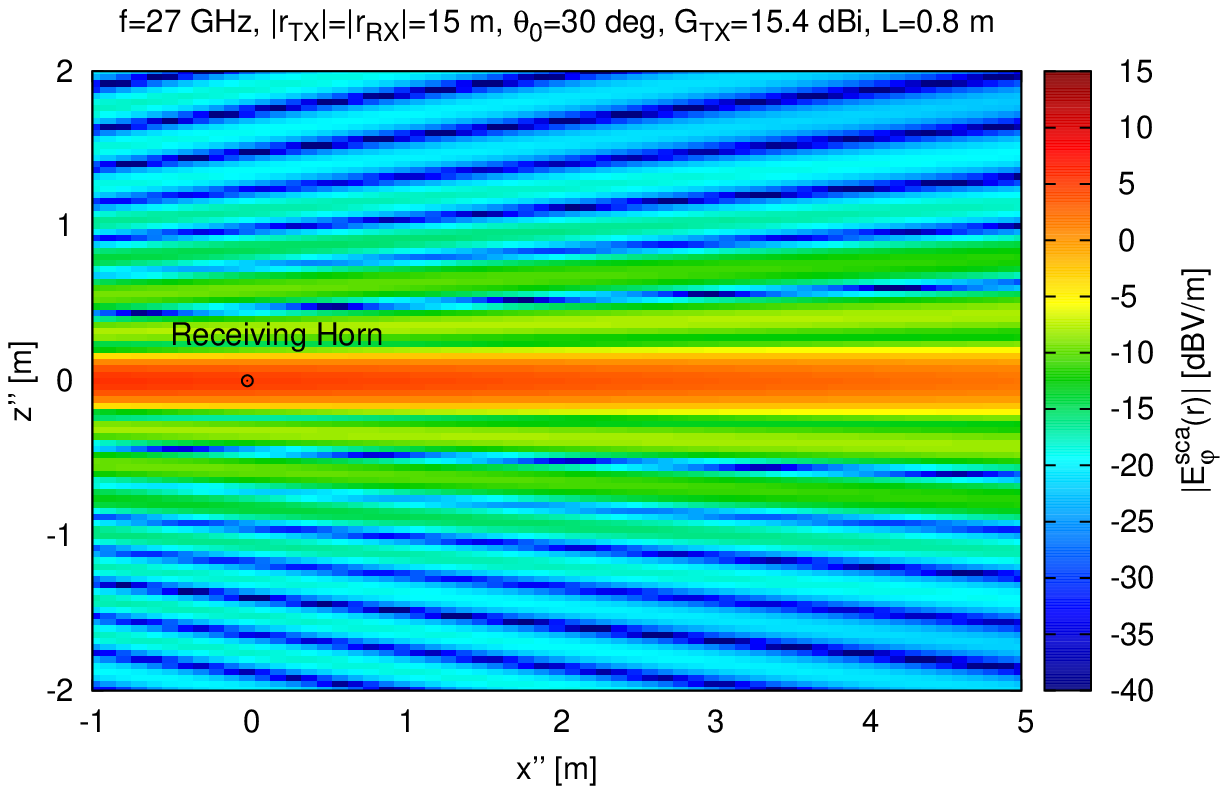}\tabularnewline
(\emph{e})&
(\emph{f})\tabularnewline
\end{tabular}\end{center}

\begin{center}~\vfill\end{center}

\begin{center}\textbf{Fig. 5 - G. Oliveri et} \textbf{\emph{al.}}\textbf{,}
\textbf{\emph{{}``}}Features and Potentialities of ... ''\end{center}

\newpage
\begin{center}~\vfill\end{center}

\begin{center}\begin{tabular}{cc}
\multicolumn{2}{c}{\includegraphics[%
  width=0.95\columnwidth]{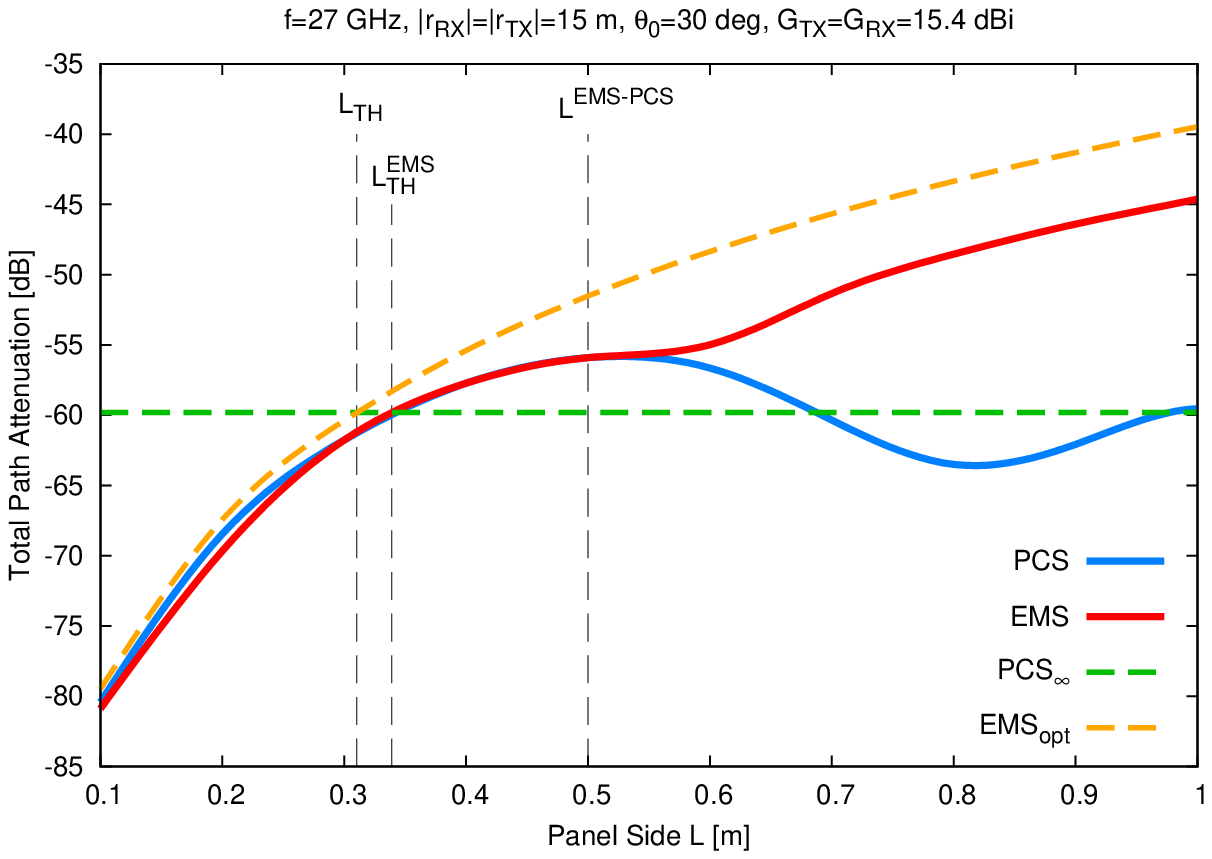}}\tabularnewline
\multicolumn{2}{c}{(\emph{a}) }\tabularnewline
$L=0.4$ {[}m{]}&
$L=1.0$ {[}m{]}\tabularnewline
\includegraphics[%
  width=0.45\columnwidth]{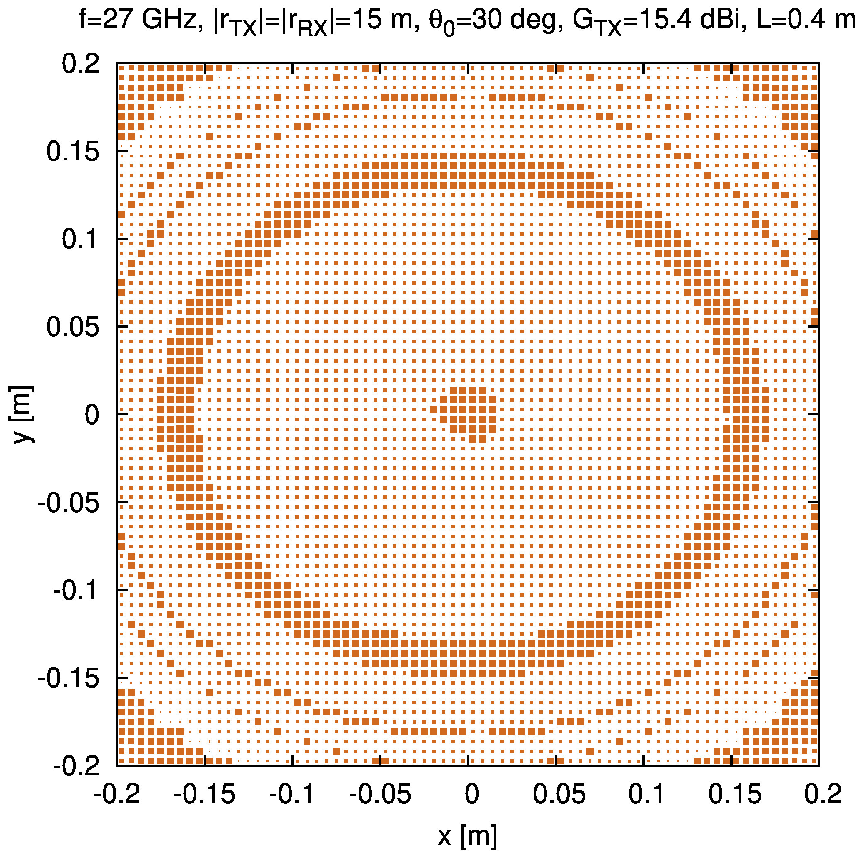}&
\includegraphics[%
  width=0.45\columnwidth]{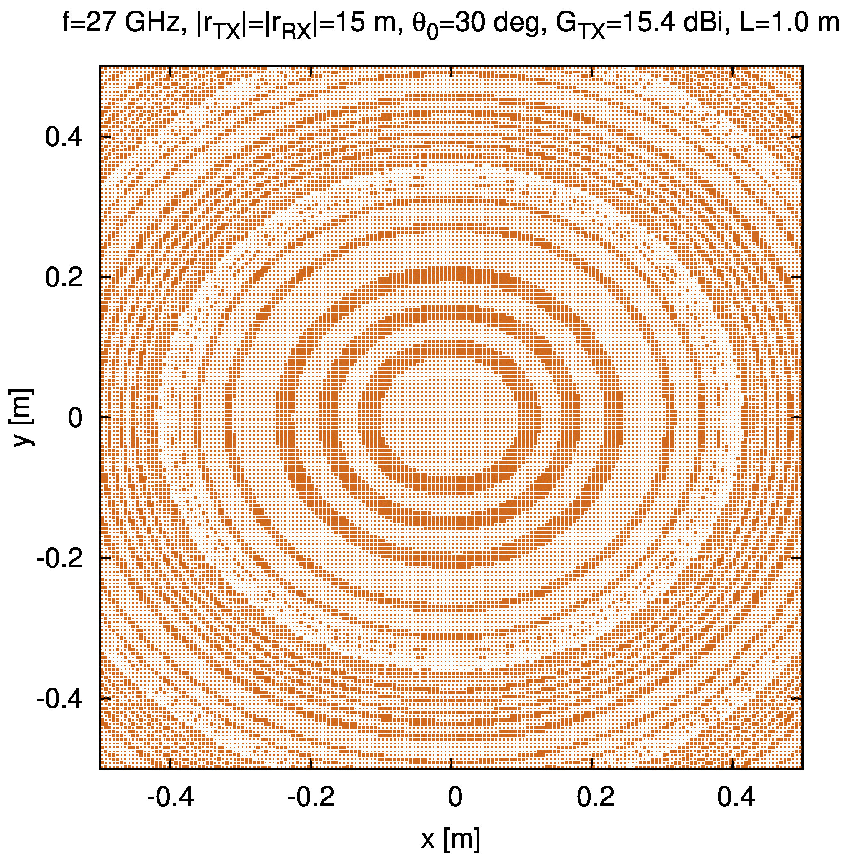}\tabularnewline
(\emph{b})&
(\emph{c})\tabularnewline
\end{tabular}\end{center}

\begin{center}~\vfill\end{center}

\begin{center}\textbf{Fig. 6 - G. Oliveri et} \textbf{\emph{al.}}\textbf{,}
\textbf{\emph{{}``}}Features and Potentialities of ... ''\end{center}

\newpage
\begin{center}~\vfill\end{center}

\begin{center}\begin{tabular}{ccc}
&
\emph{PCS}&
\emph{EMS}\tabularnewline
\begin{sideways}
~~~~~~~~~~~~~~~~~~$L=0.4$ {[}m{]}%
\end{sideways}&
\includegraphics[%
  width=0.45\columnwidth]{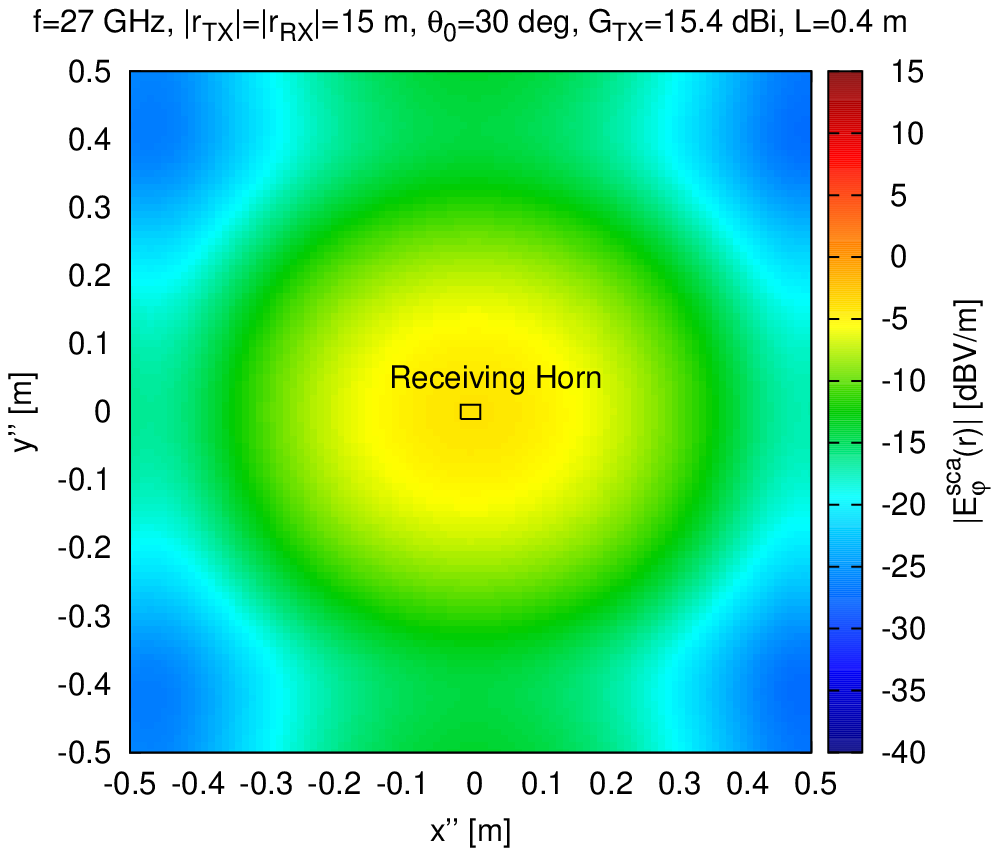}&
\includegraphics[%
  width=0.45\columnwidth]{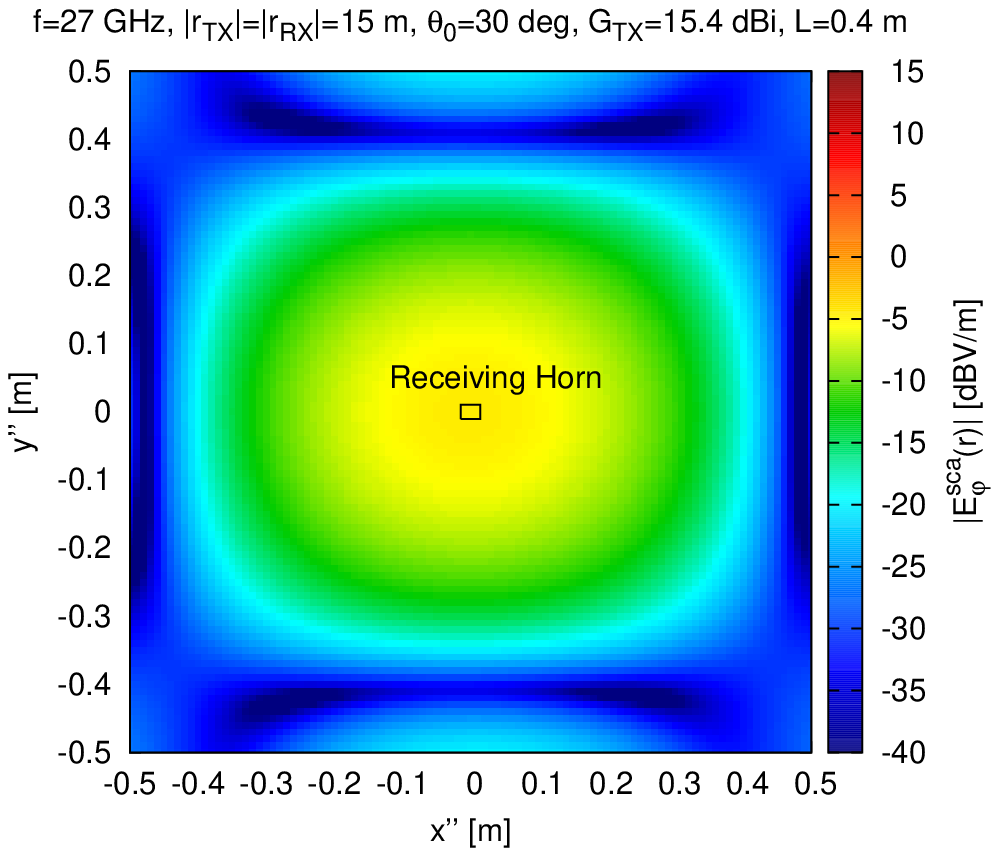}\tabularnewline
&
(\emph{a})&
(\emph{b})\tabularnewline
\begin{sideways}
~~~~~~~~~~~~~~~~~~$L=1.0$ {[}m{]}%
\end{sideways}&
\includegraphics[%
  width=0.45\columnwidth]{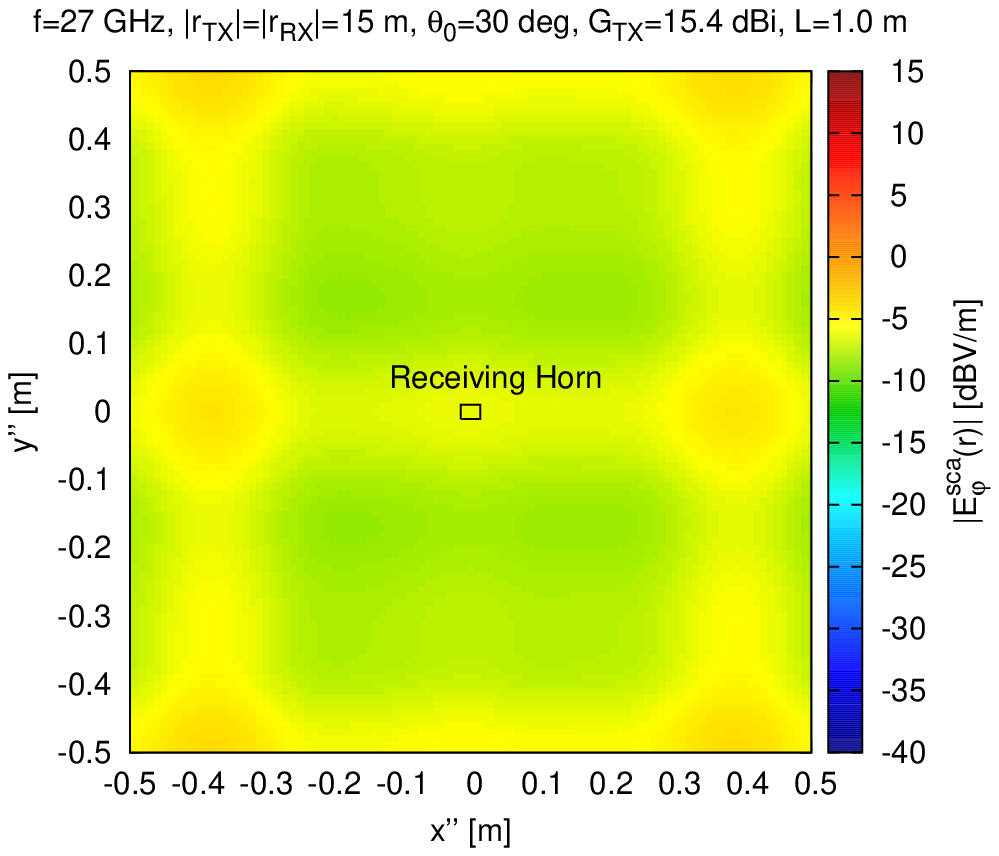}&
\includegraphics[%
  width=0.45\columnwidth]{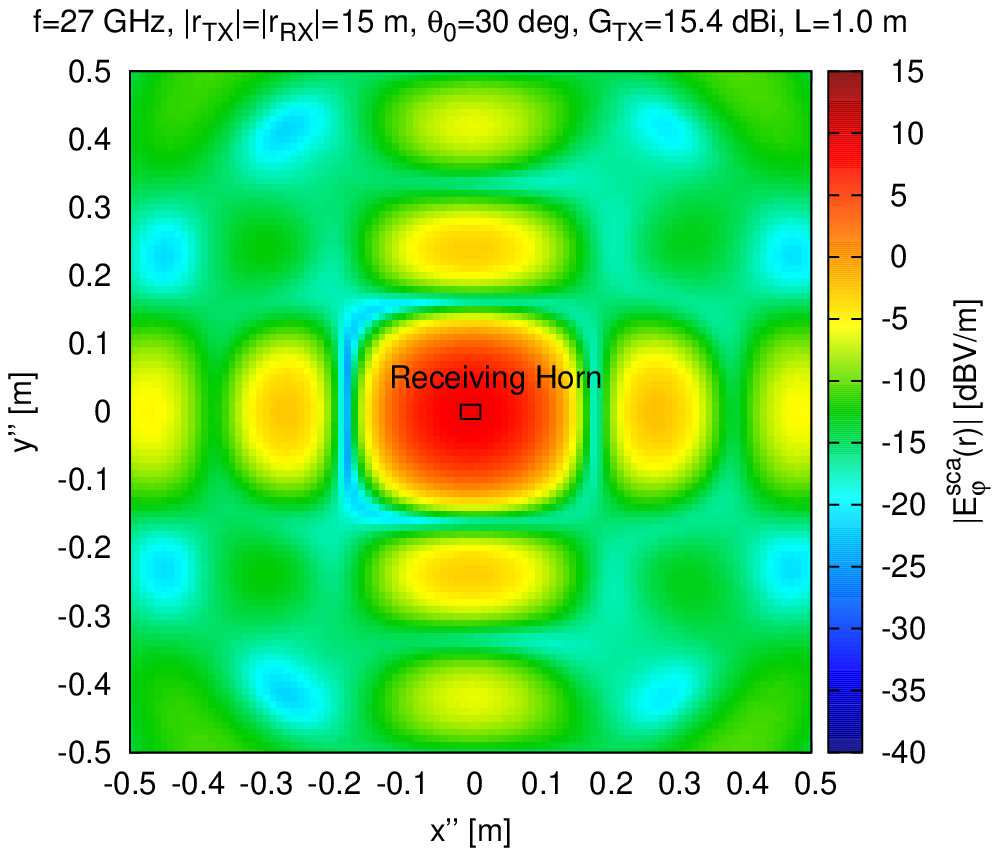}\tabularnewline
&
(\emph{c})&
(\emph{d})\tabularnewline
\end{tabular}\end{center}

\begin{center}~\vfill\end{center}

\begin{center}\textbf{Fig. 7 - G. Oliveri et} \textbf{\emph{al.}}\textbf{,}
\textbf{\emph{{}``}}Features and Potentialities of ... ''\end{center}

\newpage
\begin{center}~\vfill\end{center}

\begin{center}\begin{tabular}{ccc}
&
\emph{PCS}&
\emph{EMS}\tabularnewline
\begin{sideways}
~~~~~~~~~~~~~~~~~~$L=0.4$ {[}m{]}%
\end{sideways}&
\includegraphics[%
  width=0.45\columnwidth]{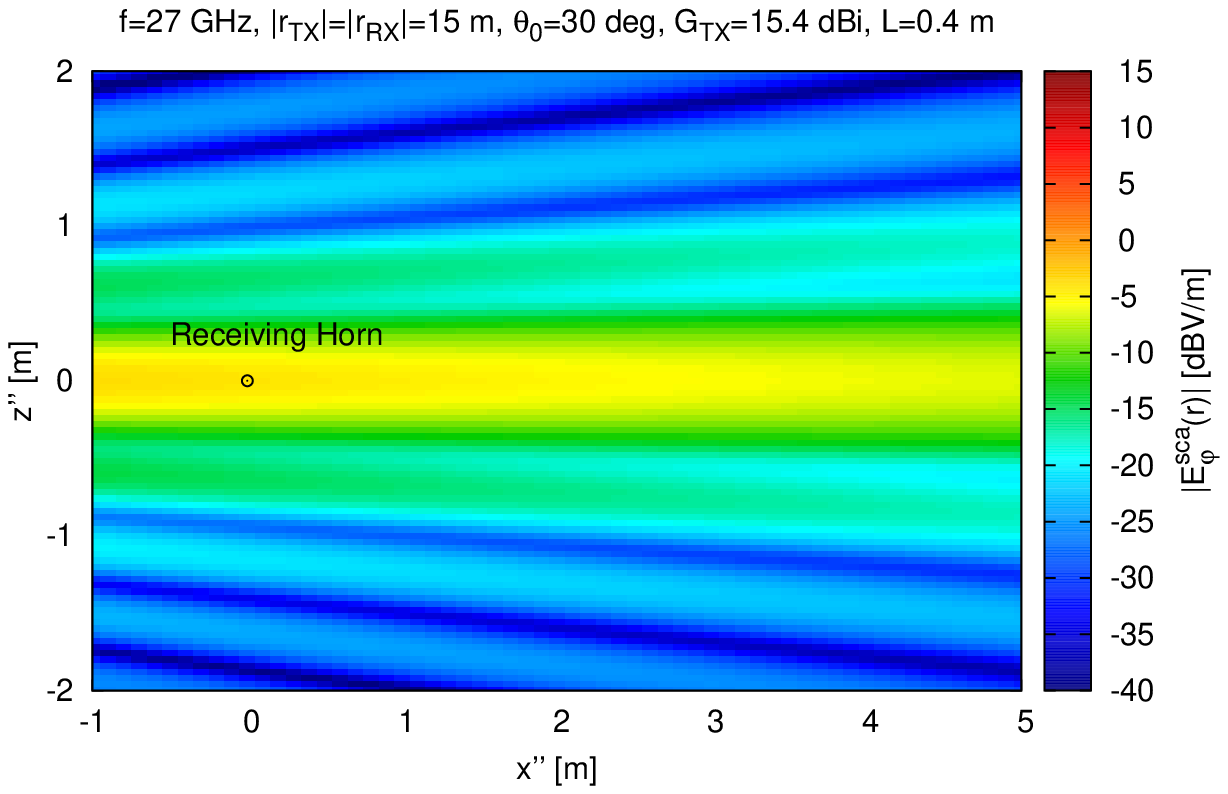}&
\includegraphics[%
  width=0.45\columnwidth]{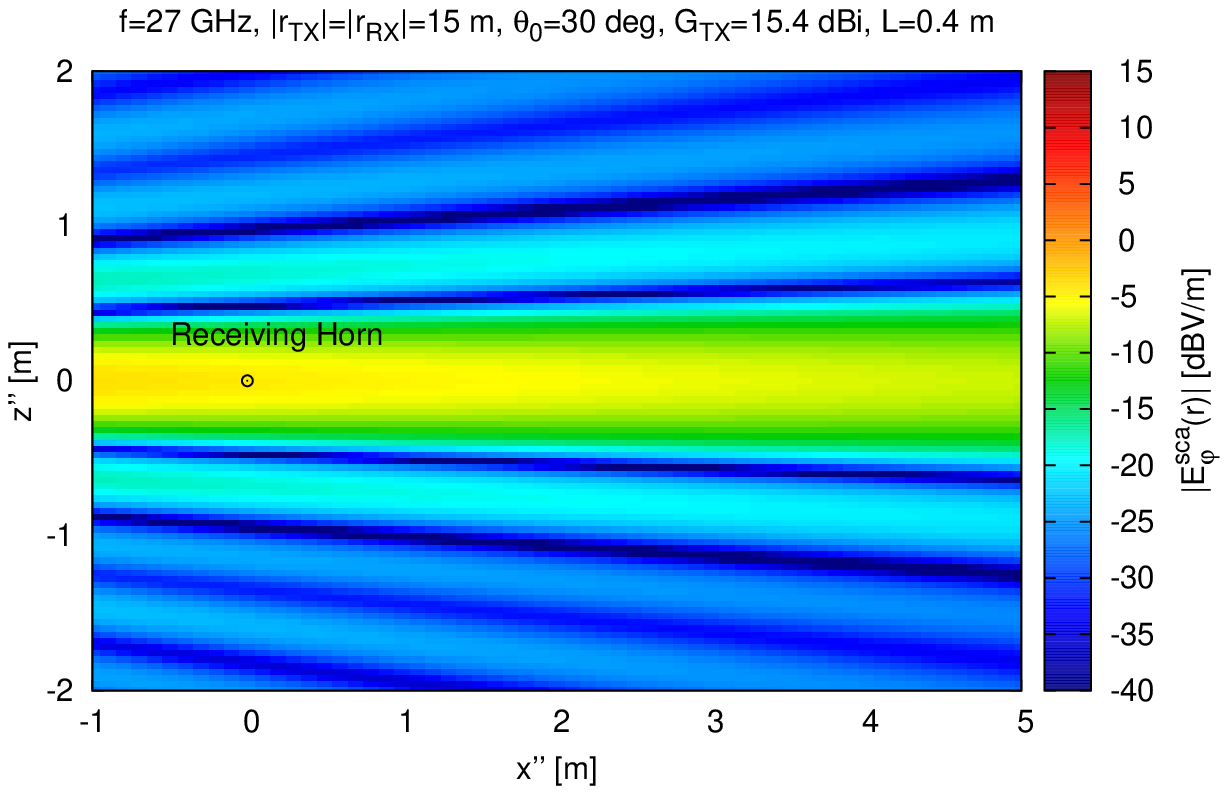}\tabularnewline
&
(\emph{a})&
(\emph{b})\tabularnewline
\begin{sideways}
~~~~~~~~~~~~~~~~~~$L=1.0$ {[}m{]}%
\end{sideways}&
\includegraphics[%
  width=0.45\columnwidth]{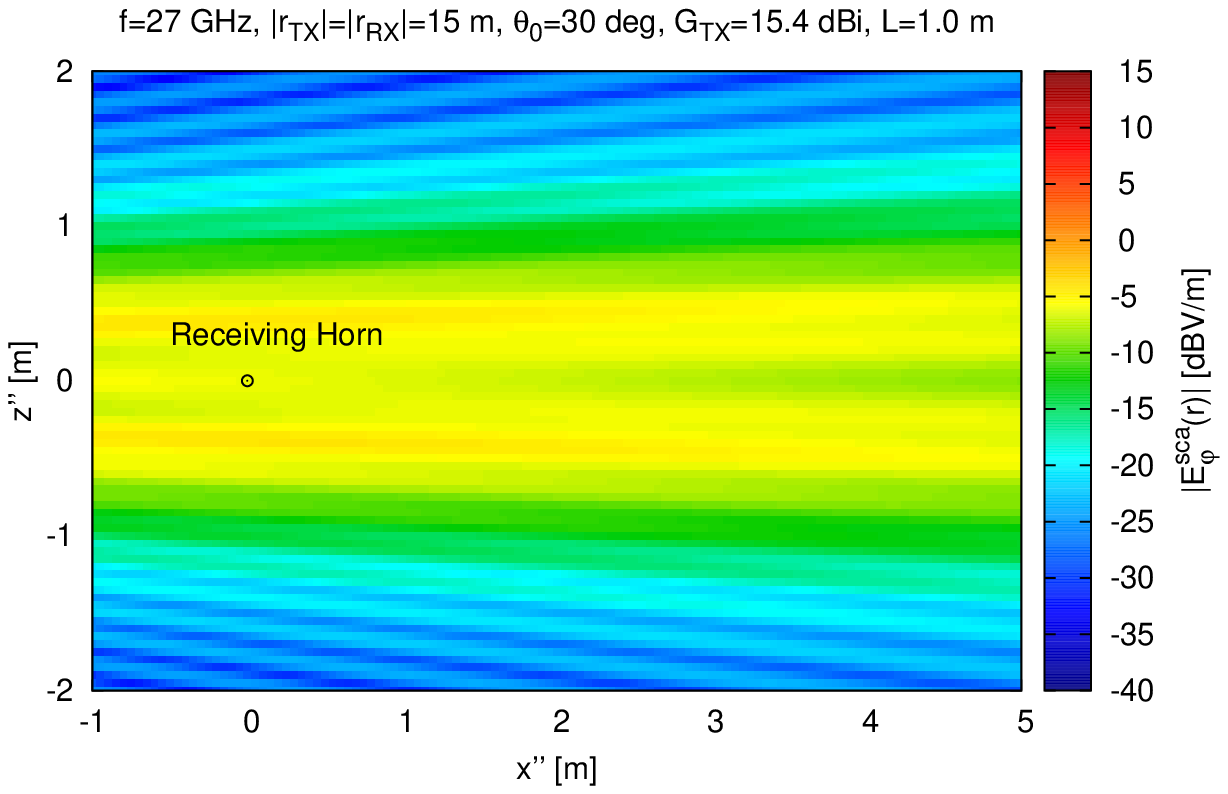}&
\includegraphics[%
  width=0.45\columnwidth]{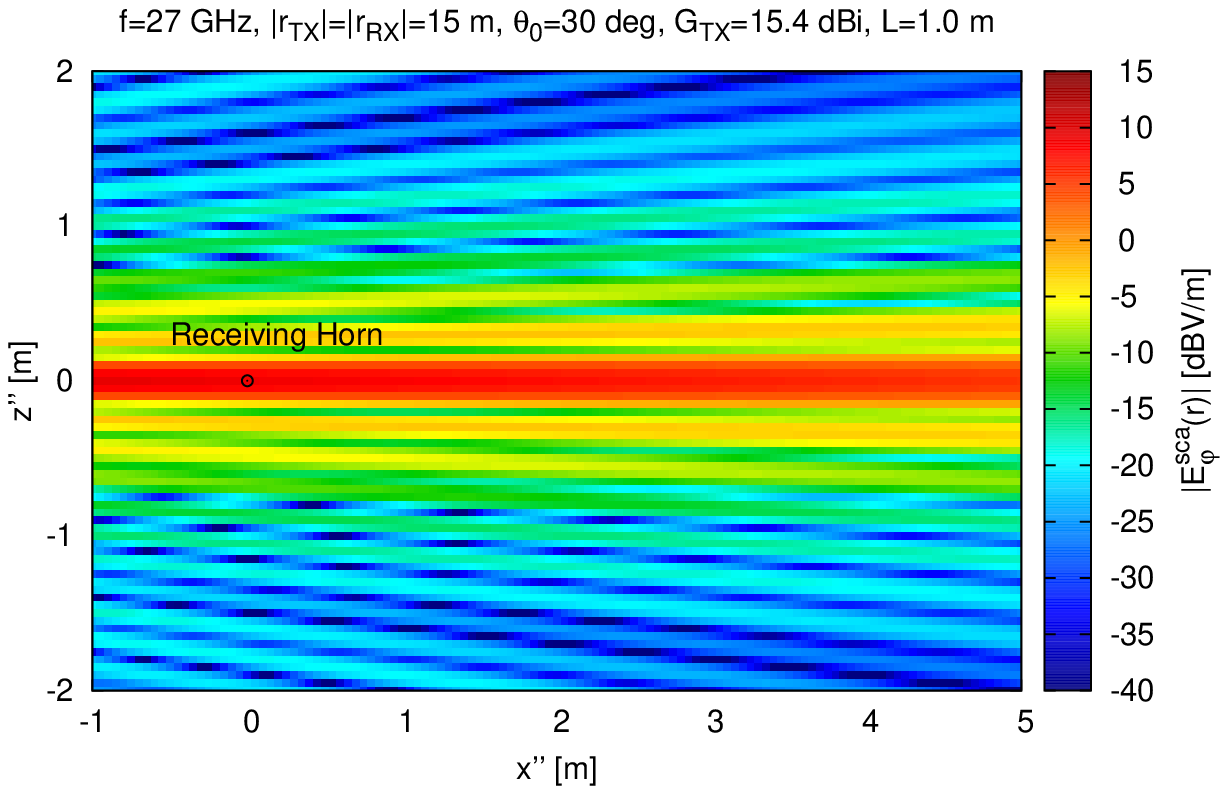}\tabularnewline
&
(\emph{c})&
(\emph{d})\tabularnewline
\end{tabular}\end{center}

\begin{center}~\vfill\end{center}

\begin{center}\textbf{Fig. 8 - G. Oliveri et} \textbf{\emph{al.}}\textbf{,}
\textbf{\emph{{}``}}Features and Potentialities of ...''\end{center}

\newpage
\begin{center}~\vfill\end{center}

\begin{center}\begin{tabular}{cc}
\multicolumn{2}{c}{\includegraphics[%
  width=0.95\columnwidth]{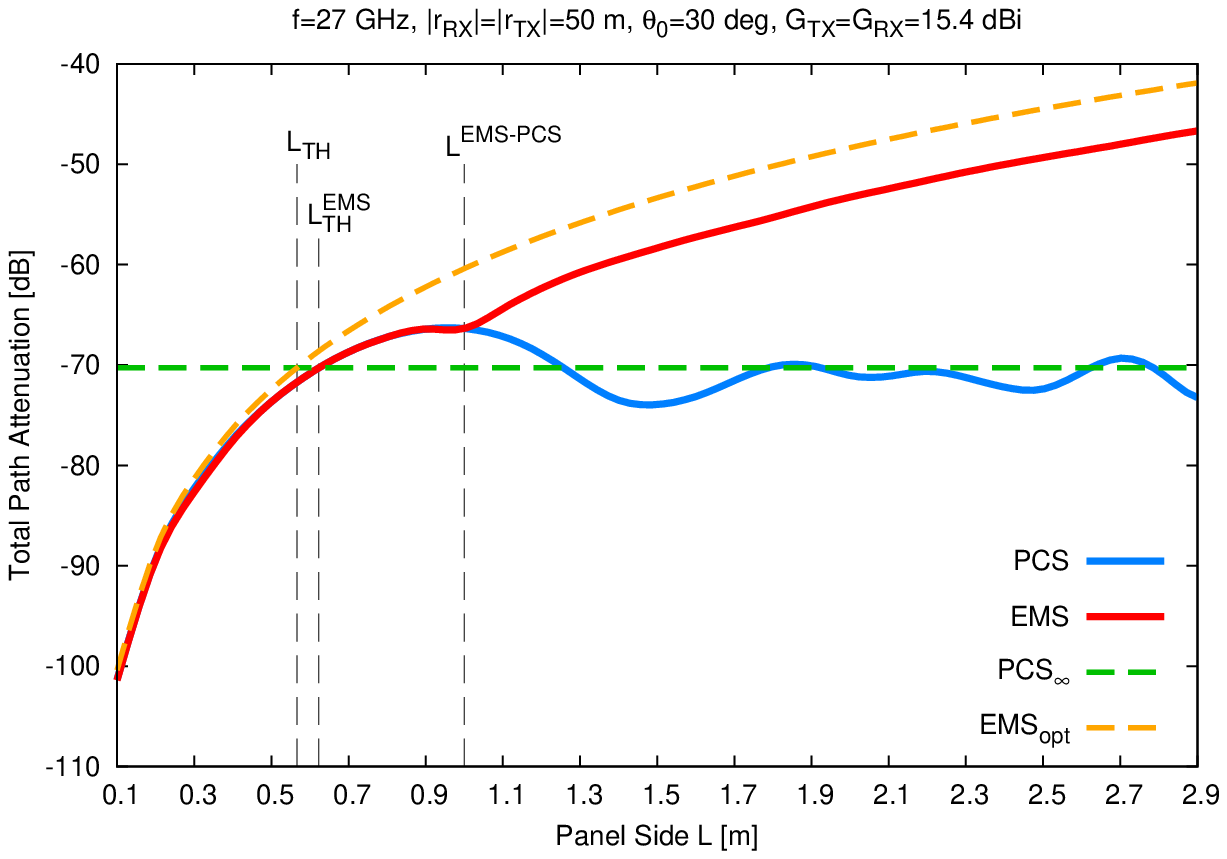}}\tabularnewline
\multicolumn{2}{c}{(\emph{a})}\tabularnewline
$L=0.5$ {[}m{]}&
$L=2.9$ {[}m{]}\tabularnewline
\includegraphics[%
  width=0.45\columnwidth]{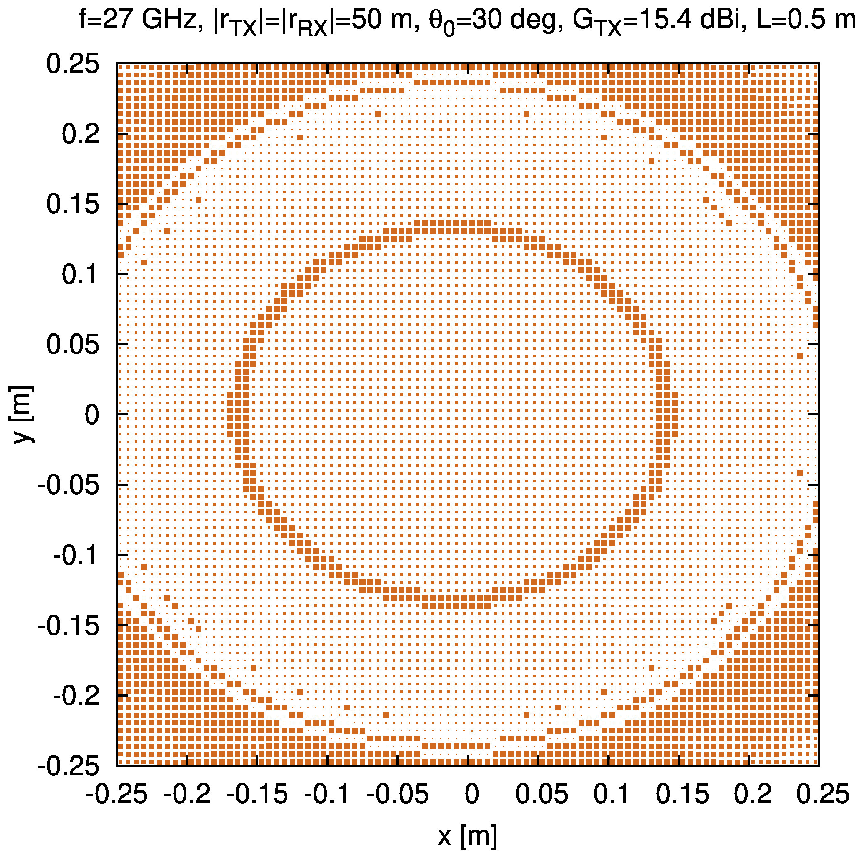}&
\includegraphics[%
  width=0.45\columnwidth]{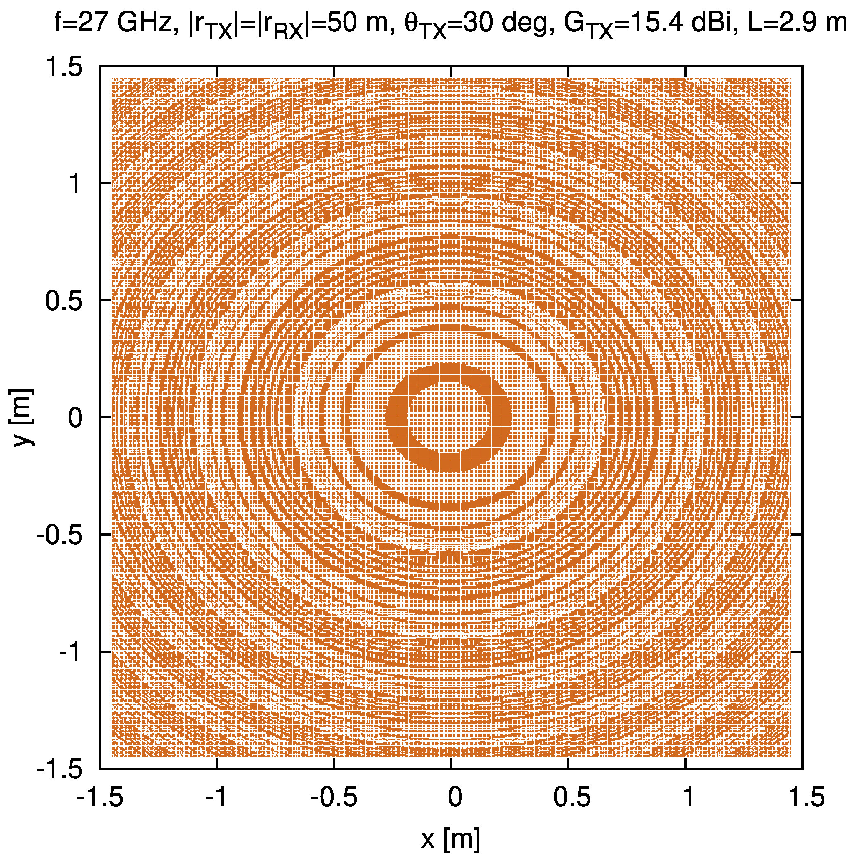}\tabularnewline
(\emph{b})&
(\emph{c})\tabularnewline
\end{tabular}\end{center}

\begin{center}\textbf{~}\vfill\end{center}

\begin{center}\textbf{Fig. 9 - G. Oliveri et} \textbf{\emph{al.}}\textbf{,}
\textbf{\emph{{}``}}Features and Potentialities of ...''\end{center}

\newpage
\begin{center}~\vfill\end{center}

\begin{center}\begin{tabular}{ccc}
&
\emph{PCS}&
\emph{EMS}\tabularnewline
\begin{sideways}
~~~~~~~~~~~~~~~~~~$L=0.5$ {[}m{]}%
\end{sideways}&
\includegraphics[%
  width=0.45\columnwidth]{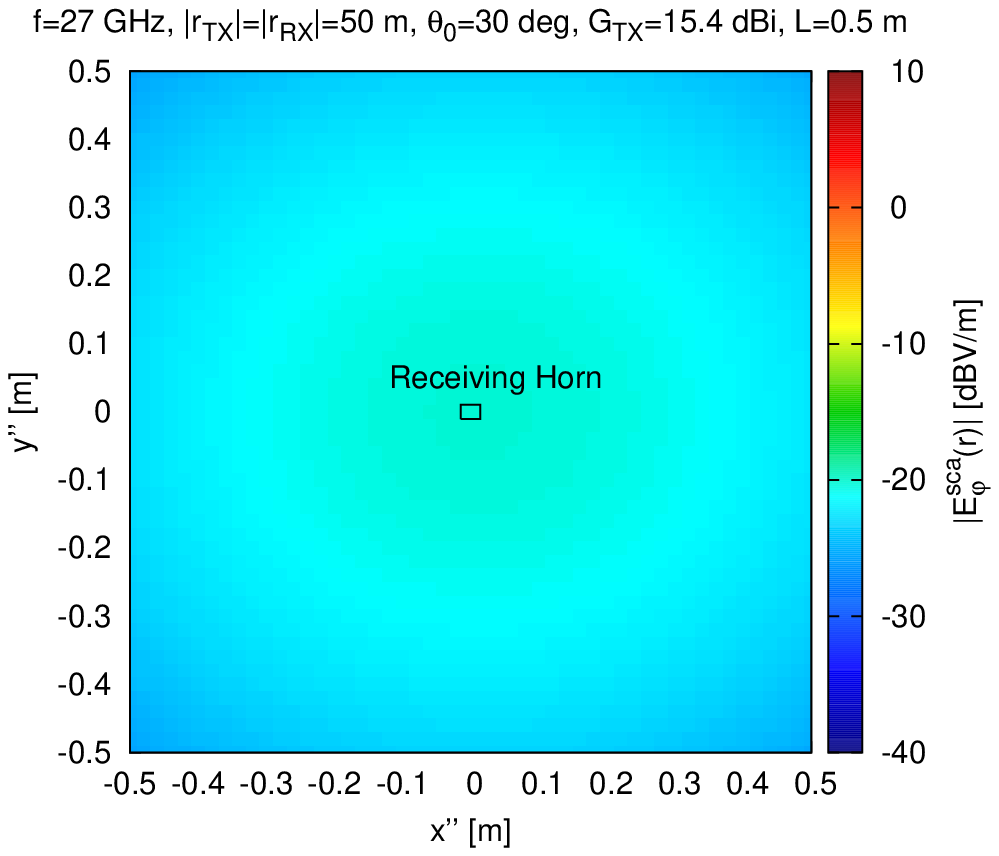}&
\includegraphics[%
  width=0.45\columnwidth]{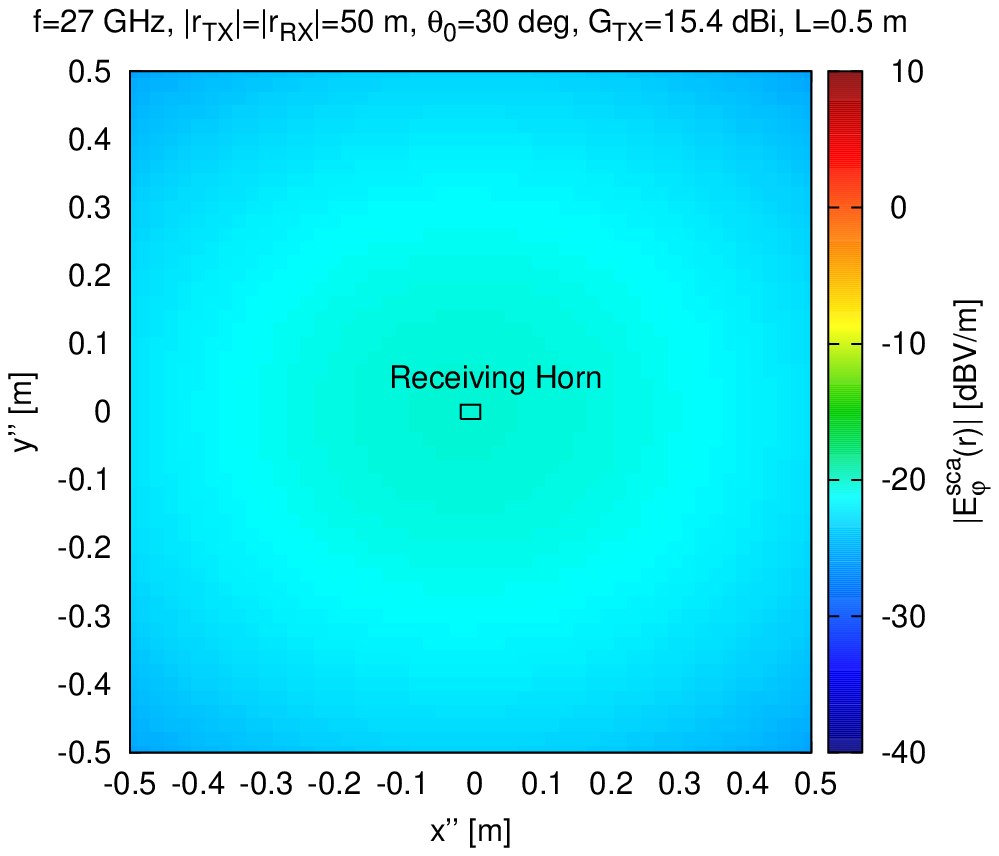}\tabularnewline
&
(\emph{a})&
(\emph{b})\tabularnewline
\begin{sideways}
~~~~~~~~~~~~~~~~~~$L=2.9$ {[}m{]}%
\end{sideways}&
\includegraphics[%
  width=0.45\columnwidth]{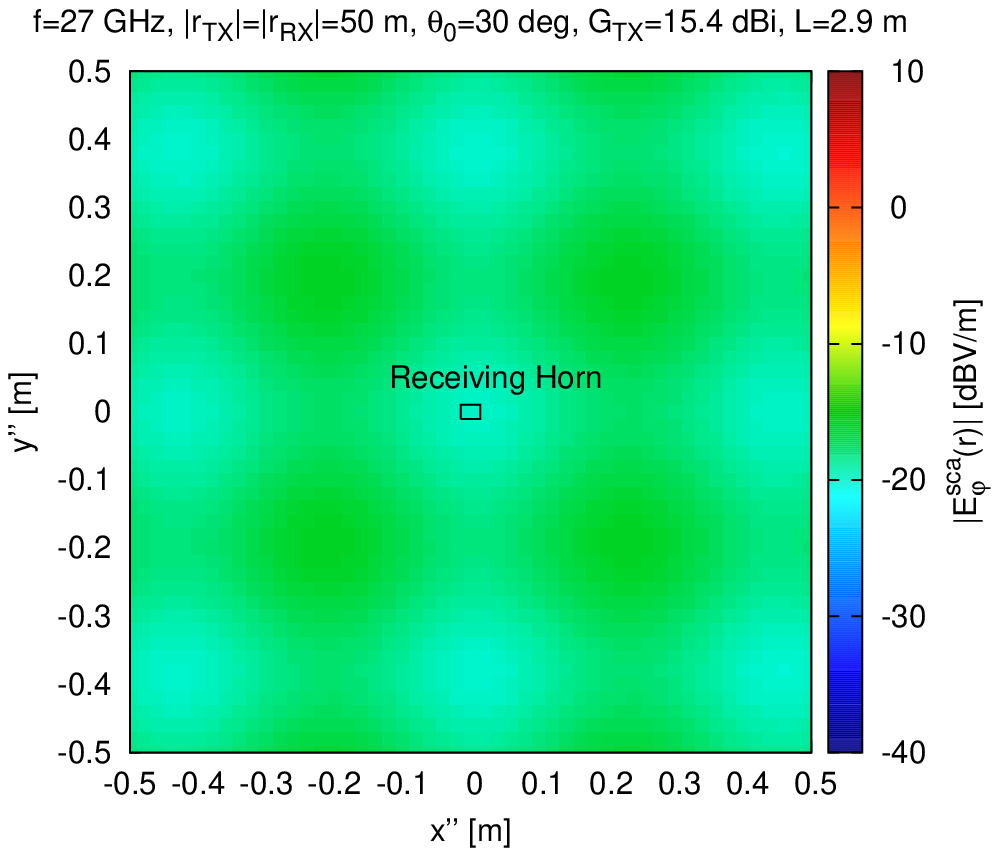}&
\includegraphics[%
  width=0.45\columnwidth]{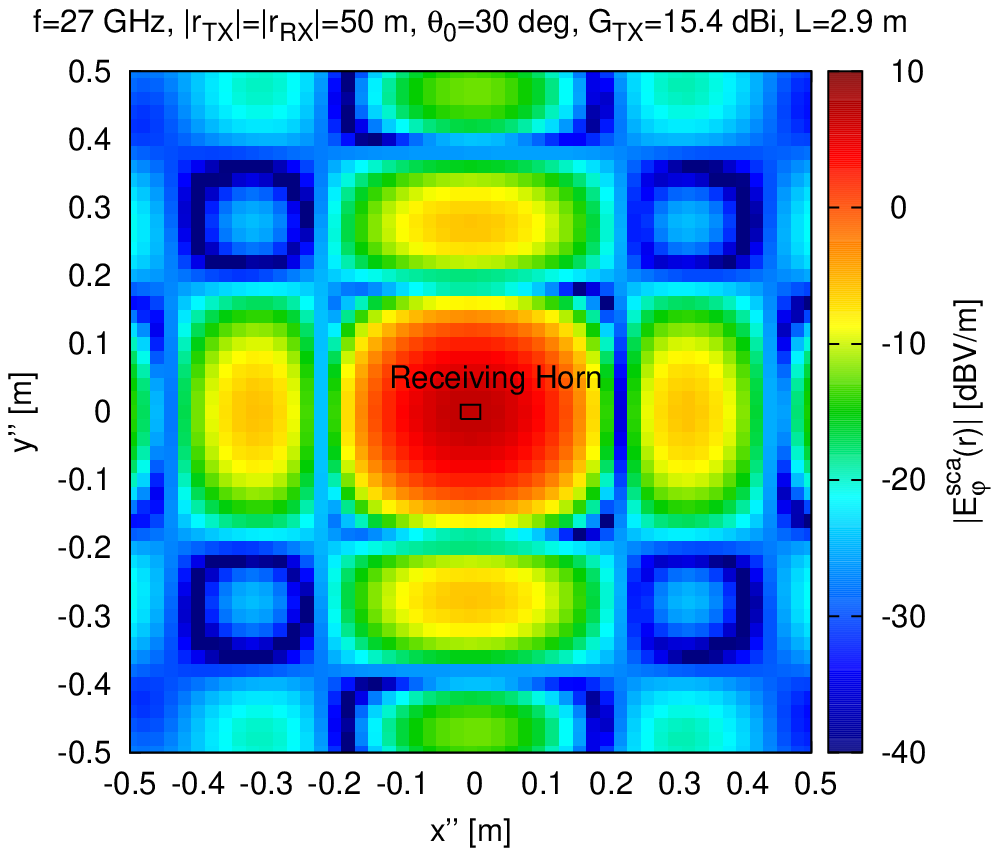}\tabularnewline
&
(\emph{c})&
(\emph{d})\tabularnewline
\end{tabular}\end{center}

\begin{center}~\vfill\end{center}

\begin{center}\textbf{Fig. 10 - G. Oliveri et} \textbf{\emph{al.}}\textbf{,}
\textbf{\emph{{}``}}Features and Potentialities of ...''\end{center}

\newpage
\begin{center}~\vfill\end{center}

\begin{center}\begin{tabular}{c}
\includegraphics[%
  width=0.90\columnwidth]{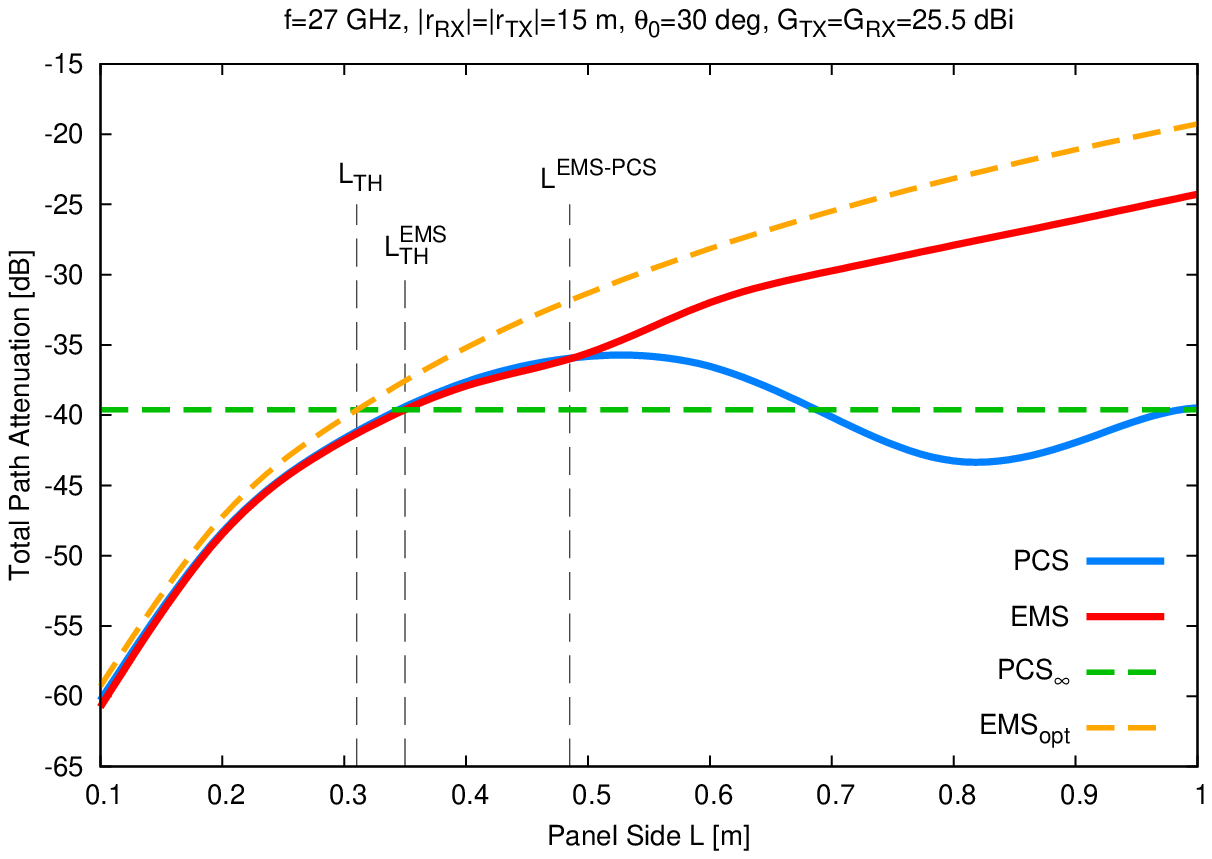}\tabularnewline
(\emph{a})\tabularnewline
\includegraphics[%
  width=0.90\columnwidth]{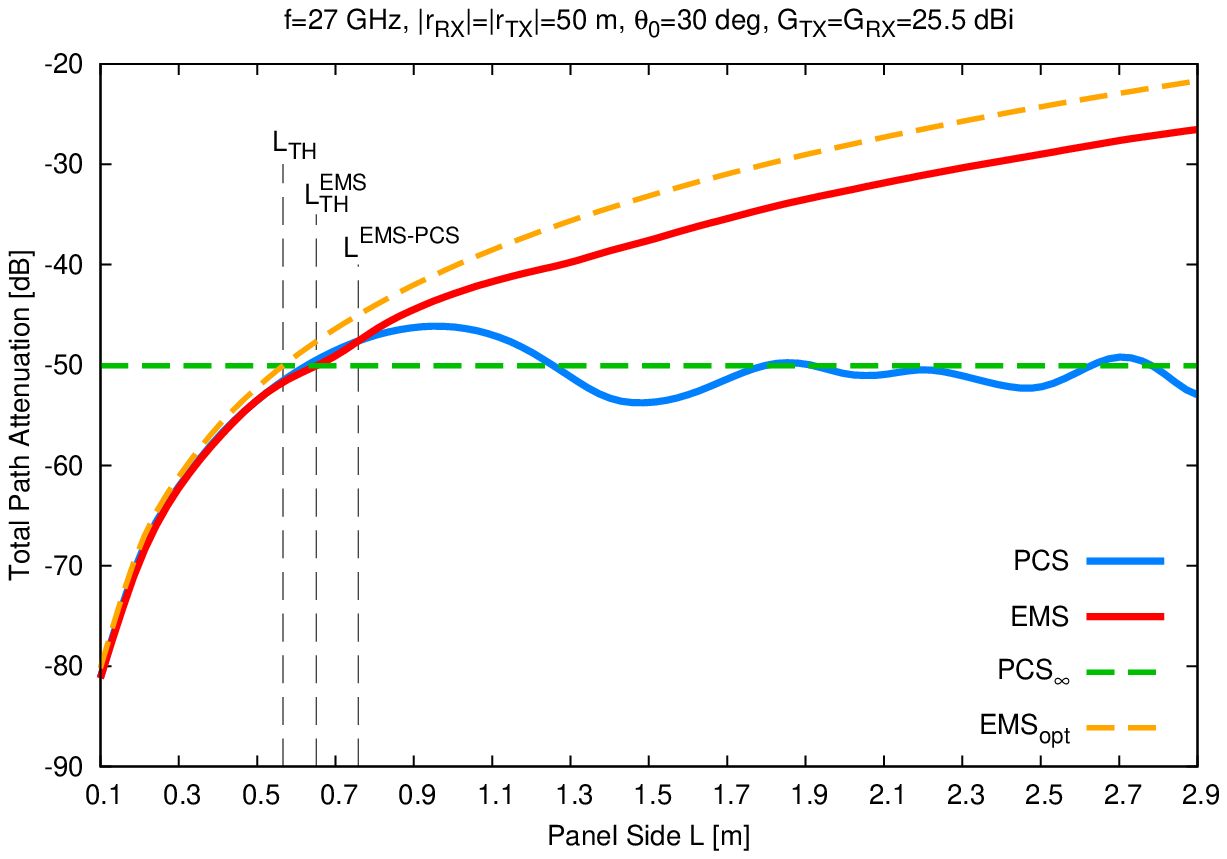}\tabularnewline
(\emph{b})\tabularnewline
\end{tabular}\end{center}

\begin{center}~\vfill\end{center}

\begin{center}\textbf{Fig. 11 - G. Oliveri et} \textbf{\emph{al.}}\textbf{,}
\textbf{\emph{{}``}}Features and Potentialities of ...''\end{center}

\newpage
\begin{center}\begin{tabular}{ccc}
&
$L=1.0$ {[}m{]}, $r_{TX}=r_{RX}=15$ {[}m{]}&
$L=2.9$ {[}m{]}, $r_{TX}=r_{RX}=50$ {[}m{]}\tabularnewline
&
\includegraphics[%
  width=0.45\columnwidth]{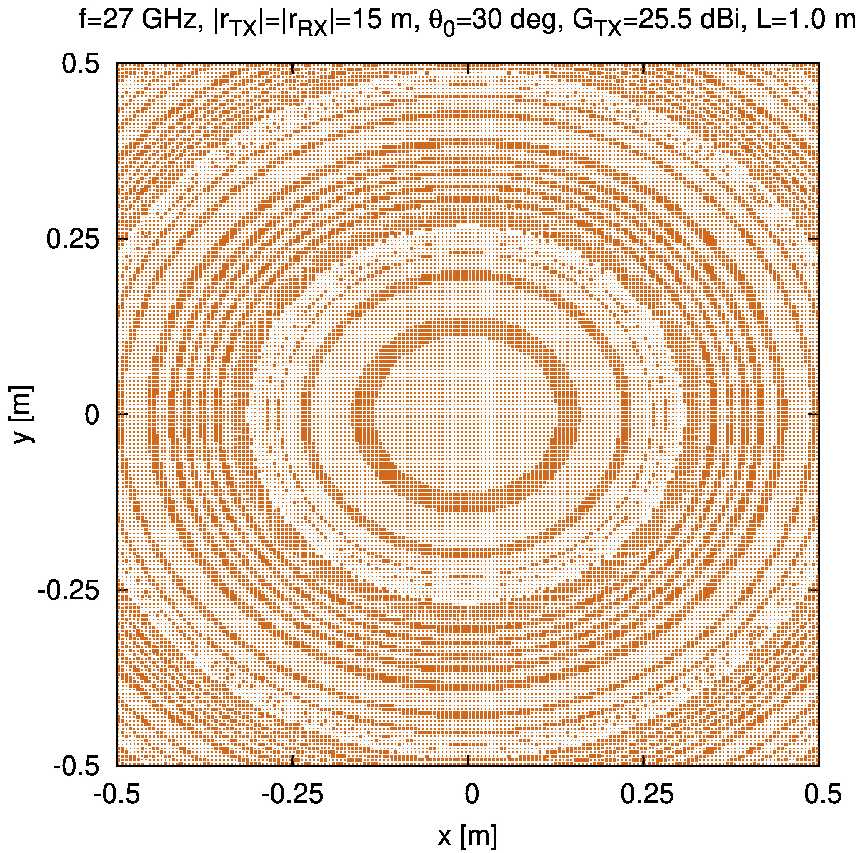}&
\includegraphics[%
  width=0.45\columnwidth]{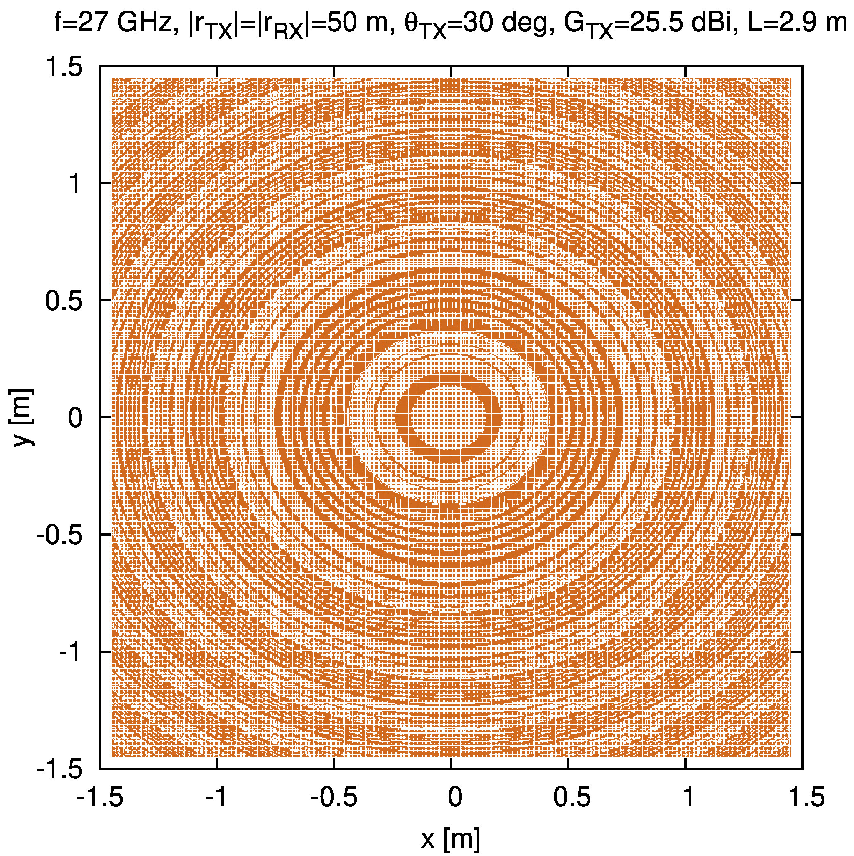}\tabularnewline
&
(\emph{a})&
(\emph{b})\tabularnewline
\begin{sideways}
~~~~~~~~\emph{~~~~~EMS}%
\end{sideways}&
\includegraphics[%
  width=0.45\columnwidth]{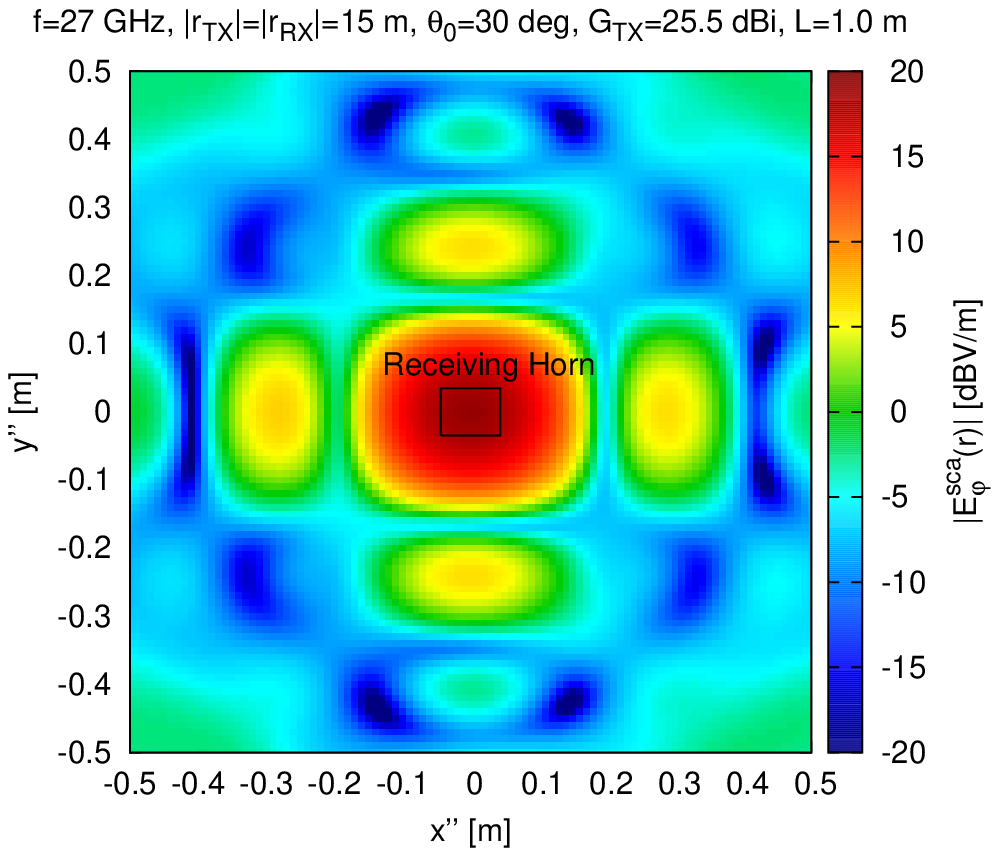}&
\includegraphics[%
  width=0.45\columnwidth]{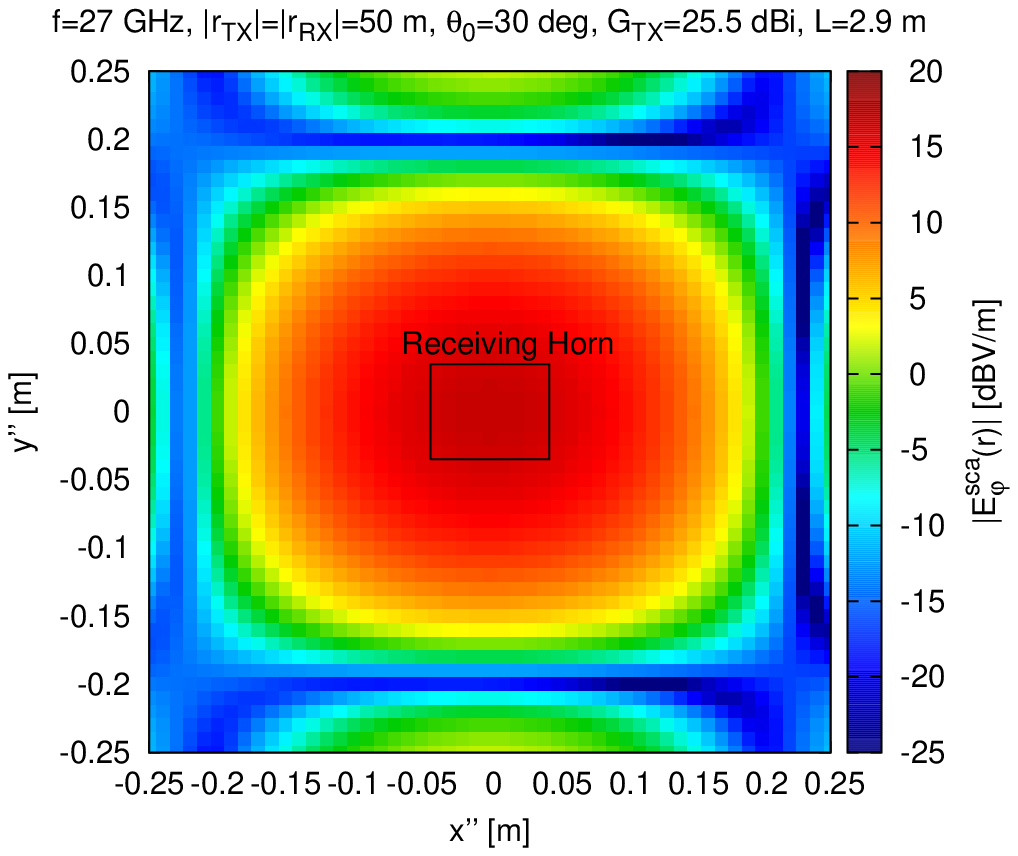}\tabularnewline
&
(\emph{c})&
(\emph{d})\tabularnewline
\begin{sideways}
~~~~~~~~~~~~~\emph{PCS}%
\end{sideways}&
\includegraphics[%
  width=0.45\columnwidth]{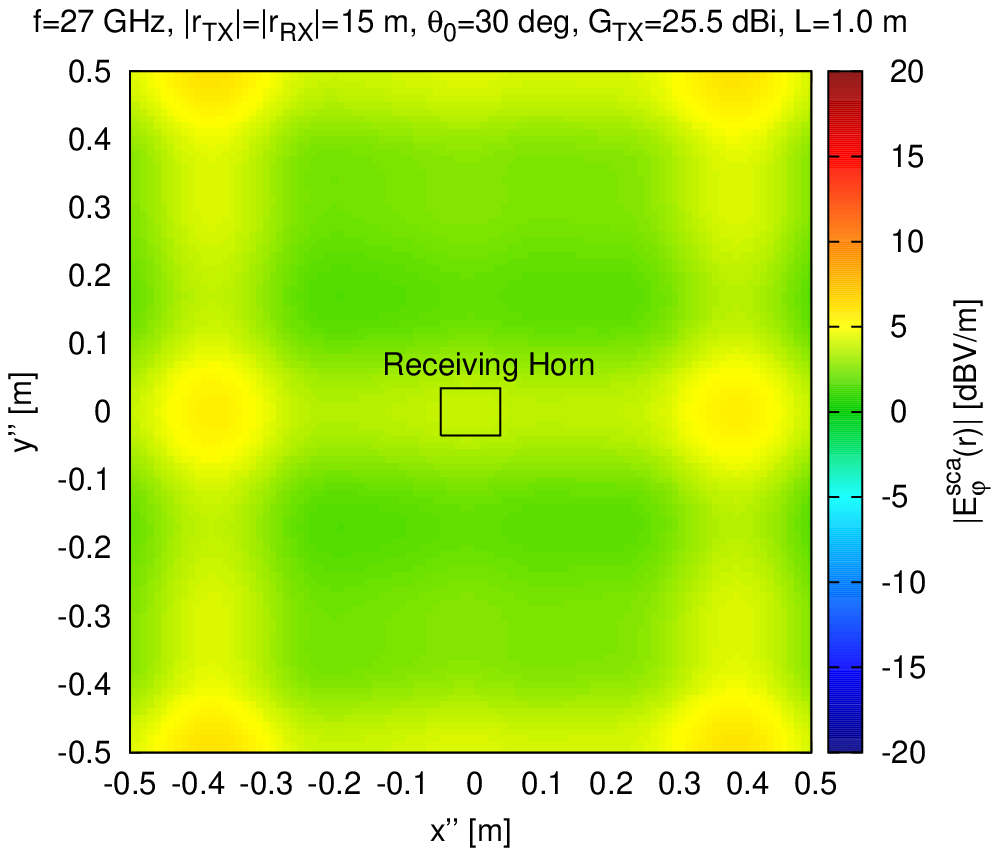}&
\includegraphics[%
  width=0.45\columnwidth]{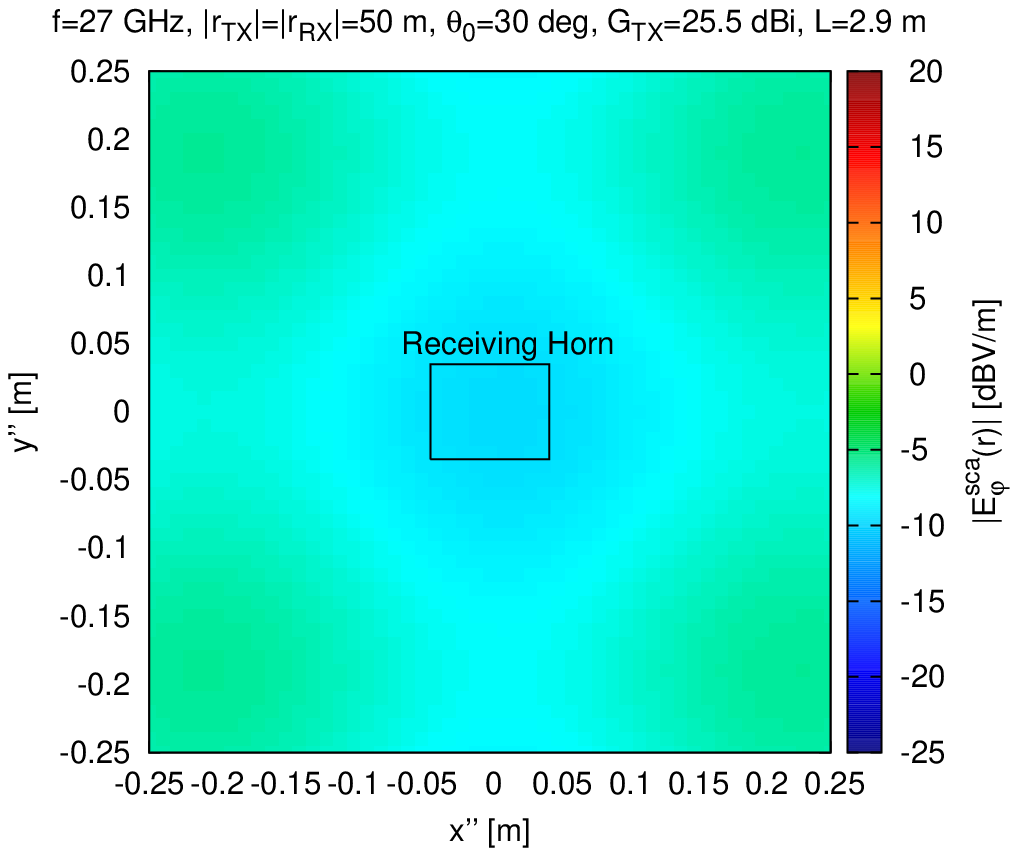}\tabularnewline
&
(\emph{e})&
(\emph{f})\tabularnewline
\end{tabular}\end{center}

\begin{center}\textbf{Fig. 12 - G. Oliveri et} \textbf{\emph{al.}}\textbf{,}
\textbf{\emph{{}``}}Features and Potentialities of ...''\end{center}

\newpage
\begin{center}~\vfill\end{center}

\begin{center}\begin{tabular}{cc}
\multicolumn{2}{c}{\includegraphics[%
  width=0.95\columnwidth]{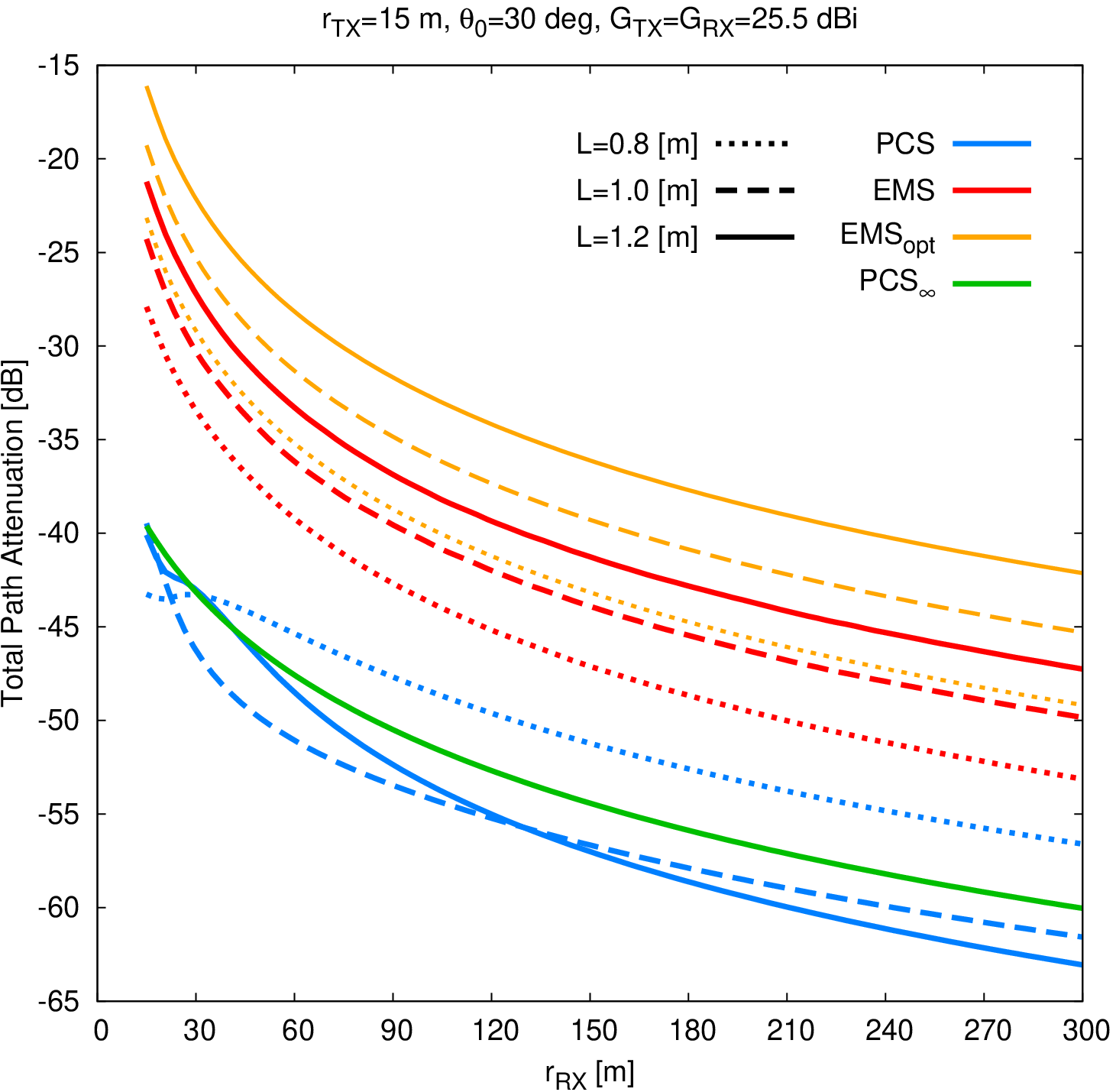}}\tabularnewline
\end{tabular}\end{center}

\begin{center}~\vfill\end{center}

\begin{center}\textbf{Fig. 13 - G. Oliveri et} \textbf{\emph{al.}}\textbf{,}
\textbf{\emph{{}``}}Features and Potentialities of ...''\end{center}
\newpage

\begin{center}\begin{tabular}{cc}
\multicolumn{2}{c}{\includegraphics[%
  width=0.60\columnwidth,
  height=0.32\paperwidth,
  keepaspectratio]{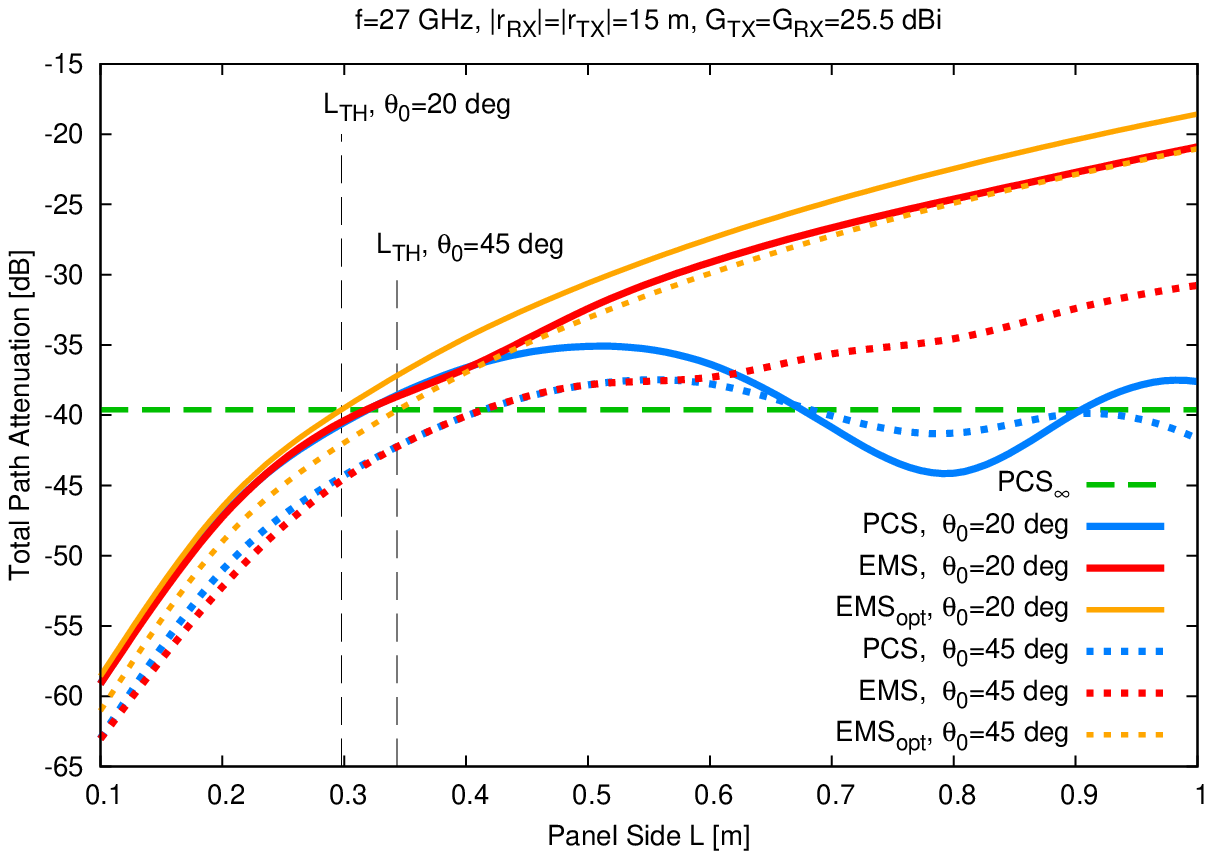}}\tabularnewline
\multicolumn{2}{c}{(\emph{a})}\tabularnewline
\multicolumn{2}{c}{\includegraphics[%
  width=0.60\columnwidth,
  height=0.32\paperwidth,
  keepaspectratio]{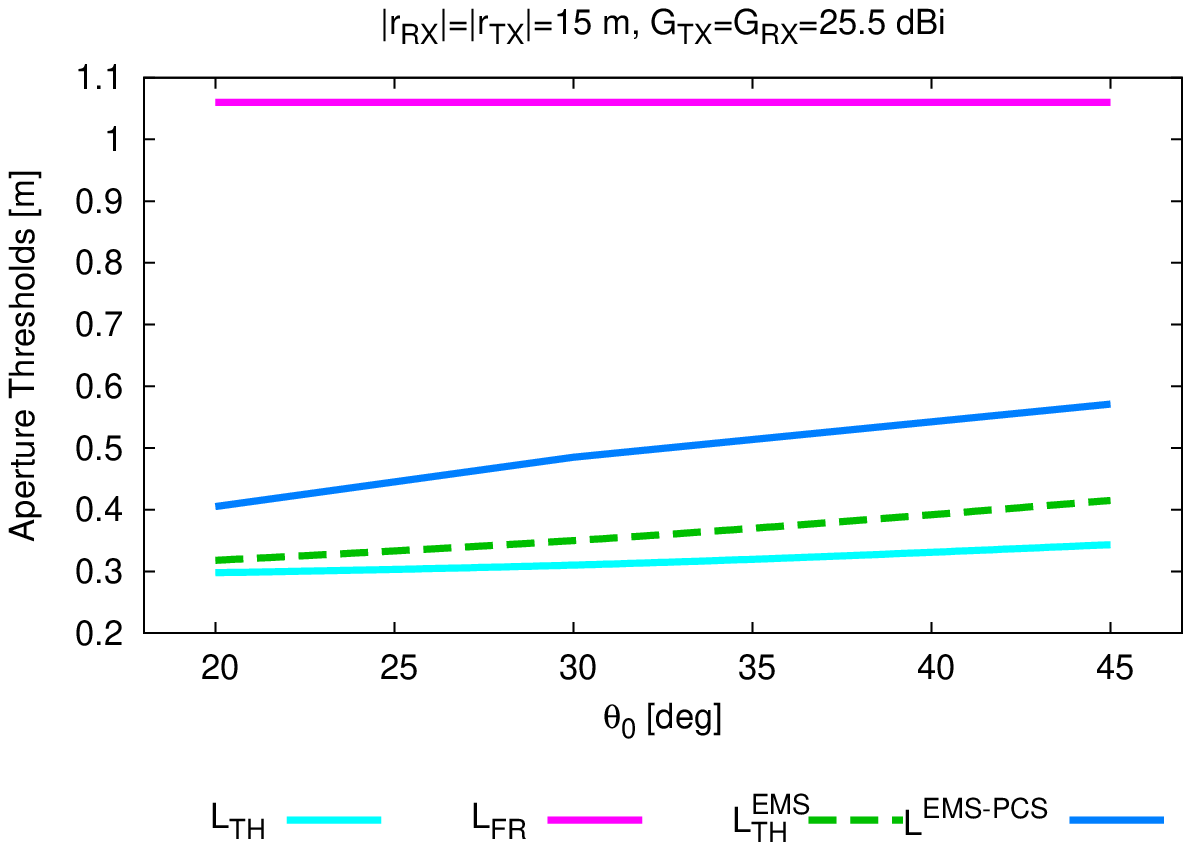}}\tabularnewline
\multicolumn{2}{c}{(\emph{b})}\tabularnewline
\multicolumn{2}{c}{\includegraphics[%
  width=0.60\columnwidth,
  height=0.32\paperwidth,
  keepaspectratio]{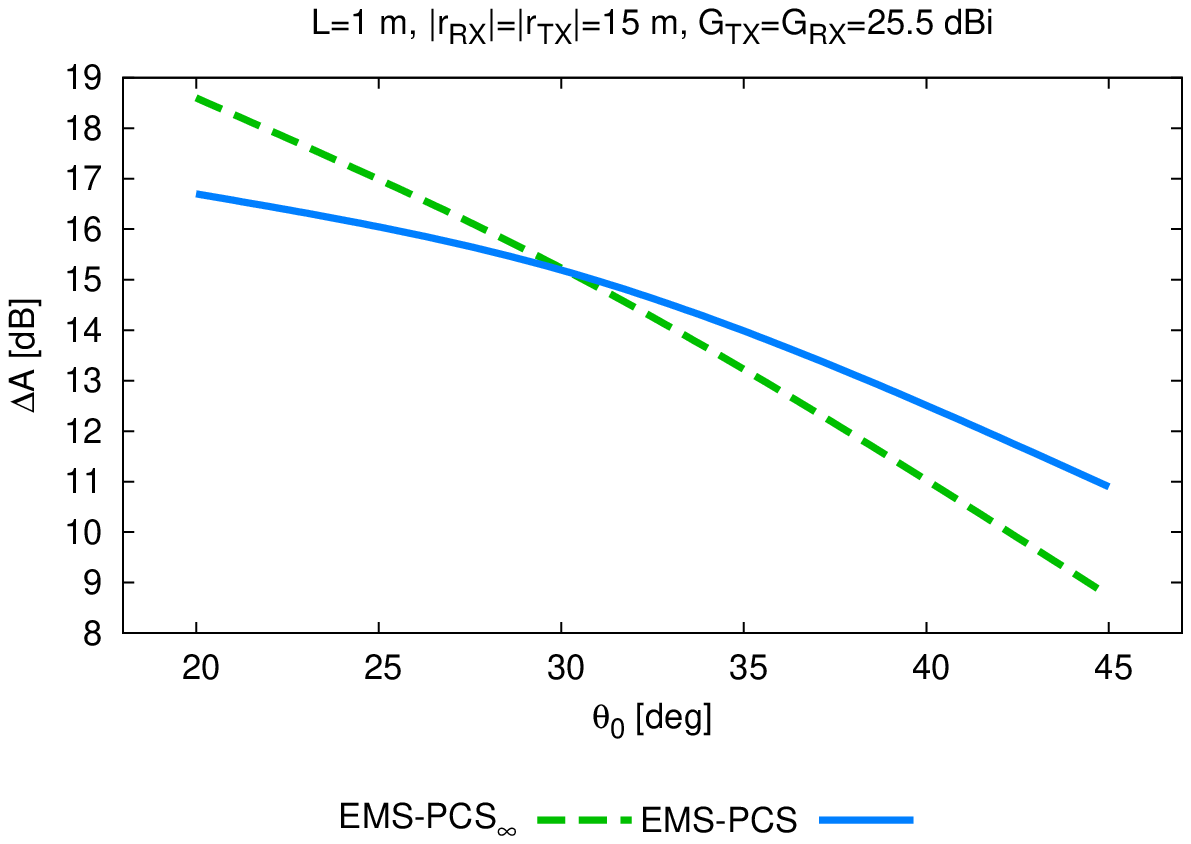}}\tabularnewline
\multicolumn{2}{c}{(\emph{c})}\tabularnewline
\end{tabular}\end{center}

\begin{center}\textbf{Fig. 14 - G. Oliveri et} \textbf{\emph{al.}}\textbf{,}
\textbf{\emph{{}``}}Features and Potentialities of ...''\end{center}

\newpage
\begin{center}~\vfill\end{center}

\begin{center}\begin{tabular}{cc}
$\theta_{0}=20$ {[}deg{]}&
$\theta_{0}=45$ {[}deg{]}\tabularnewline
\includegraphics[%
  width=0.45\columnwidth]{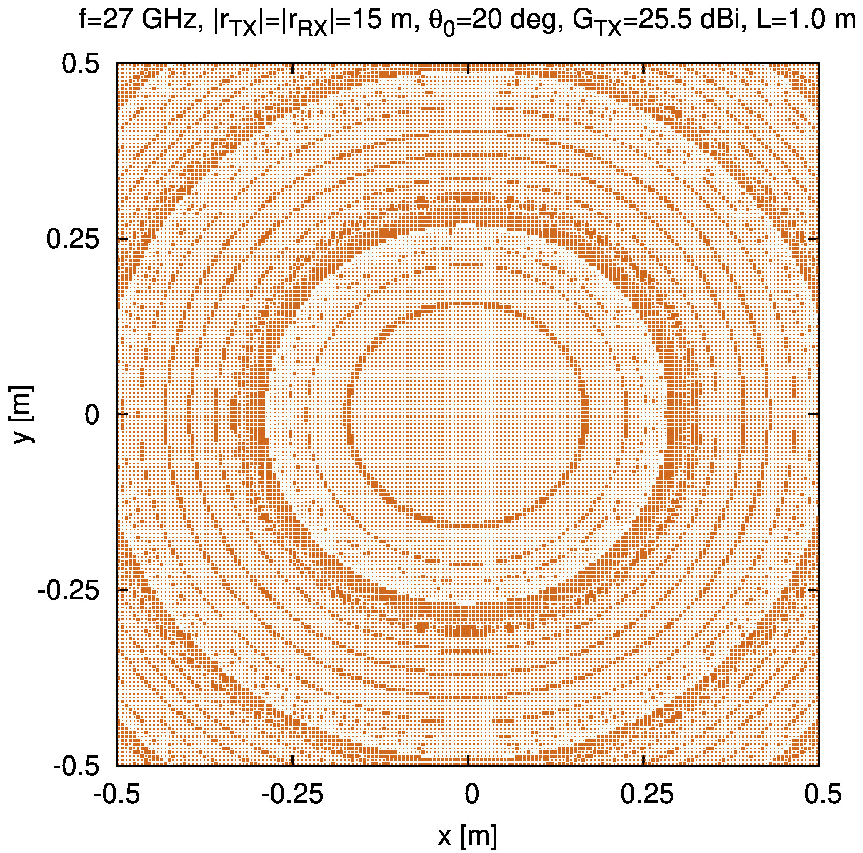}&
\includegraphics[%
  width=0.45\columnwidth]{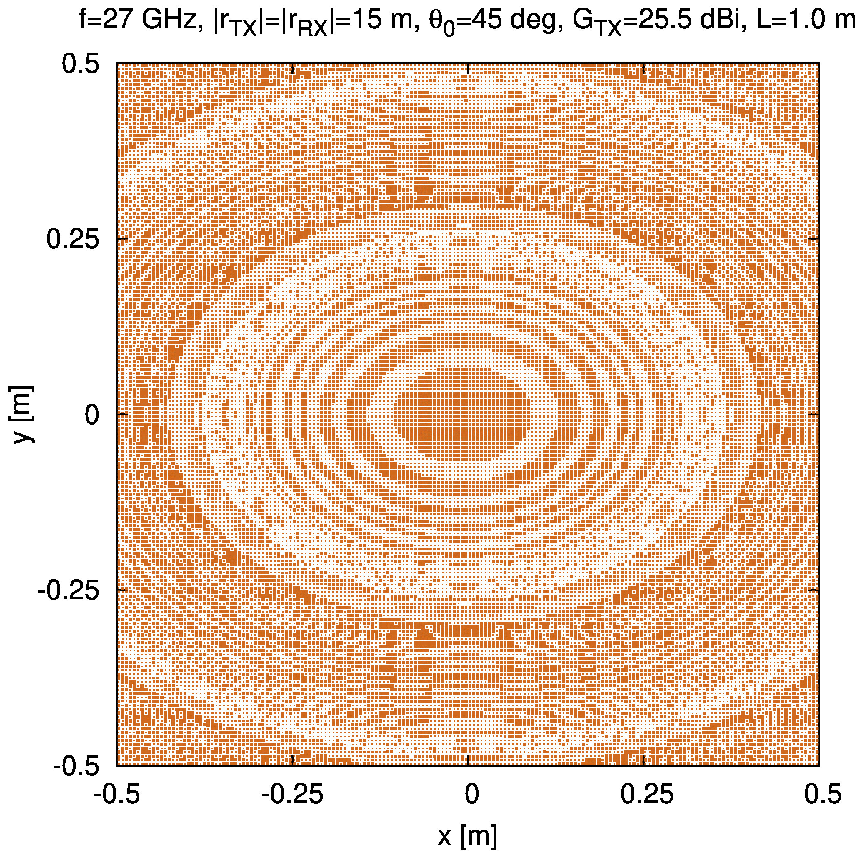}\tabularnewline
(\emph{a})&
(\emph{b})\tabularnewline
\end{tabular}\end{center}

\begin{center}~\vfill\end{center}

\begin{center}\textbf{Fig. 15 - G. Oliveri et} \textbf{\emph{al.}}\textbf{,}
\textbf{\emph{{}``}}Features and Potentialities of ...''\end{center}
\newpage

\begin{center}\begin{tabular}{cc}
\multicolumn{2}{c}{\includegraphics[%
  width=0.55\columnwidth,
  keepaspectratio]{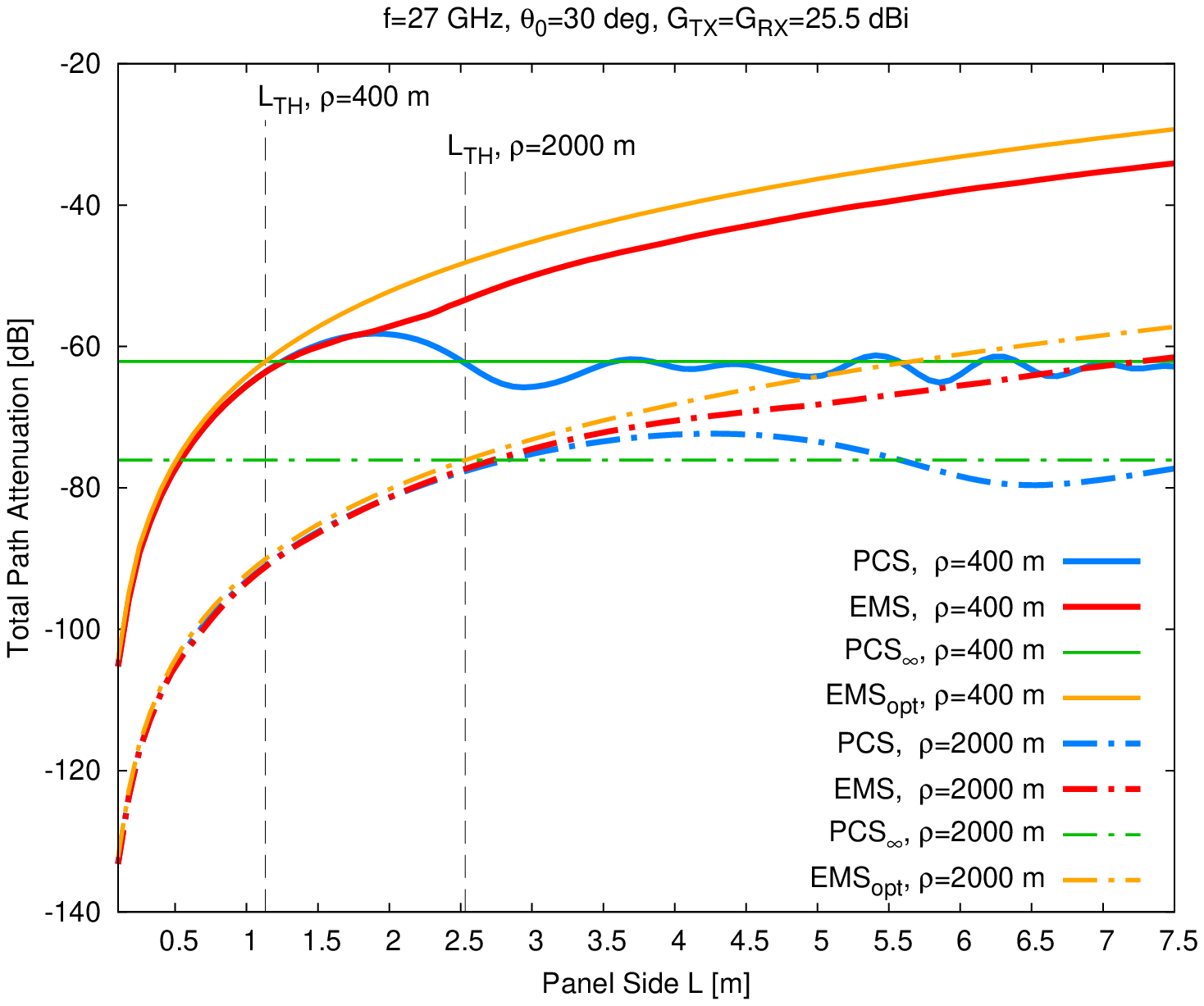}}\tabularnewline
\multicolumn{2}{c}{(\emph{a})}\tabularnewline
\multicolumn{2}{c}{\includegraphics[%
  width=0.60\columnwidth,
  keepaspectratio]{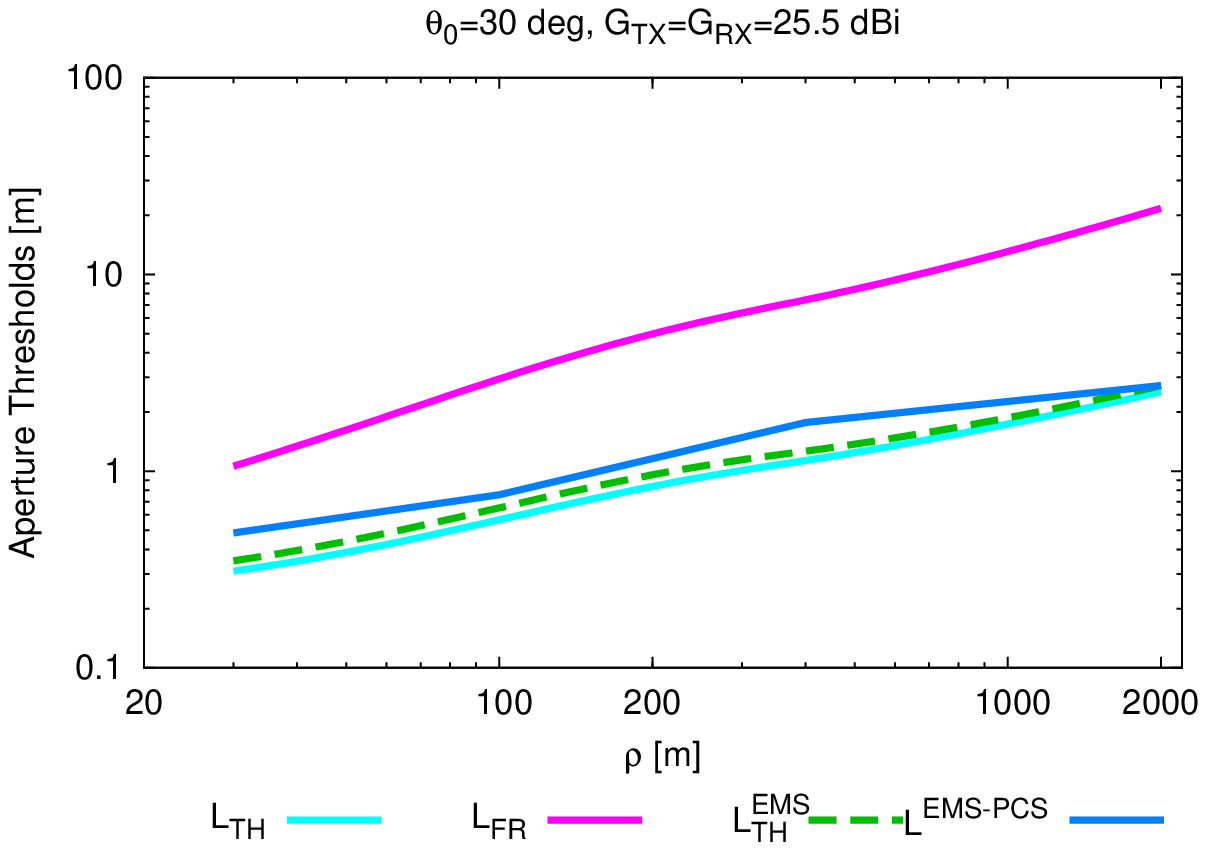}}\tabularnewline
\multicolumn{2}{c}{(\emph{b})}\tabularnewline
\multicolumn{2}{c}{\includegraphics[%
  width=0.60\columnwidth,
  keepaspectratio]{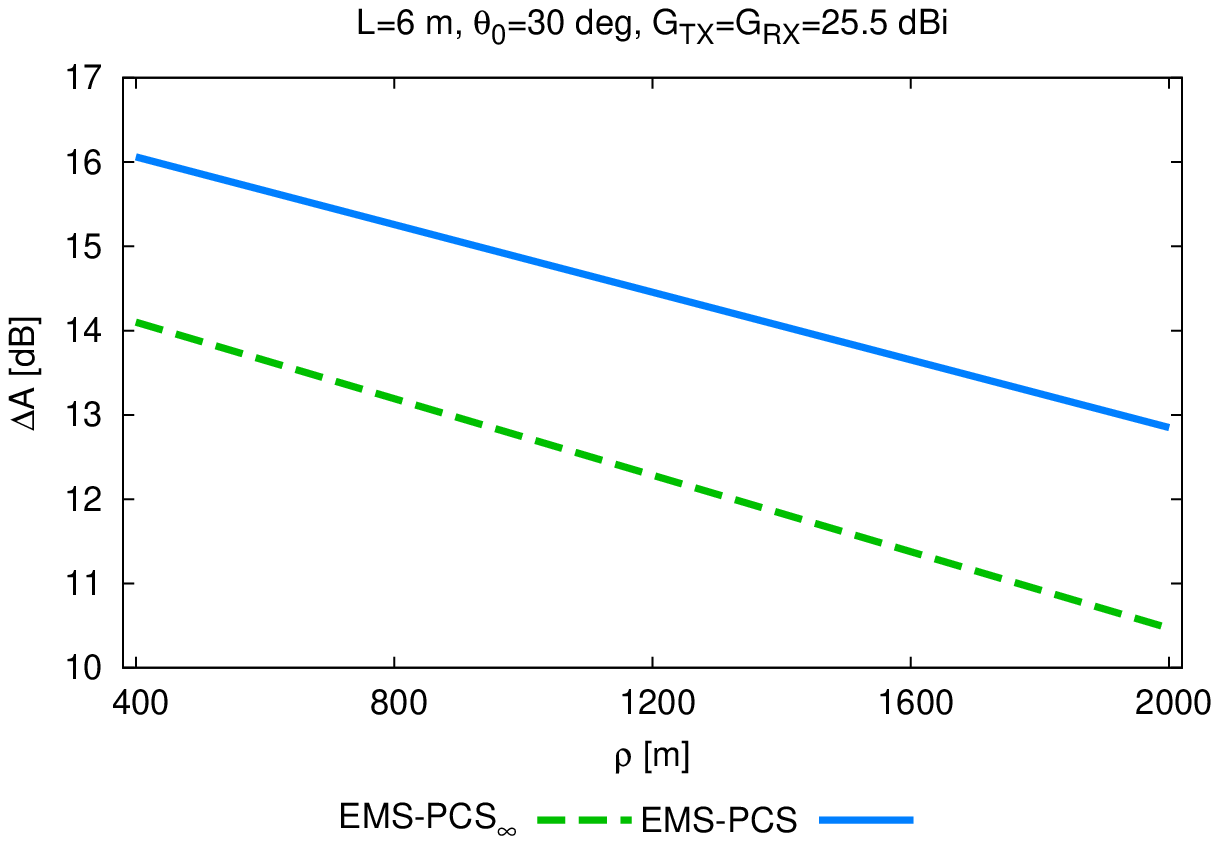}}\tabularnewline
\multicolumn{2}{c}{(\emph{c})}\tabularnewline
\end{tabular}\end{center}

\begin{center}\textbf{Fig. 16 - G. Oliveri et} \textbf{\emph{al.}}\textbf{,}
\textbf{\emph{{}``}}Features and Potentialities of ...''\end{center}

\newpage
\begin{center}~\vfill\end{center}

\begin{center}\begin{tabular}{cc}
$\rho=400$ {[}m{]}&
$\rho=2000$ {[}m{]}\tabularnewline
\includegraphics[%
  width=0.45\columnwidth]{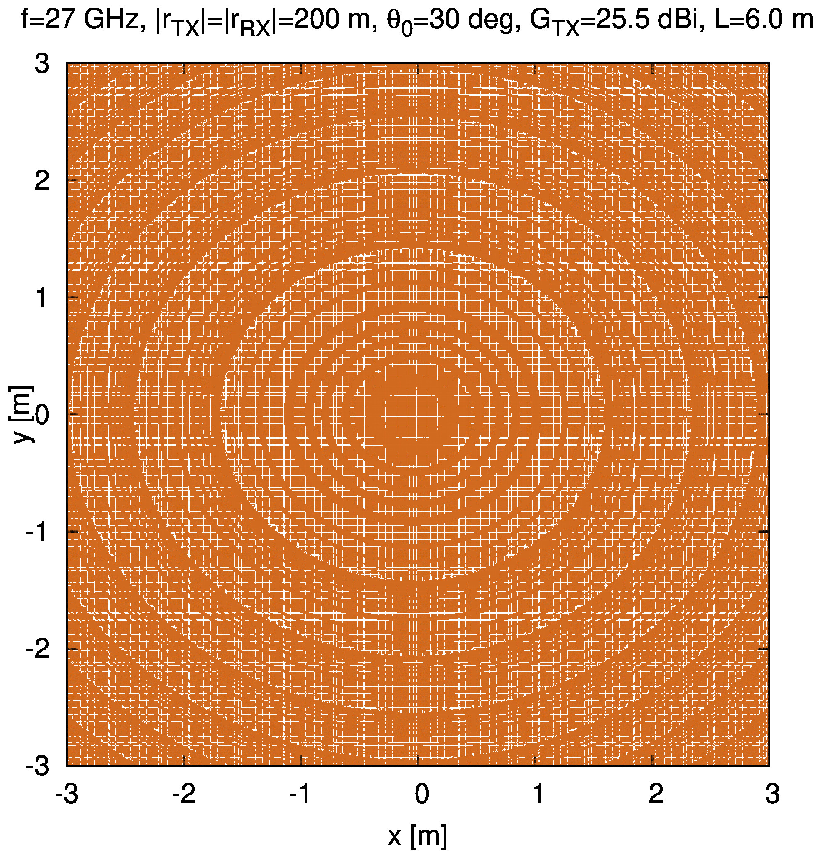}&
\includegraphics[%
  width=0.45\columnwidth]{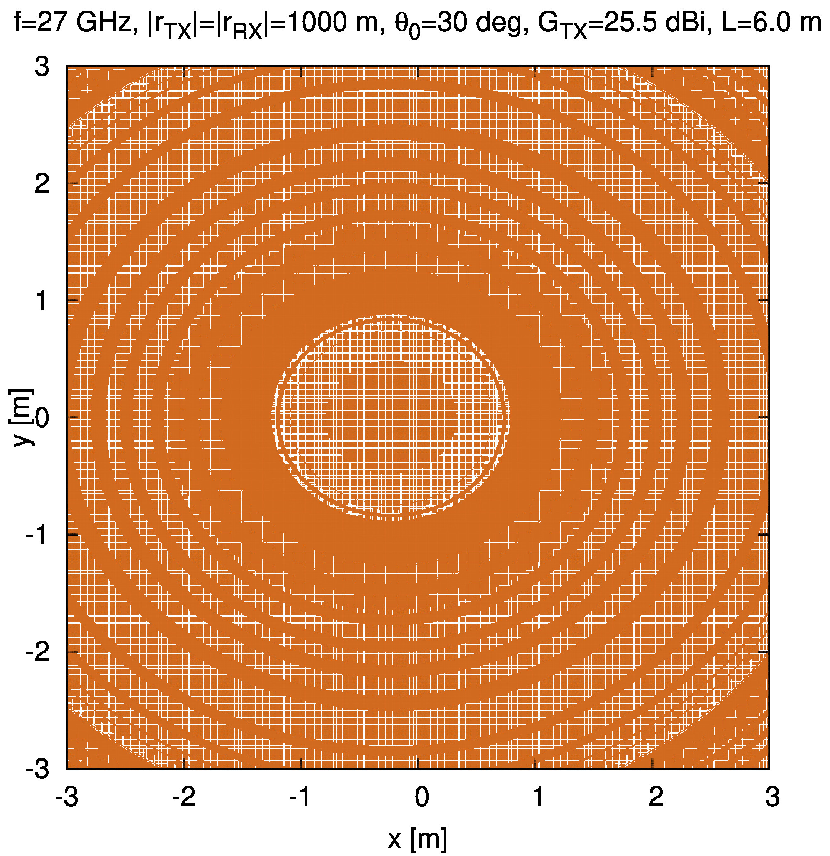}\tabularnewline
(\emph{a})&
(\emph{b})\tabularnewline
\end{tabular}\end{center}

\begin{center}~\vfill\end{center}

\begin{center}\textbf{Fig. 17 - G. Oliveri et} \textbf{\emph{al.}}\textbf{,}
\textbf{\emph{{}``}}Features and Potentialities of ...''\end{center}

\newpage
\begin{center}~\vfill\end{center}

\begin{center}\begin{tabular}{|c|c|c|}
\hline 
&
Fig. 2(\emph{b})&
Fig. 2(\emph{c})\tabularnewline
\hline
\hline 
$f$ {[}GHz{]}&
\multicolumn{2}{c|}{$27$}\tabularnewline
\hline 
$a$ {[}m{]}&
\multicolumn{2}{c|}{$8.636\times10^{-3}$}\tabularnewline
\hline 
$b$ {[}m{]}&
\multicolumn{2}{c|}{$4.318\times10^{-3}$}\tabularnewline
\hline 
$p$ {[}m{]}&
$1.3839\times10^{-2}$&
$2.0188\times10^{-1}$\tabularnewline
\hline 
$\rho_{e}$ {[}m{]}&
$2.0363\times10^{-2}$&
$2.1803\times10^{-1}$\tabularnewline
\hline 
$\rho_{h}$ {[}m{]}&
$2.4441\times10^{-2}$&
$2.2827\times10^{-1}$\tabularnewline
\hline 
$a_{1}$ {[}m{]}&
$2.8543\times10^{-2}$&
$8.7230\times10^{-2}$\tabularnewline
\hline 
$b_{1}$ {[}m{]}&
$2.1272\times10^{-2}$&
$6.9607\times10^{-2}$\tabularnewline
\hline 
$G$ {[}dBi{]}&
$15.4$&
$25.5$\tabularnewline
\hline
\end{tabular}\end{center}

\begin{center}~\end{center}

\begin{center}\vfill\end{center}

\begin{center}\textbf{Table I - G. Oliveri et} \textbf{\emph{al.}}\textbf{,}
\textbf{\emph{{}``}}Features and Potentialities of ...''\end{center}

\newpage
\begin{center}~\vfill\end{center}

\begin{center}\begin{tabular}{|c||c|c||c|c|}
\hline 
$G_{RX}$ {[}dBi{]}&
\multicolumn{2}{c||}{$15.4$}&
\multicolumn{2}{c|}{$25.5$}\tabularnewline
\cline{1-1} \cline{4-5} 
\hline 
$r_{rx}${[}m{]}&
$15$&
$50$&
$15$&
$50$\tabularnewline
\hline
\hline 
$L_{TH}$ {[}m{]}&
$0.310$&
$0.566$&
$0.310$&
$0.566$\tabularnewline
\hline 
$L_{FR}$ {[}m{]}&
$1.060$&
$2.945$&
$1.060$&
$2.945$\tabularnewline
\hline 
$L_{TH}^{EMS}$ {[}m{]}&
$0.339$&
$0.623$&
$0.350$&
$0.651$\tabularnewline
\hline 
$L_{PCS}^{EMS}$ {[}m{]}&
$0.500$&
$1.000$&
$0.485$&
$0.758$\tabularnewline
\hline
\end{tabular}\end{center}

\begin{center}~\vfill\end{center}

\begin{center}\textbf{Table II - G. Oliveri et} \textbf{\emph{al.}}\textbf{,}
\textbf{\emph{{}``}}Features and Potentialities of ...''\end{center}

\newpage
\begin{center}~\vfill\end{center}

\begin{center}\begin{tabular}{|c||c|c|c||c|}
\hline 
&
\multicolumn{3}{c||}{$L$ {[}m{]}}&
\tabularnewline
\hline
\hline 
$r_{RX}$ {[}m{]}&
$0.8$&
$1.0$&
$1.2$&
\tabularnewline
\hline
\hline 
$55$&
$8.52$&
$11.57$&
$15.30$&
\tabularnewline
$150$&
$7.31$&
$10.51$&
$13.16$&
$\Delta\mathcal{A}_{PCS_{\infty}}^{EMS}$~~{[}dB{]}\tabularnewline
$300$&
$6.91$&
$7.73$&
$12.78$&
\tabularnewline
\hline
\hline 
$55$&
$6.48$&
$14.52$&
$15.09$&
\tabularnewline
$150$&
$4.11$&
$12.75$&
$15.75$&
$\Delta\mathcal{A}_{PCS}^{EMS}$ ~~{[}dB{]}\tabularnewline
$300$&
$3.48$&
$11.47$&
$15.81$&
\tabularnewline
\hline
\end{tabular}\end{center}

\begin{center}~\vfill\end{center}

\begin{center}\textbf{Table III - G. Oliveri et} \textbf{\emph{al.}}\textbf{,}
\textbf{\emph{{}``}}Features and Potentialities of ...''\end{center}
\end{document}